\documentclass[oneside, a4paper, 11pt, openany]{book}

\usepackage[utf8]{inputenc}  

\usepackage[a-1b]{pdfx} 

\usepackage{cite}

\usepackage{acronym}
\usepackage{amsmath}
\usepackage{amssymb}
\usepackage{mathtools}
\usepackage{phaistos}
\usepackage{graphicx}
\usepackage{subcaption}
\usepackage[section]{placeins}
\usepackage{enumitem} 
\usepackage{etoolbox}
\usepackage{gensymb} 
\usepackage{bm} 
\usepackage{setspace} 
\usepackage{pdfpages}  

\usepackage{bibentry} 

\usepackage{floatpag}  
\floatpagestyle{empty}  

\usepackage{physics} 

\usepackage{hyperref} 

\usepackage{xspace}

\usepackage{cancel} 

\usepackage{comment}

\AtBeginEnvironment{align}{\setcounter{subeqn}{0}}
\newcounter{subeqn} \renewcommand{\thesubeqn}{\theequation\alph{subeqn}}%
\newcommand{\subeqn}{%
  \refstepcounter{subeqn}
  \tag{\thesubeqn}
}
\newcommand{\beginsubeqn}{\setcounter{subeqn}{0}\refstepcounter{equation}\subeqn}
\newcommand{\nn}{\nonumber}

\newcommand{\lcdm}{$\Lambda${}CDM}
\newcommand{\Msun}{M_\odot}

\newcommand{\mchirp}{\mathcal{M}_c}

\newcommand{\dm}{\text{dm}}
\newcommand{\orb}{\text{orb}}
\newcommand{\typeI}{\text{Type--I}}
\renewcommand{\b}{\text{b}}
\newcommand{\s}{\star}
\newcommand{\esc}{\text{esc}}
\newcommand{\df}{\text{df}}
\newcommand{\isco}{\text{isco}}
\renewcommand{\sp}{\text{spike}}

\newcommand{\E}{\mathcal{E}}
\newcommand{\F}{\mathcal{F}}
\renewcommand{\H}{\mathcal{H}}
\renewcommand{\v}[1] {\mathbf{#1}}
\renewcommand{\d}{\mathrm{d}}
\DeclareMathOperator{\arctantwo}{arctan2}

\DeclarePairedDelimiterX{\avg}[1]{\langle}{\rangle}{#1}
\makeatletter
\let\oldavg\avg
\def\avg{\@ifstar{\oldavg}{\oldavg*}}  
\makeatother

\renewcommand{\eqref}[1]{Eq.~(\ref{#1})}

\newcommand{\lib}[1]{\texttt{#1}}
\newcommand{\figref}[1]{Fig.~\ref{#1}}

\usepackage{scalerel,stackengine,amsmath}
\newcommand\equalhat{\mathrel{\stackon[1.5pt]{=}{\stretchto{%
    \scalerel*[\widthof{=}]{\wedge}{\rule{1ex}{3ex}}}{0.5ex}}}}

\author{Niklas Becker}
\title{Gravitational Wave Signatures of Dark Matter}

\nobibliography* 

\begin{document}
\frontmatter

\begin{titlepage}

\center 

{ \huge \bfseries \linespread{1.}\selectfont  Dancing above the abyss: Environmental effects and dark matter signatures in inspirals into massive black holes \par } 
\vfill
{\large Dissertation \\ zur Erlangung des Doktorgrades \\ der Naturwissenschaften}
\vfill
{\large vorgelegt beim Fachbereich 13 \\ der Johann Wolfgang Goethe-Universität \\  in Frankfurt am Main}
\vfill
{\large von \\ Niklas Robin Becker \\ aus Hagen}

\vfill
{\large Frankfurt (2024)}

\vfill 
\end{titlepage}

{\large Vom Fachbereich 13 der }\par 
\vfill\vspace{-5cm}
{\large Johann Wolfang Goethe-Universität als Dissertation angenommen}
\vfill
{\large Dekan: }
\vfill\vspace{-5cm}
{\large Gutachter: }
\vfill\vspace{-5cm}
{\large Datum der Disputation: }
\vfill\vspace{-3cm}

\pagenumbering{roman}
\newpage 

\includepdf{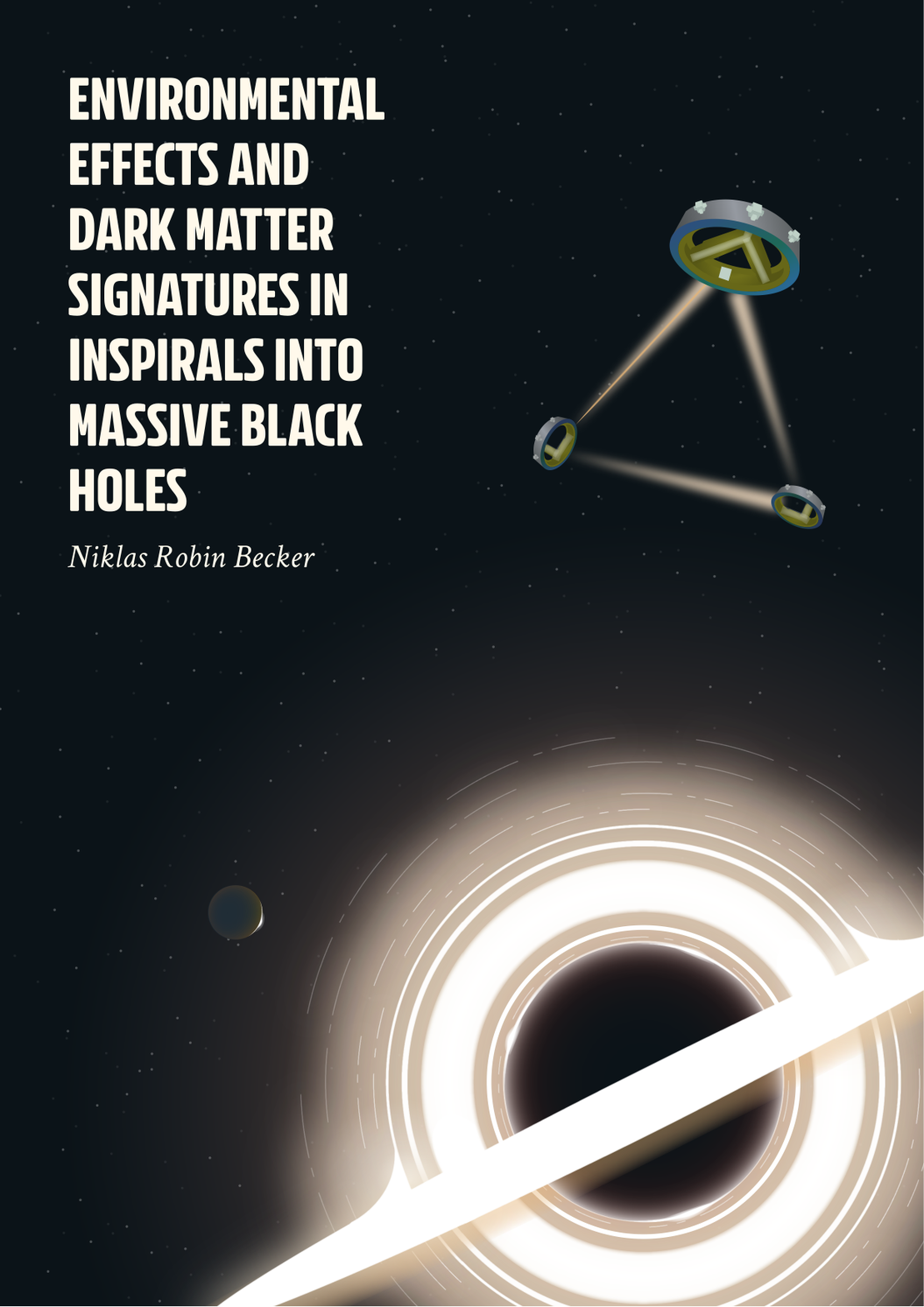}

\makebox[\linewidth]{\textbf{Abstract}} \\
In this dissertation, we look at environmental effects in extreme and intermediate mass ratio inspirals into massive black holes. In these systems, stellar mass compact objects orbit massive black holes and lose orbital energy due to gravitational wave emission and other dissipative forces. We explore environmental interactions with dark matter spikes, stellar distributions, accretion disks, and combine and compare them. We discuss the existence and properties of dark matter spikes in the presence of these environmental effects. The signatures of the environmental effects, such as the phase space flow, dephasing, deshifting of the periapse, and alignment with accretion disks, are examined. These signatures are quantified in isolated spike systems, in dry, and in wet inspirals. We generally find dark matter effects to be subdominant to the other environmental effects, but their impact on the waveform is still observable and identifiable. Lastly, the rates of inspirals and the impact of spikes are estimated. All of these results are obtained with the help of a code \lib{imripy} that is published alongside. If dark matter spikes exist, they should be observable with space-based gravitational wave observatories.
\newpage 

\makebox[\linewidth]{\textbf{Acknowledgements}} \\ 
{
First, I would like to thank my supervisor, Laura Sagunski, for giving me the opportunity to complete this PhD as part of her group. Thank you for all the exciting discussions, for the opportunities to visit schools and conferences, for your support throughout the process, and most importantly, for your unwavering and infectious enthusiasm. Thank you to J\"urgen Schaffner-Bielich as the co-supervisor of my dissertation as well.

I would like to thank my colleagues, Robin, Edwin, Daniel, and Philipp, for proofreading and giving helpful comments on this dissertation and their general existence. I also want to thank the rest of the working group, including Tamara, C\'edric, Yannik, Lea, and Jannis. It was a lot of fun working alongside all of you. We had insightful discussions, hilariously silly discussions, and overall good company. I hope we can rock more karaoke bars.

I want to thank all the people in academia who have answered not all but many of my numerous questions and given guidance. Thank you to Bradley Kavanagh, Yuri Levin, Andrea Derdzinski, Samuel Tootle, Peter Kloeden, Anton Wakolbinger, Patrick Kidger, and Ben Bar-Or. And thank you to all the people I have talked with at schools and conferences, but whose names have escaped me. I guess this is the toll of a PhD.

I would like to thank `science youtube', all the channels that provide popular science content, such as kurzgesagt, PBS Spacetime, 3Blue1Brown, Veritasium, Dr.Becky, Physics Girl, and SmarterEveryDay. Sometimes, academia can get lost in the details, but you make me see the bigger picture and keep me amazed at the world around me. Get well soon Dianna! 

I am immensely grateful to all my friends who have been part of this journey. To the old ones I do not get to see enough, and the new ones I made in Frankfurt. Thank you to my fellow students back in Aachen, our discussions about the state of academia make me feel like I'm not alone, and our vacations are balm for the soul. Thank you to my old flatmates, you guys were family during the lockdowns. 
Thank you to the Bachata community and the wonderful dances, giving inspiration for the title.
Thank you to Irene, especially for designing the title page. Thank you to my partners who have accompanied me for some of the way.
Thank you to all the wonderful people I have met who have taught me so much about friendship and love and myself.

Most importantly, I want to thank my family. I have had their unwavering support for all my life, without them, I would not be here. Even though my passion for physics came as a surprise, they have done nothing but foster it. This is for you guys.}

\newpage

\textbf{Publications relevant to this dissertation}

\begin{itemize}
    \item[\cite{Becker:2021ivq}] \bibentry{Becker:2021ivq}

    The effect of the phase space distribution of DM on the eccentricity evolution is examined. Contrary to previous findings, we see circularization due to the presence of steep spikes. We argue that the eccentricity vs semimajor axis evolution can also be a signature of DM. The implementation and writing of the publication has been performed by the author of this dissertation and cross-checked by the co-authors.
    
    In this dissertation, we expand on this idea with the help of the phase space flow.  

    \item[\cite{Becker:2022wlo}] \bibentry{Becker:2022wlo}

    The environmental effects of DM and accretion disks are compared. The different models are found to be dominant at different scales. We apply the braking index and define the dephasing index to argue that the environmental effects can be differentiated and identified.

    This is the basis for the theoretical considerations presented in the latter part of section \ref{sec:inspiral:GWsignal}, as well as some of the results presented in section \ref{sec:signatures:wet}, where we expand on this idea and include more environmental effects.
    
    \item[\cite{2023ascl.soft07018B}] \bibentry{2023ascl.soft07018B}

    The code used to obtain all the results presented in this dissertation, along with documentation and examples. It was built to be modular and easily extendable. 
    
\end{itemize}

\textbf{Publications unrelated to this dissertation}
\begin{itemize}
    \item[\cite{Diedrichs:2023trk}] \bibentry{Diedrichs:2023trk}

    \item[\cite{Genoud}] \bibentry{Genoud}
\end{itemize}

\setcounter{tocdepth}{1}
\tableofcontents

\newpage 
\mainmatter

\chapter{Introduction\label{chap:intro}}
The era of gravitational wave physics and multi-messenger astronomy has begun. The LIGO-Virgo-KAGRA (LVK) collaboration has observed dozens of compact binary coalescenses\cite{KAGRA:2021vkt}. Already, the observation of black hole and neutron star mergers is transforming our understanding of phyics, in the astrophysical sector, cosmology and the particle physics domain. The members of the  International Pulsar Timing Array (IPTA) have tentatively measured a stochastic gravitational wave background\cite{NANOGrav:2023gor}. This will allow us to explore the early history of the universe, the formation of structures, and cosmology. Once the space-based interferometers are launched, they can explore galactic cores, galaxy mergers, and a multitude of physical systems\cite{LISA:2022kgy}.

In both astrophysics and particle physics, dark matter has been one of the most pressing mysteries for decades. While astrophysical and cosmological probes give observational evidence, its underlying mechanisms or particle nature are unknown. The whole new field of astroparticle physics has spawned, but an immense worldwide effort in detection has so far remained unfruitful\cite{Bertone:2016nfn}. 

According to some hypotheses, dark matter could be found abundantly around massive black holes. As massive black holes grow, they can concentrate dark matter into \textit{spikes} around them\cite{Gondolo:1999ef}. These black holes typically reside in the centers of galaxies, and are some of the most extreme environments found in the universe. When a smaller compact object inspirals into these much larger massive black holes, forming an intermediate or extreme mass ratio inspiral (I/EMRI), it will be in the observable range of space-based interferometers, such as the Laser Interferometer Space Antenna (LISA)\cite{Babak:2017tow}. In these systems, the dark matter can affect the inspiral and leave an imprint on the gravitational waveform\cite{Eda:2014kra}. As these inspirals can be observable for months or years, even subtle effects can be measurable. Alas, dark matter is not the only thing to be found in these extreme environment. Other environmental effects include an accretion disk or a distribution of stars, as observed in the center of our Milky Way. Their presence also affects the inspiral and to track a months-long signal, accurate waveforms are essential\cite{Barausse:2014tra, Zwick:2022dih}.

Therefore, to detect dark matter, we need to disentangle all the environmental effects. Only once we understand the relevant astrophysical impacts, can we attribute the remainder to dark matter effects. While the environmental effects have been explored on their own in these mass ratio inspirals\cite{Cole:2022fir}, in this dissertation, for the first time, we combine different environmental effects and compare their impact. We look at isolated spikes, dry, and wet inspirals. Dry inspirals happen inside a stellar distribution, typically on highly eccentric orbits, while wet inspirals happen inside an accretion disk. We explore the phase space evolution, the dephasing of the waveform, (de-)shifting of the periapse angle, and alignment with accretion disks as possible indicators of dark matter. 

We find dark matter effects to be generally subdominant in these astrophysical environments, especially for physically motivated flat spikes. But there is a region in parameter space where the effects of dark matter effects should be visible within the lifetime of LISA. We discuss the current state of research, and try to give directions for future improvements. The formalism and tools developed here are applicable to any environmental effect and can be abstracted away. The code that is used to obtain these results is published alongside and can be easily adapted to look at other effects\cite{imripy}.

The structure of this dissertation is as follows: In chapter \ref{chap:background}, we summarize the relevant background for dark matter, gravitational waves, massive black holes, and introduce the concept of stochastic differential equations. In the following chapter \ref{chap:inspiral}, we present the equations and modeling of the mass ratio inspiral. In chapter \ref{chap:environment}, we discuss the environmental effects of dark matter spikes, stellar distributions, accretion disks, and relativistic effects. In chapter \ref{chap:signatures}, we show the possible signatures of dark matter in different scenarios, in an isolated dark matter spike, in a dry inspiral, and in a wet inspiral. In chapter \ref{chap:rates}, we estimate the rates of mass ratio inspirals that are observable with LISA and the impact of dark matter. Lastly, in chapter \ref{chap:discussion}, we discuss the results presented, place them into context with active research, and give an outlook.

\subsubsection{}
Throughout this dissertation, we use geometrized units with $c=G=1$.

\chapter{Theoretical Background \label{chap:background}}
\section{Dark Matter}
Dark Matter (DM) has been a mystery for a long time in astrophysics and cosmology. 
First postulated by Fritz Zwicky as a non-luminous mass to hold together galaxy clusters\cite{Zwicky:1933gu, Zwicky:1937zza}, it was later found to be necessary to explain galactic rotation curves\cite{Rubin:1970zza}, gravitational lensing\cite{Narayan:1996ba}, structure formation\cite{Huterer:2022dds}, and the early time in the universe\cite{Planck:2018vyg}. Initially, the phrase was intended to mean non-luminous stars or gas, basically baryonic matter but too dark to see. But over time, the meaning shifted to a new species of matter and it gave birth to a whole new field, astroparticle physics\cite{Bertone:2016nfn, deSwart:2017heh}.  

Today, there is a plethora of DM models, from standard model inspired weakly interacting massive particles (WIMPs)\cite{Arcadi:2017kky}, to QCD inspired axions\cite{Kim:2008hd}, and primordial BHs (PBHs)\cite{ParticleDataGroup:2022pth, HerreraMoreno:2023lvm}. Alternatively, there are models that modify the theory of gravity such as Modified Newtonian Gravity (MOND)\cite{Famaey:2011kh} or $f(R)$ gravity that can fix the missing mass problem in another way\cite{Sotiriou:2008rp}.

The current standard model of cosmology \lcdm{} invokes Cold Dark Matter (CDM), a massive, non-relativistic particle to make up DM. \lcdm{} can reproduce a lot of current observations remarkably well, but it does have some problems\cite{Perivolaropoulos:2021jda}.
It predicts that the current energy budget of the universe is made up of $\sim 23\%$ DM, compared to $\sim 5\%$ of baryonic matter, the rest being in the form of dark energy (DE), an even more elusive concept\cite{Planck:2018vyg, Tristram:2023haj, Martin:2012bt}. Even if there is $5$ times more DM than baryonic matter, there has been no direct detection of any DM particle. Either it does not interact with baryonic matter at all or too weakly to be seen\cite{ParticleDataGroup:2022pth}.

DM is one of the greatest mysteries of current astrophysics and cosmology. Here, we will quickly review the most important evidence, necessary properties, and proposed models.

\subsection{Astrophysical Evidence}
There is an overwhelming amount of evidence for an additional mass throughout the universe, starting from galactic scales and up to cosmological scales. We review the evidence in order of increasing scale following \cite{HerreraMoreno:2023lvm,Scarcella:2023voq}.

\subsubsection{Galactic Rotation Curves}
\begin{figure}
    \centering
    \includegraphics[width=0.5\textwidth]{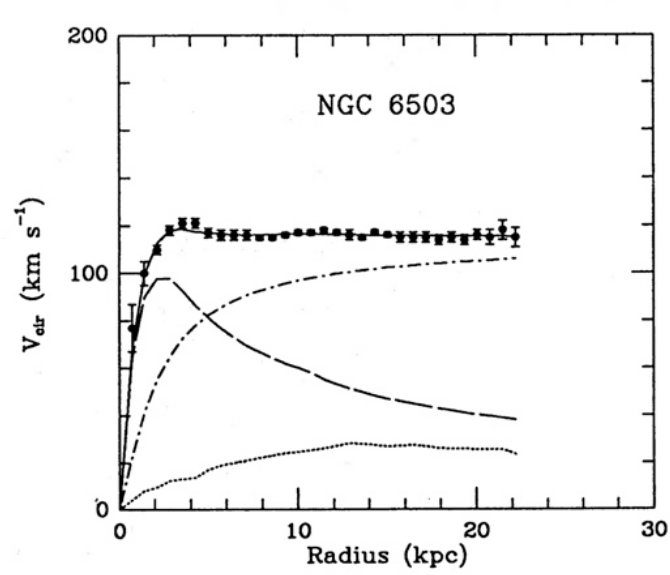}
    \caption{Measurements of the galactic rotation curve of NGC 6503. The dotted, dashed, and dash-dotted lines refer to the contributions of gas, disk, and DM, respectively. From \cite{Begeman:1991iy}.}
    \label{fig:rotation_curve}
\end{figure}
Observations show that the rotation curves of galaxies are flat and extend far further than the bulk of the visible mass\cite{Rubin:1985ze, 1973A&A....26..483R}. Following a simple Newtonian argument, the rotational velocity $v$ at a given galactic radius $r$ should depend on the enclosed mass $M(r)$ and read $v = \sqrt{M(r)/r}$. Once the bulk of the mass is included, the rotational velocity should drop as $v\sim r^{-1/2}$. Flat rotation curves (i.e. $v\equiv$ const) far past the bulk of the luminous matter require a mass profile that grows linearly $M(r)\sim r$ in the outskirts of the galaxy.

This can be explained by the presence of DM halos, made up of collisionless particles. These would have a distribution of $\rho \sim r^{-2}$, resulting in the mass profile needed. This can extend up to $10$ times the size of the galaxy and make up about $5$ times the mass of the luminous matter. An example of this is shown in \figref{fig:rotation_curve}.

\subsubsection{Galaxy Clusters}
This was the first indication for DM as derived by Zwicky. He used the virial theorem to relate the potential energy $U$ and average kinetic energy $T$ as $2T= -U$ of the galaxies inside the Coma cluster. The potential energy can be estimated as $\abs{U} = \frac{M^2}{R}$, the average kinetic energy as $T = \frac{3}{2}M \avg{v_\parallel^2}$, where R is the radius of the cluster, and $v_\parallel$ the tangential velocity of the galaxies. Measuring the tangential velocity, Zwicky found that the required mass is about 10 times larger than the visible matter\cite{Zwicky:1937zza}. 

This initial estimate was later supported by gravitational lensing. The presence of mass deflects the trajectories of photons, therefore measuring the deflection allows one to infer the mass\cite{Narayan:1996ba}.

\begin{figure}
    \centering
    \includegraphics[width=0.5\textwidth]{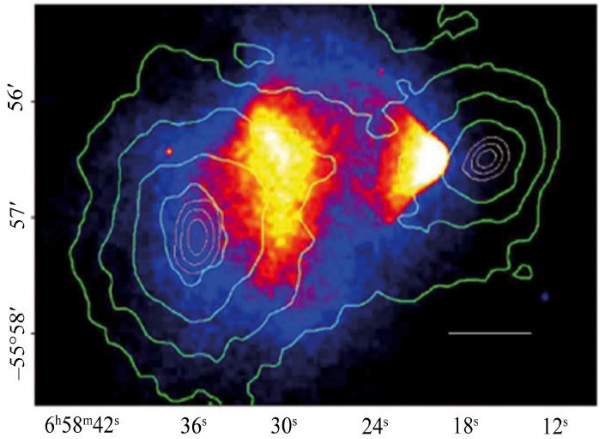}
    \caption{The combined mapping of the mass and thermal emission of the Bullet Cluster\cite{Clowe:2006eq}. The green contours show the reconstruction of the mass profile through gravitational lensing, the color gradient is the map of thermal X-ray emissions. The centers of the two distributions do not line up, hinting at a collisional and collisionless component. }
    \label{fig:bullet_cluster}
\end{figure}

Another hint for DM comes from the observation of the Bullet Cluster\cite{Clowe:2006eq}, see \figref{fig:bullet_cluster}. Here, two clusters of galaxies have collided and moved through each other. Reconstructing the mass profile shows that most of the matter remains inside the individual clusters, while thermal emissions show that the baryonic gas remains in the collision region. This also gives a hint that most of the mass is made up of collisionless particles.

\subsubsection{Large Scale Structure}
Simulating the formation of large scale structure with a DM component reproduces its observed properties very well. There have been large N-body simulations that show structure formation happens from the bottom-up. Here, initial perturbations in the distribution of DM are the seeds for the first structures, as the distribution of baryonic matter follows the dominant DM distribution\cite{Angulo:2012ep}. These then coalesce into the large cosmic web. Without DM, our models would not reproduce the observations. For a review on structure formation see \cite{Huterer:2022dds, Wechsler:2018pic}.

\subsubsection{Cosmic Microwave Background}
\begin{figure}
    \centering
    \includegraphics[width=\textwidth]{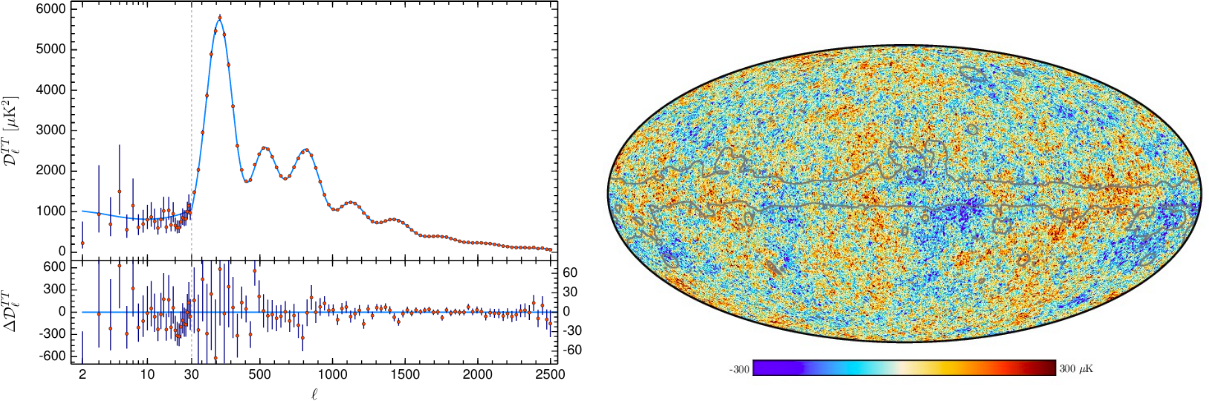}
    \caption{The fit to the power spectrum of the two-point correlation function the temperature perturbations of the CMB\cite{Planck:2018vyg}. The shape of the spectrum reveals the properties of the primordial baryon-photon fluid and the DM content. }
    \label{fig:cmb}
\end{figure}

In the \lcdm{} model, the early universe was dominated by radiation. Late during radiation domination, density perturbations oscillate due to the interplay of gravity and pressure inside the photon-baryon fluid. These perturbations can grow until the universe cools down enough such that the first atoms can form and the universe becomes transparent, the period of recombination. The light that has traveled throughout the universe ever since has redshifted and now forms the cosmic microwave background (CMB). Analyzing its statistical properties and two-point correlation functions gives information about the primordial photon-baryon fluid. The presence of DM is vital to explain the observed structure of the correlation functions, as it depends on the amount of gravitating fluids in the primordial plasma\cite{Planck:2018vyg}.

The measurement of the CMB gives the most accurate measurement of DM on cosmological scales\cite{Planck:2018vyg}. This gives a DM abundance today 
\begin{equation}
    \Omega_\dm h^2 = 0.120\pm 0.001,
\end{equation}
where $h$ is the reduced Hubble constant $h=67.4\pm0.5$, which is about 5 times higher than the baryon abundance today
\begin{equation}
    \Omega_b h^2 = 0.0224 \pm 0.0001.
\end{equation}
Here, $\Omega_\text{b,dm}$ represents the energy density of the species $\Omega= \frac{\rho}{\rho_c}$ with regard to the critical density of the universe, $\rho_c = 3H_0^2/8\pi$. This is in accordance with the DM content at smaller scales.

\subsection{Properties}
With all these observations it is possible to infer some basic properties of DM. To fulfill its role, DM needs to be massive, stable throughout the age of the universe, mostly collisionless, and barely interacting with baryonic matter. Of course, one can cook up models that circumvent these requirements, or assume the existence of an extended dark sector with several distinct species. But generally, and to invoke Occam's razor, a successful candidate must meet these requirements. See also \cite{Taoso:2007qk} for a review.

\textbf{Stable}. To both explain CMB observations (originating at redshift $z\sim 1100$) and today's galactic rotation curves, DM must be stable throughout the lifetime of the universe.

\textbf{Cold}. DM must be cold in the sense that its constituents are non-relativistic and have low average kinetic energy. Relativistic (and, to a degree, warm) DM has a larger free-streaming length that can suppress structure formation, washing out small-scale structure\cite{Benson:2012su}.

\textbf{Collisionless.} To be compatible with structure formation and the Bullet cluster observation, DM needs to be mostly collisionless. In \lcdm{} it is modeled as a collisionless fluid. There are models of self-interacting DM (SIDM) for which observations can place bounds on the cross-section $\sigma$. These are in the range of $\sigma / m <$ cm$^2$/g. SIDM models can mostly alter the properties of small-scale structure in the universe, see \cite{Tulin:2017ara} for a review.

\textbf{Dark}. While it does not interact greatly with itself, it also cannot interact significantly with baryonic matter, especially electromagnetically. There is a multitude of direct detection experiments that can constrain many interaction channels of DM\cite{ParticleDataGroup:2022pth}, as well as cosmological observations\cite{Becker:2020hzj}.

\textbf{Mass range}. Assuming a particle nature, the de Broglie wavelength can help constrain the DM mass range. The typical scale should be $\lambda \lesssim 1$kpc, smaller than dwarf-spheroidal galaxies. Paired with a velocity dispersion of $\sim100$km/s, one can obtain a lower bound on the mass $m>10^{-22}$eV. In this range, DM would exhibit collective quantum properties. For fermionic particles, the Tremaine-Gunn bound excludes particles with mass $\leq 1$MeV\cite{Tremaine:1979we}. An upper limit can be given by constraints on massive compact halo objects (MACHOs) to $\sim 10^{-7}\Msun$\cite{Brandt:2016aco}.

\subsection{Models}
None of the particles known in the standard model (SM) have the aforementioned properties. Therefore, if DM is a particle, it needs to extend the standard model. There is a multitude of proposed models for DM.  Here, we quickly want to summarize the most popular ones\cite{Profumo:2019ujg, HerreraMoreno:2023lvm}.

\subsubsection{WIMPs}
WIMP stands for weakly interacting massive particle. As the name suggests, it interacts only through the weak interaction, is stable, and has a mass around the electroweak scale ($\sim 200$GeV). They rose to prominence as part of the \textit{WIMP miracle}, as the relic abundance from thermal decoupling of the weak scale corresponds to the abundance of DM. It also appears naturally in supersymmetric extensions of the SM. But a decades long search has proven unfruitful in collider and direct detection experiments\cite{Arcadi:2017kky}, and the model has since fallen out of favor.

\subsubsection{Sterile Neutrinos}
In the SM, neutrinos do not have mass. To fix this, a heavy Majorana partner to the neutrinos can be assumed, which explains the small masses with the see-saw mechanism. These sterile neutrinos can be DM candidates\cite{Dodelson:1993je, Drewes:2016upu, Boyarsky:2018tvu}. They can decay into three neutrinos, or a neutrino and a photon. The photon emission would be monochromatic, and the absence of such an astrophysical observation allows for tight constraints\cite{Horiuchi:2013noa}. However, the model can be tweaked to evade current constraints\cite{Bringmann:2022aim}.

\subsubsection{Axions}
According to the SM, neutrons should have an electric dipole moment, due to a CP violating term in the QCD Lagrangian. However, all observations show that CP is not violated in QCD, resulting in the strong CP problem. To mitigate this, Peccei \& Quinn have proposed an additional global $U(1)$ symmetry with a complex scalar field that, with spontaneous symmetry-breaking, nullifies the offending terms\cite{Peccei:1977hh}. This process also results in a pseudo-Goldstone boson, termed the axion\cite{Weinberg:1977ma}. 
Similar extensions of the SM are usually called axion like particles (ALPs) and have been investigated as potential DM candidates. The parameter space remains largely unconstrained for light axions, with obervational bounds giving a mass $< 1$eV\cite{Irastorza:2018dyq}.

\subsubsection{Primordial Black Holes}
As BHs meet most of the criteria described above, they have been a longterm candidate for DM. This requires their existence in the early universe, possibly formed during the collapse of overdensities in the primordial plasma\cite{Chapline:1975ojl}. These models are called Primordial BH (PBH). As BHs experience Hawking radiation, they need a minimal size to be stable throughout the universe's lifetime, giving a minimal mass for PBHs. A large chunk of the heavier mass parameter space has been ruled out by microlensing observations and missing accretion signatures\cite{Carr:2016drx}, but asteroid sized PBH are not ruled out as of yet. Even if they do not constitute all of DM, their existence has intriguing consequences, and has been used recently to explain a BBH merger with a mass gap BH\cite{Carr:2020xqk}.

\subsection{\lcdm}
The current standard model of cosmology, \lcdm{}, invokes the existence of two \textit{dark} components, dark matter and dark energy. It uses cold DM (CDM), a non-relativistic, collisionless fluid. $\Lambda$ refers to the cosmological constant, playing the role of dark energy and causing an accelerated expansion of the universe. Together with baryonic matter, radiation, and assuming an inflationary period in the early universe, the model aims to explain all of cosmic history.

Its most striking success has been the measurement and explanation of the CMB\cite{Planck:2018vyg}. The measurement of the large scale structure also gives hints for the Baryon Acoustic Oscillation (BAO) in the early universe, starting the era of "precision cosmology".

The model is not without fault though. The largest problem is that of the Hubble tension, which has been termed a \textit{crisis in cosmology}\cite{Abdalla:2022yfr}. Briefly, the current expansion rate of the universe, quantified by the Hubble parameter, can be inferred from the distance and redshift measurements of type IA supernovae. These "local" measurements do not agree with the number predicted by the best model fit to the cosmological probes (CMB, BAO, BBN). The SH0ES measurement of the aforementioned supernovae gives $H_0 = 73.04 \pm 1.04$km/s/Mpc\cite{Riess:2021jrx}, while the best fit value of the Planck team gives $H_0 = 67.4 \pm 0.5$km/s/Mpc\cite{Planck:2018vyg}. These are in disagreement at more than $5\sigma$. Many solutions have been proposed, but none has conclusively been accepted\cite{DiValentino:2021izs}.

Other problems relate to the growth of structure, quantified by $\sigma_8$. This is the rms variance of the linear density field smoothed on a scale of $8$Mpc/h, loosely speaking the amount of clustering on these scales. Larger $\sigma_8$ means stronger clustering of matter on these scales. The parameter that observations can constrain is $S_8 = \sigma_8 \sqrt{\Omega_m/0.3}$. Weak lensing measurements give $S_8=0.745\pm0.039$\cite{Hildebrandt:2016iqg}, while the best fit value of Planck gives $S_8=0.834 \pm 0.016$\cite{Planck:2018vyg}, which disagree at more than $2\sigma$. For a review of problems in \lcdm{} see \cite{Perivolaropoulos:2021jda}.

More recently, the observations of the James Webb Space Telescope (JWST) have revealed early massive structures that might be in tension with \lcdm{} as well\cite{Coley:2023jju, Silk:2024rsf}.

We will primarily focus on the DM description of \lcdm{} in this dissertation, since it is the prevalent model. Most of the techniques here are applicable to other DM models and can be used for further study. In this dissertation, we will look for DM around massive black holes.

\section{Black Holes}
Black Holes (BHs) are another intriguing prediction of GR. It took a long time for physicists to think them real and their existence to be observationally supported. But as of today, we have imaged two supermassive BHs, one at the center of M87 and one at the center of our own galaxy\cite{EventHorizonTelescope:2019dse, EventHorizonTelescope:2022wkp}, and seen the inspiral signal of about a hundred binary BH (BBH)\cite{LIGOScientific:2016aoc, KAGRA:2021vkt}. Additionally, there is kinematic evidence\cite{Gillessen:2008qv} and X-ray observations in support of BHs\cite{Celotti:1999tg}.

BHs are actually quite simple objects. According to the no-hair theorem, they can be described by $3$ properties: mass, spin, and charge. In an astrophysical environment, we can expect charge neutrality, leaving mass and spin the defining factors of BHs, described by the Kerr metric. Any deviations or perturbations are quickly radiated away, leaving all BHs with the same parameters to have the same properties. The spacetime around these BHs is complex and has interesting features. Of course, there is the event horizon, beyond which there is no return to the outside. For a spinning BH there are also frame dragging effects, where the spacetime itself rotates around the BH and forces all objects to orbit in the same direction. For a review of the properties of (astrophysically relevant) BHs, see \cite{Hughes:2005wj}.

Their existence being certain, the next question is their abundance and mass distribution in the universe. We will quickly summarize the different BH populations.

\subsection{Stellar Mass Black Holes}
Stellar Mass Black Holes (sBH) are created primarily through stellar collapse at the end of their lifetime. These generally have a mass $5-100\Msun$. There have been several candidate detections of sBHs through their X-ray emissions or as part of a binary system\cite{Ozel:2010su,El-Badry:2022zih}. At the same time, LVK is mapping out the distribution of sBHs and their mergers\cite{LIGOScientific:2018mvr, LIGOScientific:2021djp}. 


\subsection{Supermassive Black Holes}
\begin{figure}
    \centering
    \includegraphics[width=0.5\textwidth]{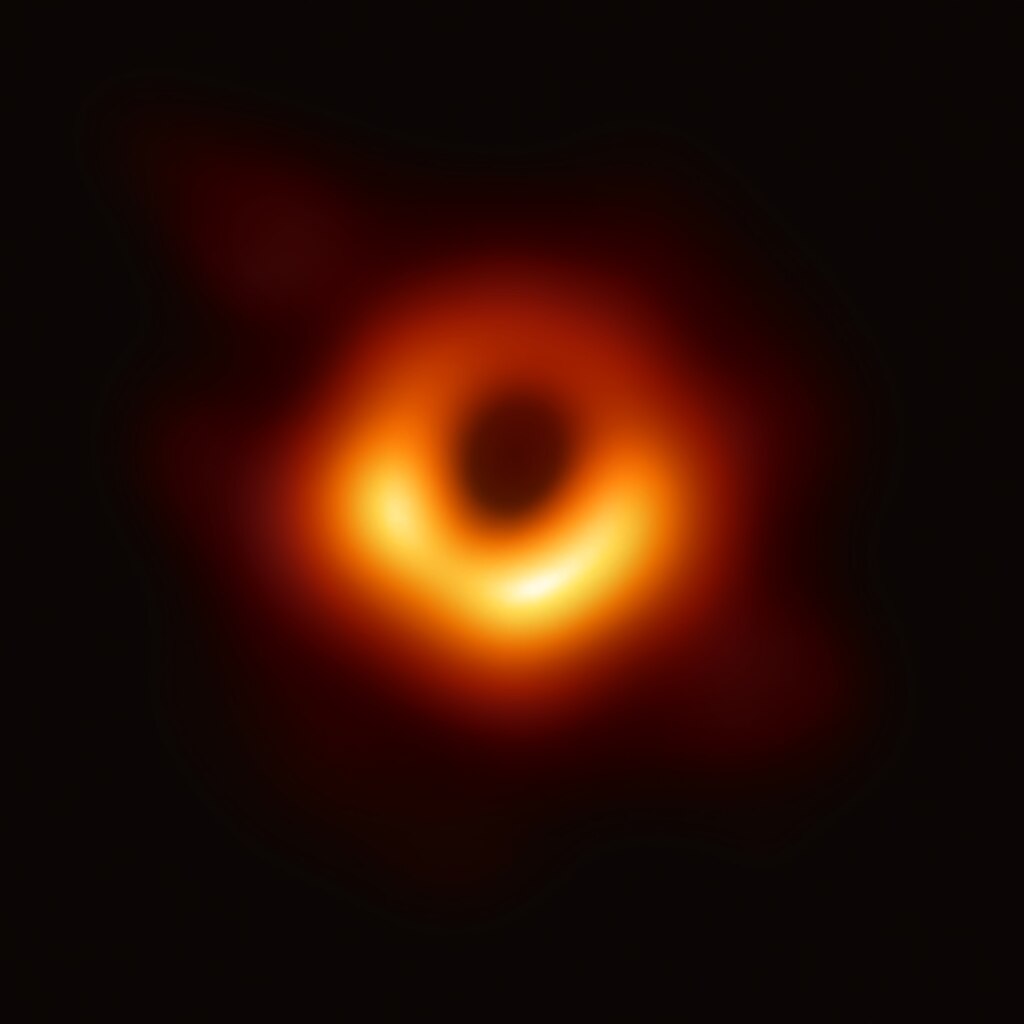}
    \caption{The reconstructed image of the SMBH M87*\cite{EventHorizonTelescope:2019dse}.}
    \label{fig:M87}
\end{figure}
There is hard evidence for the existence of supermassive Black Holes (SMBHs). We have imaged two SMBHs with very-long baseline interferometry\cite{EventHorizonTelescope:2019dse, EventHorizonTelescope:2022wkp}, see \figref{fig:M87}. And we have kinematic evidence for an SMBH at the center of the Milky Way\cite{Genzel:2003cn}. They are believed to reside in the centers of most galaxies and power Active Galactic Nuclei (AGNs)\cite{Padovani:2017zpf}. They are generally defined to have a mass $>10^6 \Msun$.

While their existence is widely accepted, their origin is still debated, as they seem to have formed improbably quickly during structure formation in the early universe\cite{Banados:2022mrl,Matsuoka:2016pho,Maiolino:2023bpi, Izquierdo-Villalba:2023tyv}. The possible seeds for these SMBHs include the collapse of early massive Population III stars, direct collapse of gas clouds, efficient mergers, or PBHs, see \cite{Volonteri:2021sfo} for a recent review. The growth of these seeds into SMBHs is an active area of research. Low mass seeds have to consistently grow at super-Eddington rates to reach the observed masses, while heavy mass seeds are difficult to form with known mechanisms. The accretion strongly depends on the host system and feedback processes \cite{Ricarte:2018mzn}. This is also indicated by the strong correlations between the central SMBH and its galaxy\cite{Ferrarese:2000se, Izquierdo-Villalba:2023tyv}.

\subsection{Intermediate Mass Black Holes}
Black holes in the intermediate mass range, IMBH, are generally defined in a mass range $100-10^5\Msun$, see \cite{Greene:2019vlv} for a recent review. Observationally, they are a rare find compared to the SMBH. There are a few candidate observations, but none have been conclusive\cite{Mezcua:2017npy, Baldassare_2015, Lin:2018dev, Chilingarian:2018acs, Pechetti_2022, Micic:2022xci}. Since their population is not well understood, the mass range has to be taken with a grain of salt.

But if the universe contains SMBHs, at some point, there need to have been IMBHs, as there is no known mechanism to create them `from scratch'. They are thought to be the link between stellar mass and supermassive black holes, and might be `leftover' seeds\cite{Madau:2001sc}. Additionally, they can be formed from successive mergers of stellar mass black holes\cite{Fragione:2022avp, Atallah:2022toy, Rose:2021ftz}.  

Depending on the formation scenario, the estimations of the distribution of IMBHs vary wildly. They might be at the center of dwarf galaxies\cite{Chilingarian:2018acs}, ejected from clusters after mergers and therefore \textit{rogue}\cite{2023arXiv231008079P}, or around larger SMBHs in AGNs \cite{Fragione:2022egh, DiMatteo:2022gak, Strokov:2023kmo}. The different populations would be subject to vastly different environments. In this dissertation, we will focus on the first option, where IMBHs are the dominant force in their local environment. There are semi-analytic models that follow the estimated evolution of IMBHs through structure formation. We will discuss these in more detail in chapter \ref{chap:rates}.

The massive IMBH and lighter SMBH are prime targets for the LISA observatory, as an inspiral of a solar mass type object will be in its frequency band. Once the detector is operational, it will uncover more about the population and history of MBH. 

\subsection{Black Holes and Dark Matter}
Both BHs and DM have a long history in astrophysics and cosmology. While BHs were originally thought to be mathematical artifacts, it took a long time for the community to believe in their existence. Detecting and collecting proof of them is an ongoing effort. DM, on the other hand, comes from observational necessity. Its properties are modeled to our needs to make sense of astrophysical and cosmological considerations, and mathematical description follows observation\cite{Bertone:2016nfn, deSwart:2017heh}.

Despite these foundational differences, due to their exotic nature, both of them continue to fascinate and stir up imagination. Combining them also has a long history, first with the proposal of (P)BH as DM candidates. When this was largely ruled out, DM was put into the spacetime around BH. This can be done in the form of a DM field, and provide additional `hair' to the BH, or create atom like structures -- gravitational atoms\cite{Hannuksela:2019vip}. Or this can be done in the form of a highly concentrated DM density, a so called \textit{spike}\cite{Gondolo:1999ef}. Close to the BH, the DM spike can have really large densities, affecting the environment around them. Spikes and their effects are central to this dissertation and we will discuss them in detail in section \ref{sec:environment:spike}.

As mentioned, we observe these BH inspirals with the help of Gravitational Waves.

\section{Gravitational Waves}
One of the most prominent predictions of General Relativity is the existence of Gravitational Waves (GWs). These are transversal waves in 4D spacetime, traveling at the speed of light and squeezing and stretching the fabric of space. They have two polarizations, $+$ and $\cross$\cite{Maggiore:2007ulw}.

They have, for the first time, been directly observed in 2015 by the LIGO observatories as part of the LIGO-Virgo collaboration, about 100 years after the publication of GR. This opens up a completely new window to study the universe\cite{LIGOScientific:2016aoc}. 

\subsection{Derivation}
By assuming a flat background with a perturbation $g_{\mu\nu} = \eta_{\mu\nu} + h_{\mu\nu}$, one can derive a wave equation with the linearized Einstein equations \cite{Maggiore:2007ulw}
\begin{equation}
    \Box \bar{h}_{\mu\nu} = 16\pi T_{\mu\nu}.
\end{equation}
Here, $\Box=\partial_\mu \partial^{\mu}$ is the flat space d'Alembertian, and $\bar{h}_{\mu\nu}$ is the trace reversed metric perturbation
\begin{equation}
    \bar{h}_{\mu\nu} = h_{\mu\nu} - \frac{1}{2} \eta_{\mu\nu} h_{\mu\nu}
\end{equation}
in the Lorentz gauge $\partial^{\nu}\bar{h}_{\mu\nu} = 0$.

In the vacuum case $T_{\mu\nu} = 0$, one can choose the transverse-traceless gauge $h^{0\mu} =0, h^i_i = 0, \partial^{j}h_{ij}=0$, where $i,j$ refer to the spatial indices. The wave equation is then solved by
\begin{equation}
    h^{TT}_{ij} = \begin{bmatrix}
        h_+ & h_{\cross} & 0 \\
        h_{\cross} & - h_+ & 0 \\
        0 & 0 & 0\\
    \end{bmatrix}
    \cos(\omega(t-z))
\end{equation}
giving a plane wave solution with two polarizations in the $z$-direction at the speed of light.

\subsection{Sources}
GWs can be generated through a changing mass quadrupole. The quadrupole formula, first derived by Einstein in 1918, gives the rate of gravitational waves emitted by a system
\begin{equation}
    \bar{h}_{ij} = \frac{2}{r} \ddot{I}_{ij}(t-r), \label{eq:quadrupole}
\end{equation}
where $I_{ij}$ is the mass quadrupole moment. This means that the emitting system needs to be anisotropic.

\begin{figure}
    \centering
    \includegraphics[width=\textwidth]{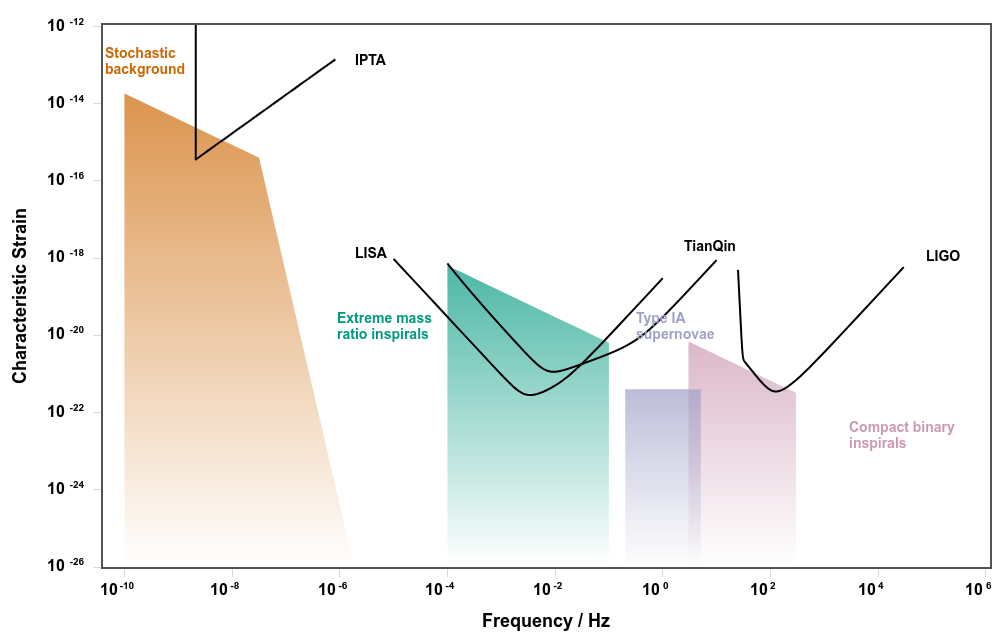}
    \caption{The range of astrophysical GW sources and the observatories. Plot created with \url{gwplotter.com}\cite{Moore:2014lga}.}
    \label{fig:gwplotter}
\end{figure}

Binary configurations of compact objects (COs) -- stars, neutron stars (NS) and BHs -- are prime candidates for the emission of GWs, the higher the mass density, the better. The emitted GW frequency is a harmonic of the orbital frequency, which in turn depends on the masses of the COs. This can be seen in \figref{fig:gwplotter}, where lower mass binaries give off higher frequency GWs. These GWs carry away energy from the system, leading to a contraction of the orbit, an \textit{inspiral}, and eventual \textit{coalescence}. A number of these CO binary coalescences, either binary BHs, binary NS, or NS-BH, have been observed with current GW detectors\cite{LIGOScientific:2021djp}. The inspiral of SMBHs during structure formation of the universe can also produce a stochastic background of GWs, with a tentative detection recently\cite{NANOGrav:2023gor}. While not a specific target of this dissertation, binary SMBHs can also be modeled with similar techniques as described in the following, and valuable information can be gained about structure formation, DM, and MBH seeds. This dissertation focuses on compact binaries with a sufficiently large mass ratio. We will explore the dynamics of these systems further in chapter \ref{chap:inspiral}.

Another system that can emit GWs is a supernova. Stellar collapse is a powerful event that is difficult to model with current technology. But there are mechanisms like rotating core collapse and bounce that can effectively generate GWs in the $100$Hz range \cite{Maggiore:2007ulw}. These have not been observed as of yet.

A prominent area of research is also that of first order phase transitions (FOPTs) in the early universe. This is a process in which a field permeating the universe undergoes a first order phase transition from a false to a true vacuum state. This happens stochastically throughout the universe. Where a true vacuum is formed -- nucleated -- it expands outward in the form of a bubble. These bubbles of true vacuum eventually collide until the whole field is in the energetically favorable state\cite{Hogan:1983ixn}. This process is by nature highly anisotropic and can emit GWs, which would also form a stochastic GW background. These signals might even reach us from further out than the CMB, as GWs can propagate through the radiation dominated universe unhindered\cite{Kamionkowski:1993fg, Sagunski:2023ynd}.

\subsection{Detectors}
As the quadrupole waves squeeze and stretch spacetime, the most obvious detection method is measuring the geodesics of photons. This can efficiently be achieved through an interferometer. When sending two photon beams down two perpendicular paths of equal length, the photons should be in phase when they return. A passing quadrupolar GW will stretch the spacetime in one direction and squeeze it in the other, thereby altering the path lengths of the photons, which is observable in their interference pattern. 

There are several ground based observatories that implement this technique. The main ones are the LIGO observatories, the Virgo observatory, and the KAGRA observatory. Together, they form the LVK cluster. There are several advancements planned, such as the Einstein Telescope (ET). Ground-based observatories generally have arm lengths in the order of $\sim$km, and are subject to seismic noise, making them sensitive to GWs in the frequency range $10-10^3$Hz\cite{Moore:2014lga}.

There are also space-based interferometers planned, such as the Laser Interferomenter Space Antenna (LISA). This mission would have three satellites forming multiple interferometers at less than right angles. These would be placed in a solar orbit with arm lengths of several million km, making them sensitive to much lower frequencies in the mHz range. It is planned to launch in the late 2030s\cite{LISA:2017pwj}. A similar project is TianQuin\cite{TianQin:2015yph}.

A different approach is that of Pulsar Timing Arrays (PTAs). Here, the periodic signals of pulsars and their spatial distribution are used. A passing GW can change the time of arrival (ToA) of the light pulses. The presence of an SGWB would cause a distinctive correlation in the spatial distribution of the ToA, the Hellings-Downs curve\cite{Hellings:1983fr}. The scales here are of $\sim$pc and upward, at even lower frequencies. This has been measured for the first time recently\cite{NANOGrav:2023gor}.

Together, these detectors span a broad frequency range, with some speculated and confirmed sources as shown in \figref{fig:gwplotter}.

Lastly, for high frequency GWs, microwave cavities might be able to detect them. These microwave cavities use the interaction between the GWs and electromagnetic fields, where the former can induce oscillations in the latter\cite{Berlin:2021txa}. These can in principle on the orders of $\sim$ meters, making them much more compact. While there are no definite astrophysical processes that can produce GWs at these frequencies, there are some proposed mechanisms, for example PBHs or axion DM\cite{Berlin:2021txa}.

\subsection{Signal}
The GW signal emitted by the binary inspirals will be discussed in chapter \ref{chap:inspiral}. Here, we want to briefly summarize the detection of the signal, following \cite{Moore:2014lga}. Of course, any detector has an intrinsic noise floor from which the GW signal has to be extracted. If we assume that the output of the detector is given by a noise term and a signal
\begin{equation}
    s(t) = n(t) + h(t),
\end{equation}
then we can characterize the noise with its Power Spectral Density (PSD) $S_n(f)$
\begin{equation}
    \avg{\tilde{n}(f)\tilde{n}^*(f')} = \frac{1}{2} \delta(f-f') S_n(f),
\end{equation}
where $\tilde{n}$ is the Fourier transform of the noise signal, and the brackets symbolize an ensemble average.

To extract the signal, we can use the Wiener optimal filter $K(t)$\cite{wiener1964extrapolation}, by convolving the signal with the filter
\begin{equation}
    (s * K)(\tau) = \int_{-\infty}^{\infty} \d t \,(h(t) + n(t))K(\tau-t) \approx \mathcal{S} + \mathcal{N}.
\end{equation}
The signal contribution $\mathcal{S}$ is defined as the expectation of this convolution, while the expectation value of the noise $\mathcal{N}$ is zero. These can be calculated as 
\begin{align}
    \mathcal{S} &{}= \int_{-\infty}^{\infty} \d f \tilde{h}(f) \tilde{K}^*(f), \\
    \mathcal{N}^2 &{} = \int_{-\infty}^{\infty} \d f \frac{1}{2} S_n(f) |\tilde{K}(f)|^2 ,
\end{align}
where the tilde refers to the Fourier transform again.

The signal-to-noise ratio (SNR) is then defined as $\rho^2 = \frac{\mathcal{S}^2}{\mathcal{N}^2}$. This is maximized by choosing the Wiener optimal filter to be $\tilde{K}(f) = \frac{\tilde{h}(f)}{S_n(f)}$. Thus, extracting the signal requires knowing the wavefrom a priori.
Then, the SNR is given by
\begin{equation}
    \rho^2 = \int_0^{\infty} \d f \frac{4\abs{\tilde{h}(f)}^2}{S_n(f)}.
\end{equation}
Alternatively, we can define the characteristic strain $h_c$ and the noise amplitude $h_n$
\begin{align}
    h_c(f)^2 ={}& 4f^2 \abs{\tilde{h}(f)}^2, \\
    h_n(f)^2 ={}& fS_n(f),
\end{align}
and the SNR is given by 
\begin{equation}
    \rho^2 = \int_0^{\infty} \d\log f \abs{\frac{h_c(f)}{h_n(f)}}^2. \label{eq:SNR}
\end{equation}
The larger the SNR, the more easily a signal is detected. Many publications take an SNR $>20$ as detection threshold\cite{Babak:2017tow}. 

The noise amplitude for the detectors is plotted in \figref{fig:gwplotter}. If the characteristic strain of a signal is above the noise amplitude, it should be extractable given the known waveform. We will assume the PSD for LISA as given in Eq. (13) of \cite{Robson:2018ifk}.

Therefore, to extract the signal, the waveform has to be very well known to maximize the SNR. For BBHs this problem has been mostly resolved with PN expansions and numerical simulations. For I/EMRIs, the problem is more complex, especially because these signals can be visible for months or years, requiring accurate waveforms to track them throughout\cite{Baghi:2022ucj,LISAConsortiumWaveformWorkingGroup:2023arg}. 

There is also an expected stochastic background of signals, where no individual source can be mapped out\cite{Bonetti:2020jku,Oliver:2023xan, Pozzoli:2023kxy, Seoane:2024nus}. This further increases the need to both model an individual signal more accurately and try to model the background so that it can be separated from the noise. 

\section{Stochastic Differential Equations}
The environment around MBH can be a violent place, subject to strong variations and hard to predict occurences. A lot of these processes can be described in a stochastic manner. In this dissertation, we will describe the interactions with surrounding stars as a Brownian process, with the help of Stochastic Differential Equations (SDEs). These are closely related to the Fokker-Planck equations.

As it is not part of every physicist's education, here, we will briefly introduce SDEs, following \cite{evans2012introduction}.
\subsection{Motivation}
First, consider an ordinary differential equation (ODE)
\begin{equation}
    \v{\dot{x}}(t) = \v{f}(\v{x}(t)), \text{  with } \v{x}(0) = \v{x}_0
\end{equation}
The solution is a trajectory $\v{x}(t):[0,\infty)\to \mathbb{R}^n$. In many applications, either the measurement is uncertain and stochastic or the system itself contains stochastic mechanisms, such as in the case of Brownian motion.

One could modify the evolution equations and add an additional \textit{white noise} $\v{\xi}(t)$
\begin{equation}
    \v{\dot{X}}(t) = \v{f}(\v{X}(t)) + \bm{\sigma}(\v{X}(t)) \v{\xi}(t),
\end{equation}
where $\bm{\sigma}(\v{X}(t))$ is an n$\cross$m matrix and $\v{\xi}$ an m-dimensional white noise.
We can formally write this with the help of a \textit{Wiener Process} or \textit{Brownian motion} $\v{W}(\cdot)$ such that
\begin{equation}
    \v{\dot{W}}(\cdot) = \v{\xi}(\cdot)
\end{equation}
Technically, white noise is nowhere differentiable. This equation has to be taken in the sense of a weak formulation, involving integration. This results formally in the equations
\begin{equation}
    d\v{X}(t) = \v{f}(\v{X}(t)) dt + \bm{\sigma}(\v{X}(t)) d\v{W}(t)
\end{equation}
This is a stochastic differential equation. Intuitively, the second term on the rhs is the randomness added to the solution. But what does it mean to solve these equations? We can integrate both sides and find
\begin{equation}
    \v{X}(t) = \v{X}_0 + \int^t \v{f}(\v{X}(s)) ds + \int^t \bm{\sigma}(\v{X}(s)) d\v{W} \label{eq:sample_sde}
\end{equation}
Therefore, we need to construct the stochastic integral $\int d\v{W}$ to meaningfully define SDEs.

\subsubsection{Stochastic Integral}
We can outline the definition here in the one-dimensional case. A real valued stochastic process $W$ is called Brownian motion iff
\begin{align*}
    (i)\,& W(0) = 0 \text{ almost surely}\\
    (ii)\,& W(t) - W(s) \sim N(0, t-s) \text{for all } t \geq s \geq 0 \\
    (iii)\,& \text{for all times }0 < t_1 < t_2 < ... < t_n, \\
          & W(t_1), W(t_2) - W(t_1), ..., W(t_n)-W(t_{n-1}) \text{ are independent}
\end{align*}
With this, we have $E(W(t)) = 0$ and $E(W(t)^2) = t$.

How can the integral $\int dW$ be interpreted? We first construct the special case of $\int WdW$ analogously to the Riemann integral. Let $P = {0 < t_1 < ... < t_m < T}$ be a partition of the interval $[0,T]$, $0\leq\lambda\leq 1$, and $\tau_k + (1-\lambda)t_k  + \lambda t_{k+1}$ a collection of points inside the partition intervals. Then we define, similarly to the Riemannian integral 
\begin{equation}
    R(P,\lambda) = \sum_{k=0}^{m-1} W(\tau_k) (W(t_{k+1}) - W(t_k) )
\end{equation}

It can be shown that 
\begin{equation}
    \lim_{\abs{P}\to 0} R(P,\lambda) = \frac{W(T)^2}{2} + (\lambda-\frac{1}{2}) T
\end{equation}
in the limit of vanishing partition mesh size $\abs{P} = \max_k (t_{k+1}-t_k)$.

In contrast to the simple Riemannian integral, the result depends on the choice of midpoint $\lambda$. There are two common definitions: the choice $\lambda = 0$ corresponds to the Itô integral, while $\lambda = \frac{1}{2}$ corresponds to the Stratonovich integral. The differences can heuristically be described as follows: The Itô integral is \textit{causal} in the sense that the future only depends on the past, i.e., the leftmost value of $W(t_k)$ the interval. Unfortunately, the ordinary chain rule does not hold, resulting in the Itô formula. The Stratonovich integral, on the other hand, conserves the ordinary chain rule but is not causal. The two different interpretations are generally marked with $dW$ and $\circ dW$ for the Itô and Stratonovich descriptions, respectively. In this dissertation, we will stick to the Itô description. 

Another limit that is of interest
\begin{equation}
    \sum_{k=0}^{m-1} \underbrace{(W(t_{k+1}) - W(t_k) )^2}_{\equalhat \Delta W^2} \to W(T)^2.
\end{equation}
Taking the expectation value gives $E(W(T)^2)=T$, and thus, in a sense, $\d W^2 = \d t$. This definition of the stochastic integral allows precise definition of solutions to the SDEs above. This can be generalized to higher dimensional Brownian motion.

\subsection{Itô formula and Fokker-Planck}
As mentioned, in the Itô calculus, the chain rule needs to be modified. Conceptually, this can be derived with the Taylor series expansion. In the ordinary case, terms of order $\d\v{X}^2$ would be neglected. But as $\d\v{W}^2$ can be of order $\d t$, these terms need to be included in the linear approximation.
This gives the Itô formula. If $\v{X}(t)$ is an Itô process, then a scalar function $\phi(\v{X}(t), t)$ has the differential \cite{särkkä_solin_2019}
\begin{equation}
    \d\phi = \pdv{\phi}{t}\d t + \sum_i \pdv{X_i}{t} \d X_i + \frac{1}{2} \sum_{i,j} \pdv[2]{\phi}{X_i}{X_j} \d X_i \d X_j . \label{eq:ito_formula}
\end{equation}
For vector valued functions, the equation is valid componentwise.

Closely related to SDEs are the Fokker-Planck equations. These have found lots of applications in astrophysics. For a process solving the Itô SDE 
\begin{equation}
    \d\v{X}(t) = \v{f}(\v{X}(t), t) \d t + \bm{\sigma}(\v{X}(t)) \d\v{W}(t) \text{ with } \v{X}(0) \sim p(\v{X}(0))
\end{equation}
where $\v{W}$ is an $n$-dimensional Brownian motion. We can denote the corresponding probability density of $\v{X}(t)$ as $p(\v{X},t)$. The probability density function solves the Fokker-Planck equation \cite{särkkä_solin_2019}
\begin{equation}
    \pdv{p}{t} = - \sum_i \pdv{}{X_i}[f_i p] + \frac{1}{2} \sum_{i,j} \pdv[2]{}{X_i}{X_j}\left[\left(\bm{\sigma}\bm{\sigma}^T\right)_{ij} p\right] \label{eq:fokker_planck}
\end{equation}
where the $(\v{X}(t),t)$ dependencies were left out. This can be derived by applying the Itô formula to the expectation value $E(\phi) = \int \phi p \d\v{X}$ and using integration by parts. 

\begin{figure}[ht]
    \centering
    \includegraphics[width=0.8\textwidth]{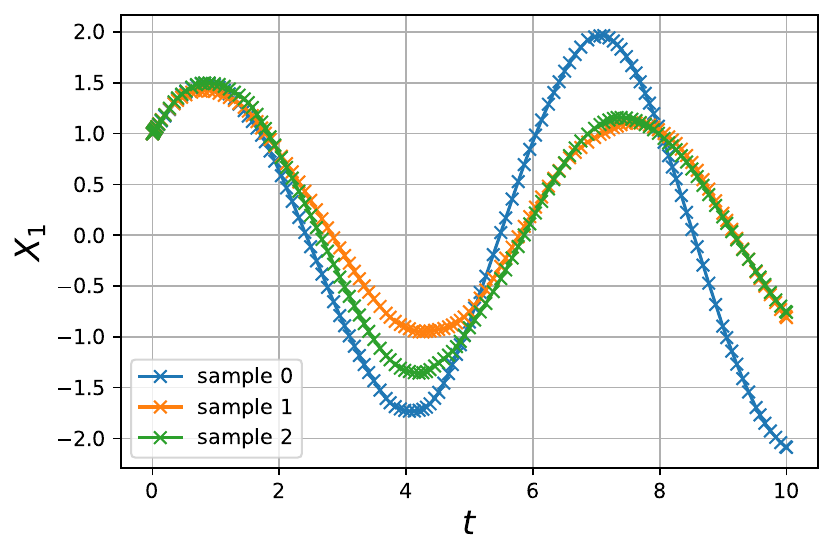}
    \caption{Example solutions to an SDE that models a harmonic oscillator with a stochastic contribution to the velocity 
    \leavevmode\\\begin{minipage}{\linewidth}
    \begin{equation*}
    \d\v{X}(t) = \begin{bmatrix}
            X_2 \\
            - X_1
        \end{bmatrix} \d t + 
        \begin{bmatrix}
            0 & 0 \\
            0 & X_1/5
        \end{bmatrix} \d\v{W} \text{ with } \v{X}(0) = \begin{bmatrix}
            1 \\
            1
        \end{bmatrix} 
    \end{equation*} \end{minipage}
    }
    \label{fig:example_sde}
\end{figure}

For a more thorough introduction, see \cite{evans2012introduction, doi:10.1137/1.9781611976434, särkkä_solin_2019}. An example SDE is shown in \figref{fig:example_sde}.

\FloatBarrier

\chapter{Mass Ratio Inspiral\label{chap:inspiral}}
Intermediate and Extreme Mass Ratio inspirals (I/EMRIs) are prime targets for the LISA mission\cite{Babak:2017tow, LISAConsortiumWaveformWorkingGroup:2023arg}. Due to their long signals, they allow for minute measurements of the background spacetime and environment. This allows for a large number of tests for GR, the host's environment, history, and origin, and possibly even subtle DM effects. In this section, we give a basis for the modeling of these systems.

I/EMRIs do not have a clear cut definition in the literature. We will use them in the sense of a stellar mass type compact object $m_2$ (BH, NS, star) inspiraling into an MBH $m_1 > 10^3\Msun$\cite{Mezcua:2017npy}. This gives a mass ratio of $q = m_2/m_1 \leq 10^{-2}$). Other uses of the phrase might include exotic objects such as heavy primordial black holes inspiraling into each other\cite{Cole:2022ucw}, or an IMBH inspiraling into an SMBH. In this dissertation, we will focus on the first interpretation, but this section is generally applicable to other interpretations.

The central MBH $m_1$ will be called the \textit{primary}, and the stellar mass type object $m_2$ the \textit{secondary}. 

We assume our inspiral to happen \textit{adiabatically}. In this approximation, at any given instance, the orbit of the secondary is given by a Keplerian orbit, i.e., an ellipse. Over secular timescales, due to dissipative effects, the orbit will shrink and slowly inspiral. The secular timescale of the inspiral is always much larger than the orbital timescale. 

\section{Keplerian Orbit \label{sec:inspiral:kepler_orbit}}
A classic Keplerian orbit is an ellipse that can be parameterized with the semi-major axis $a$ and eccentricity $e$. The eccentricity parameter is $0\leq e < 1$, where $e=0$ is a circular orbit.

\begin{figure}
    \centering
    \includegraphics[width=0.7\textwidth]{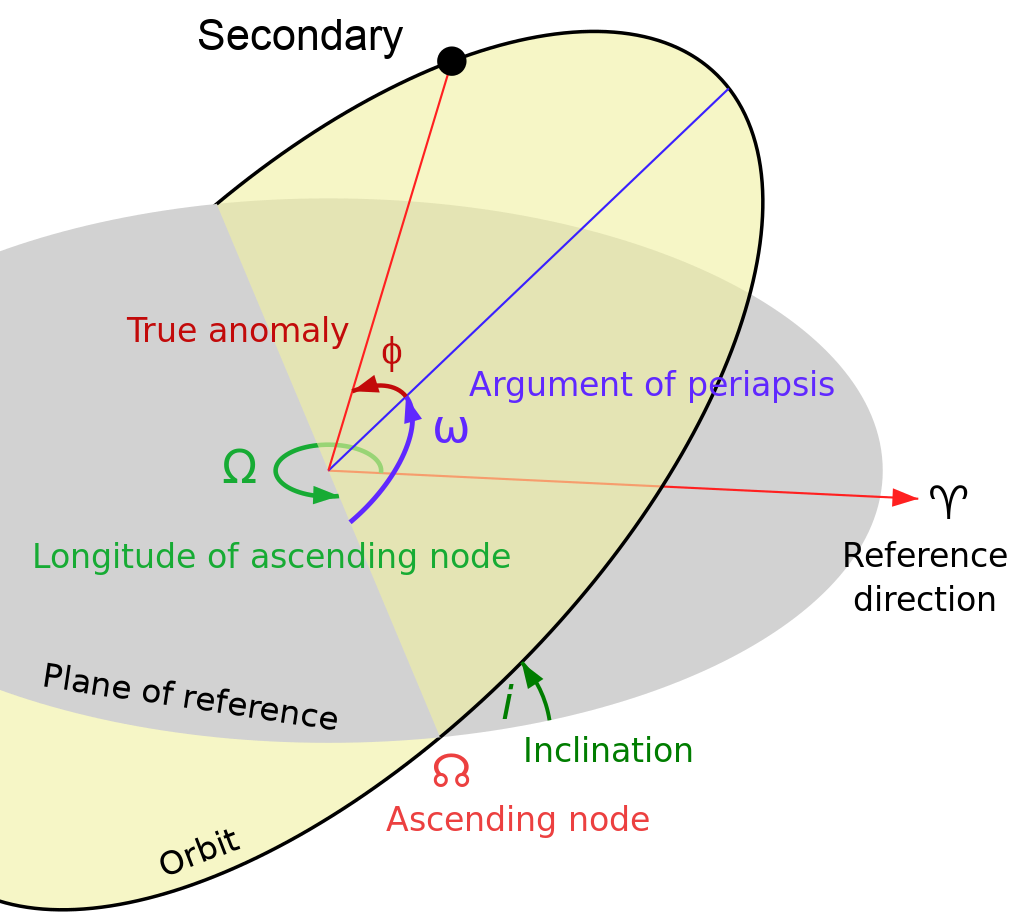}
    \caption{Parametrization of the Kepler orbit\cite{lasunncty}.}
    \label{fig:kepler_orbit}
\end{figure}

The orientation of this ellipse is defined with regard to some reference plane, the \textit{fundamental plane}, with several parameters, depicted in \figref{fig:kepler_orbit}. The fundamental plane is arbitrary, it could be given by the angular momentum of the primary, the accretion disk, or the disk of the surrounding galaxy. The plane of the orbit crosses the fundamental plane at a line called the \textit{line of nodes} at an \textit{inclination} angle $\iota$. The \textit{ascending node} is the point at which the orbit crosses the fundamental plane from below. Given a reference direction of the fundamental plane, the \textit{longitude of the ascending node} $\Omega$ is the angle formed by the reference direction and the line of nodes. The orbital ellipse has a \textit{periapse} (also called pericenter) --  the point at which the secondary is closest to the primary -- and an \textit{apoapse} -- the point where the secondary is furthest from the primary. The angle formed by the line of nodes and the periapse is called the longitude or \textit{argument of periapse} $\omega$ measured in the orbital plane. Lastly, the position of the secondary on the orbit is given by the \textit{true anomaly} $\phi$, the angle formed by the periapse and the direction of position inside the orbital plane. A Keplerian orbit is therefore defined by the five elements $(a,e, \iota, \Omega, \omega)$, and $\phi$ is needed to solve any initial value problem of the Kepler problem\cite{poisson_will_2014}.

We also define the following variables used throughout the text: the total mass is $m=m_1 + m_2$, the reduced mass $\mu = \frac{m_1 m_2}{m}$, and the mass ratio $q = \frac{m_2}{m_1}$. While we are close to the primary, the contribution of other mass distributions (i.e., DM spike, accretion disk, star distribution) can be neglected for the Keplerian orbit.

The orbital energy $E_\text{orb}$ of the secondary's orbit is given by \cite{Maggiore:2007ulw}
\begin{equation}
    E_\text{orb} = -\frac{m\mu}{2a}, \label{eq:E}
\end{equation}
and the angular momentum $L_\text{orb}$ by 
\begin{equation}
    e^2 - 1 = \frac{2 E_\text{orb} L_\text{orb}^2}{m^2 \mu^3}. \label{eq:e}
\end{equation}

Assuming to be in the orbital plane, throughout one orbit, the position $r$ and velocity $v$ are given by
\begin{align}
    r = {}& \frac{a(1-e^2)}{1 + e \cos\phi}, \\
    v^2 = {}& m\, \left(\frac{2}{r} - \frac{1}{a} \right).
\end{align}

The mean orbital frequency is given by
\begin{equation}
    \F = \frac{1}{2\pi}\sqrt{\frac{m}{a^3}}. \label{eq:kepler_f}
\end{equation}

\section{Osculating Elements \label{sec:inspira:osculating}}
The above equations are a solution to the Kepler planetary equations of motion
\begin{equation}
    \v{a} = -\frac{m}{r^2}\, \v{n},
\end{equation}
with the acceleration $\v{a}$ between the two bodies, the separation $\v{r} = \v{r}_1 - \v{r}_2$, and $\v{n} = \frac{\v{r}}{r}$.
When an additional force is acting on the secondary,
\begin{equation}
    \v{a} = -\frac{m}{r^2}\, \v{n} + \v{f},
\end{equation}
the solution is not necessarily given by a Keplerian orbit anymore.

If the force $\v{f}$ (technically an acceleration here) is small, we can assume the Keplerian orbit to hold approximately and look at its perturbations. To this end, it is useful to have a vectorial basis for the secondary on the orbit. We will follow Gauß and decompose the perturbing acceleration as
\begin{equation}
    \v{f} = S \v{n} + P\v{m} + W \v{k}
\end{equation}
with the separation vector $\v{n}$ given in the fundamental frame\cite{2013degn.book.....M}
\begin{equation}
    \v{n} = \begin{bmatrix}
           \cos(\phi + \omega) \cos \Omega - \sin(\phi+\omega) \sin \Omega \cos \iota \\
           \cos(\phi + \omega) \sin \Omega + \sin(\phi+\omega) \cos \Omega \cos \iota \\
           \sin(\phi + \omega)\sin \iota 
         \end{bmatrix}
\end{equation}
and 
\begin{equation}
    \v{m} = \pdv{\v{n}}{(\phi+\omega)} \qquad \v{k} = \frac{1}{\sin(\phi+\omega)}\pdv{\v{n}}{\iota}.
\end{equation}
Here, $\v{k}$ points in the direction of angular momentum, and $\v{m}$ is perpendicular to $\v{n}$ in the orbital plane. Together, they form an orthonormal basis at any point of the orbit. Therefore, any perturbing force can be decomposed into its constituents.

Lagrange's planetary equations now give the effect of the perturbing force on the Keplerian parameters as follows\cite{poisson_will_2014}:
\begin{align}
    \dv{a}{t} ={}& \frac{2}{n\sqrt{1-e^2}} \left( Se \sin \phi+ P\frac{p}{r} \right), \\
    \dv{e}{t} ={}& \frac{\sqrt{1-e^2}}{na} \left( S \sin \phi+ P(\cos \phi + \cos E) \right), \\
    \dv{\iota}{t} ={}& \frac{r \cos(\omega +f)}{na^2 \sqrt{1-e^2}}W,  \label{eq:dinclination_angle}\\
    \dv{\omega}{t} ={}& -\cos \iota \dv{\Omega}{t} + \frac{\sqrt{1-e^2}}{nae} \left( -S \cos \phi + P(1+\frac{r}{p}) \sin \phi \right), \label{eq:oe_omega}\\
    \dv{\Omega}{t} ={}& \frac{r\sin(\phi+\omega)}{na^2\sin\iota \sqrt{1-e^2}}W .
\end{align}
Here, $n = 2\pi \F$, the semilatus rectum $p=a(1-e^2)$, and $\cos E = \frac{e+\cos\phi}{1+e\cos\phi}$ is the eccentric anomaly.

If we assume the perturbing force is small, we can calculate the secular changes for each of these parameters as
\begin{equation}
    \avg{\dv{\omega}{t}} = \frac{\Delta\omega}{T} = \frac{1}{T} \int_0^T \dv{\omega}{t} \d t
\end{equation}
where $T = 1/\F$ is the orbital period. This integral can be evaluated more easily with the following identity \cite{Maggiore:2007ulw}
\begin{equation}
    \int_0^T \frac{\mathrm{d}t}{T} G(r(t),v(t)) = (1-e^2)^{\frac{3}{2}}\int_0^{2\pi} \frac{\mathrm{d}\phi}{2\pi}\frac{G(r(\phi), v(\phi))}{(1 + e \cos{\phi})^{2}}  ,
\end{equation}
valid for any integrable function $G$.

Therefore, for any small perturbing force, we can calculate its impact on the Keplerian orbit with the help of these osculating elements.

\section{Inspiral \label{sec:inspiral:inspiral}}
In the adiabatic approximation, over secular timescales, the orbit changes due to the perturbative forces. If the forces are \textit{dissipative}, i.e., they dissipate energy and angular momentum, the orbit will decay and eventually inspiral into the primary.
To describe such an inspiral, we will track the energy and angular momentum, the semimajor axis and eccentricity, the periapse, and the inclination angle of the system. They form a system of differential equations. The equations have been implemented in \lib{imripy}\cite{2023ascl.soft07018B, imripy} and solved numerically. These results form the basis for most of this dissertation.

\subsection{Energy and Angular Momentum Loss}
To describe these dissipative forces, we can use the osculating elements from the previous sections, or we can alternatively use energy and angular momentum considerations. The latter part is physically more enlightening, and we will quickly describe it here.

For a given dissipative force $\v{F}(r,v)$, the energy and angular momentum loss for the Keplerian orbit are given by \cite{Secunda:2020cdw}
\begin{align}
    \avg{\dv{E_\text{orb}}{t}} ={}& \int_0^T \frac{\d t}{T} \dv{E}{t} = \int_0^T \frac{\d t}{T}\, \v{v} \cdot \v{F}(r,v) \beginsubeqn \\ 
    \avg{\dv{L}{t}} ={}& \int_0^T \frac{\d t}{T} \dv{L}{t} = \int_0^T \frac{\d t}{T}\, r \, \v{k}\cdot \v{F}(r,v). \subeqn
\end{align}
where $\v{k}$ points in the direction of angular momentum. If the force is antiparallel to the velocity $\v{F}=-\v{v}$, these equations simplify to those in \cite{Dai:2021olt,Becker:2021ivq}
\begin{align}
    \avg{\dv{E_\text{orb}}{t}} ={}& - \int_0^T \frac{\d t}{T} F(r,v)v, \beginsubeqn\\ 
    \avg{\dv{L}{t}} ={}& -\sqrt{ma(1-e^2)} \int_0^T \frac{\d t}{T} \frac{F(r,v)}{v}. \subeqn
\end{align}

Due to the change in energy and angular momentum, the semimajor axis $a$ and eccentricity $e$ also change. Their evolution is given by 
\begin{align}
    \dv{a}{t} ={}&  \frac{2a^2}{m\mu} \dv{E_{\text{orb}}}{t} , \label{eq:da_dt} \beginsubeqn \\ 
    \dv{e}{t} ={}& -\frac{1-e^2}{2e} \left(\dv{E_{\text{orb}}}{t}/E_{\text{orb}} + 2\dv{L_{\text{orb}}}{t}/L_{\text{orb}} \right), \subeqn \label{eq:de_dt}
\end{align}
where we have dropped the averaging symbols. This has to be understood on secular timescales.

These are the principal equations (\ref{eq:da_dt})\&(\ref{eq:de_dt}) we solve throughout this dissertation. We will use different dissipative forces that reduce $E_\orb, L_\orb$ and assess their impact on the evolution of the eccentricity and semimajor axis. We can also track the changes in the periapse and inclination with \eqref{eq:oe_omega} and \eqref{eq:dinclination_angle} as described in the following. 

But first, the dissipative force that we apply in all our inspiral models.

\subsubsection{Gravitational Waves}
The emission of GWs is a dissipative effect, which of course gives us the ability to observe the system with the GW detectors. The form of the GW signal will be described in \ref{sec:inspiral:GWsignal}. Here, we quote the energy and angular momentum loss induced by GW emission at leading order \cite{Maggiore:2007ulw}
\begin{align}
    \avg{\dv{E_{\text{gw}}}{t}} =& - \frac{32}{5} \frac{\mu^2 m^3 }{a^5} \frac{1 + \frac{73}{24}e^2 +\frac{37}{96}e^4}{(1-e^2)^{7/2}}, \label{eq:dE_gw} \beginsubeqn\\
    \avg{\dv{L_{\text{gw}}}{t}} =& - \frac{32}{5} \frac{\mu^2 m^{5/2} } {a^{7/2}} \frac{1 + \frac{7}{8}e^2}{(1-e^2)^2} .\label{eq:dL_gw} \subeqn
\end{align}
This can be calculated from the quadrupole formula \eqref{eq:quadrupole} for the binary system on Keplerian orbits or from a PN expansion of the two-body system as the 2.5PN correction. This is the lowest dissipative correction, and we will include it in all our inspiral models.

\subsubsection{Capture}
Black holes can capture objects. There are \textit{capture orbits} that inevitably progress past the event horizon of a BH. This is an inherently relativistic process, so it must be solved in a Schwarzschild or Kerr background. A relativistic calculation gives a minimum specific angular momentum $J_{lc} = 4m_1$ for the Schwarzschild spacetime\cite{2013degn.book.....M}, below which the secondary will imminently be captured. For large eccentricities ($e\to1$), this corresponds to a minimal periapse of $r_p = 8m_1$. On a circular orbit, this is the radius of the \textit{innermost stable circular orbit} (ISCO), $r_\isco = 6m_1$. The collection of capture orbits is commonly called the \textit{loss cone}. We evolve the differential equations until the minimum angular momentum is reached.

\subsubsection{Orbital Precession}
The argument of periapse $\omega$ describes the angular position of the periapse, the lowest point in the orbit. The change in periapse is given by \eqref{eq:oe_omega}.

The first effect that we will consider is \textit{mass precession}, the precession due to a mass distribution $m(r)$ around the MBH. Even if the potential is subdominant, the distribution can have effects on the argument of periapse. 
If we assume the perturbing force to be \cite{Dai:2021olt}
\begin{equation}
    \v{f} = -\frac{m(r)}{r^2} \, \v{n},
\end{equation}
the osculating orbital perturbations give 
\begin{equation}
    \avg{\dv{\omega}{t}} = \frac{1}{eT}\int_0^{2\pi} \cos \phi \frac{m(r)}{m} \d\phi .\label{eq:mass_precession}
\end{equation}
The integral is generally negative for monotonically decreasing mass distributions, resulting in a retrograde precession.

The second effect is \textit{relativistic} or \textit{Schwarzschild} (SS) \textit{precession}. This is the only relativistic effect we will consider at this stage because it is the most prominent one. This secular precession is given by \cite{2013degn.book.....M}
\begin{equation}
    \avg{\dv{\omega}{t}} = 6\pi \frac{m}{pT}. \label{eq:relativistic_precession}
\end{equation}
This precession is prograde, counter to the mass precession. This was famously one of the earliest successes of GR, explaining the observed precession of Accretion Disks vs Spikes (Copy)Mercury's orbit\cite{Trageser2018}.

\subsubsection{Inclination Change}
As derived in \eqref{eq:dinclination_angle}, we can also track the change in inclination of the orbital plane. This is most prominently important in the presence of an accretion disk. If the secondary's orbit is inclined with regard to the disk, frequent interactions will align the orbital plane with the disk\cite{Fabj:2020qqc, Nasim:2022rvl}. 
Assuming a dissipative force $\v{F}$ acting on the secondary $m_2$, this gives 
\begin{equation}
    \avg{\dv{\iota}{t}} = \frac{1-e^2}{\sqrt{ma} m_2} \int_0^{2\pi}\frac{\d\phi}{2\pi} \frac{r\cos(\omega+\phi)}{(1+e\cos\phi)^2} \, \v{F} \cdot \v{k}.
\end{equation}

\subsection{Stochastic Forces}
We describe the dissipative forces with the help of their energy and momentum loss, but the differential equations are posed with regard to the semimajor axis $a$ and eccentricity $e$ as in equations (\ref{eq:da_dt}),(\ref{eq:de_dt}). If we have stochastic contributions to the differential equations, resulting in SDEs, we need to apply the Itô formula \eqref{eq:ito_formula} to accurately describe the evolution. 

If we have an SDE for the energy and angular momentum as
\begin{equation}
    \d\begin{bmatrix}
        E \\
        L
    \end{bmatrix} = 
    \begin{bmatrix}
        D_E \\ 
        D_L
    \end{bmatrix} \d t + \\ 
    \bm{\sigma} \d\v{W}
\end{equation}
with 
\begin{equation}
    \v{\sigma }\v{\sigma}^T = \v{D} = 
    \begin{bmatrix}
        D_{EE} & D_{EL} \\ 
        D_{EL} & D_{LL} \\ 
    \end{bmatrix},
\end{equation}
the application of the Itô formula results in an SDE for $a,e$ as
\begin{align}
\label{eq:sde}
    \d\begin{bmatrix}
        a \\
        e
    \end{bmatrix} = {}&
    \left(
    \mathcal{J}
    \begin{bmatrix}
        \dv{E_{orb}}{t} \\ 
        \dv{L_{orb}}{t}
    \end{bmatrix} +  \frac{1}{2}\, \v{D} : \H \begin{bmatrix}
        a \\
        e
    \end{bmatrix} \right) \d t \\ 
    & + \mathcal{J} \bm{\sigma} \d\v{W} \nn
\end{align}
with the Jacobian $\mathcal{J} = \pdv{(a,e)}{(E,L)}$ calculated from \eqref{eq:E}, (\ref{eq:e})
\begin{equation}
    \pdv{(a,e)}{(E,L)} = \begin{bmatrix}
        -a/E  &  0 \\
        \frac{e^2-1}{2eE} & \frac{e^2-1}{eL}
    \end{bmatrix}
\end{equation}
and the operator
\begin{equation}
    \v{D} : \H = \sum_{i,j} D_{ij} \pdv[2]{}{x_i}{x_j}
\end{equation}
where $x_{i,j} = {E,L}$. This can be interpreted as the sum over the weighted Hessian $\H a, \H e$ componentwise. The Hessians are given by
\begin{equation}
    \H a = \begin{bmatrix}
        2a/E^2  &  0 \\
        0 & 0
    \end{bmatrix},
    \quad
     \H e = \frac{2}{\mu^3 m^2} \begin{bmatrix}
        \frac{-1}{2\mu^3 m^2} \frac{L^4}{e^3}   &  L\frac{1+e^2}{2 e^3} \\
        L\frac{1+e^2}{2 e^3} & \frac{E}{e^3}
    \end{bmatrix} \label{eq:Hessians}
\end{equation}
While this section is quite abstract, it will become relevant in section \ref{sec:environment:stars}.

\section{Gravitational Wave Signal \label{sec:inspiral:GWsignal}}
The system emits gravitational waves as a result of the change in the mass quadrupole moment, see \eqref{eq:quadrupole}. For an observer, the orientation of the orbit can be described by two angles, $\iota'$ and $\beta'$. The inclination angle $\iota'$ describes the inclination of the plane of the orbit to the plane of the sky, and $\beta'$ describes the angle of the major axis with regard to the direction of the observer in the orbital plane (see Fig. 1 of \cite{Moreno-Garado:1995msd}).

While in the previous sections we discussed how the orbital frame relates to the fundamental frame, the transformation of the orbital frame through the fundamental frame to the observer frame is explained in appendix \ref{sec:appendix:orbtoobs}.

The gravitational strains of the two polarization modes for a Keplerian orbit are \cite{Martel:1999tm, poisson_will_2014}
\begin{align}
    h_+ = & - \frac{m\mu}{p D_L} \Bigg[  \left( 2 \cos(2\phi + 2\beta') + \frac{5}{2}e \cos(\phi + 2\beta') + \frac{1}{2}e \cos(3\phi + 2\beta') + e^2 \cos(2\beta')   \right) (1+\cos^2\iota') &&\nn\\
    & \quad\quad\quad + (e\cos\phi + e^2) \sin^2\iota' \Bigg] \label{eq:h_+} &&\\
    h_{\cross}  = & - \frac{m\mu}{p D_L} \Bigg[ 4\sin(2\phi + 2\beta') + 5e\sin(\phi + 2\beta') + e\sin(3\phi + 2\beta') + 2e^2 \sin(2\beta')  \Bigg] \cos\iota'  \label{eq:h_x}&&
\end{align}
for a system at luminosity distance $D_L$.

This can be decomposed into the harmonics of the mean orbital frequency $\F$ \cite{Moreno-Garado:1995msd, Moore:2018kvz}
\begin{equation}
\label{eq:h_harmonics}
    h_{+,\cross} = \mathcal{A} \sum_{n=1}^{\infty} \left(C_{+,\cross}^{(n)} \cos(n\ell) + S_{+,\cross}^{(n)} \sin(n\ell)\right)
\end{equation}
with the \textit{mean anomaly}
\begin{equation}
    \ell(t) = 2\pi\int^t \mathrm{d}t\,   \F ,
\end{equation}
and the amplitude
\begin{equation}
    \mathcal{A} = -\frac{\mchirp}{D_L} (2\pi\mchirp \F)^{2/3}.
\end{equation}
with the \textit{chirp mass} $\mchirp = \mu^{3/5}m^{2/5}$.

The coefficients $C_{+,\cross}^{(n)}, S_{+,\cross}^{(n)}$ can be calculated as described in \cite{Moreno-Garado:1995msd, PhysRevD.80.084001, Moore:2018kvz}, which gives for arbitrary eccentricity \cite{Chandramouli:2021kts} 
\begin{subequations}
\begin{align}
    C_{+}^{(n)}  = &  \Big[ 2s_{\iota'}^2 J_n(ne) + \frac{2}{e^2}(1 + c_{\iota'}^2)c_{2\beta'} \left( 
        (e^2 - 2)J_n(ne) + ne(1-e^2) (J_{n-1}(ne) - J_{n+1}(ne))\right) \Big],  \\
    S_{+}^{(n)}  = & - \frac{2}{e^2} \sqrt{1-e^2} (1+c_{\iota'}^2)s_{2\beta'} \Big[ -2(1-e^2)nJ_n(ne) + e(J_{n-1}(ne) - J_{n+1}(ne)) \Big], \\
    C_{\cross}^{(n)}  = & - \frac{4}{e^2} c_{\iota'} s_{2\beta'} \Big[ (2-e^2)J_n(ne) + ne(1-e^2)(J_{n-1}(ne) - J_{n+1}(ne))  \Big], \\
    S_{\cross}^{(n)}  = & - \frac{4}{e^2} \sqrt{1-e^2} c_{\iota'} c_{2\beta'} \Big[ -2(1-e^2)nJ_n(ne) + e(J_{n-1}(ne) - J_{n+1}(ne))\Big],
\end{align}  
\end{subequations}
where the $J_n$ are Bessel functions of the first kind, and using shorthands $c_{\beta'}=\cos (\beta')$, $s_{\beta'}=\sin(\beta')$.

\subsubsection{Stationary Phase Approximation}
To calculate the signal in the frequency domain, the Stationary Phase Approximation (SPA) is commonly applied. Since the amplitude varies on secular timescales, the Fourier transform over rapidly oscillating sinusoidal terms is negligibly small, except at the stationary phase condition
\begin{equation}
    n\F(t^*_n) = f
\end{equation}
at a given time $t_n^*$. Thus, the Keplerian orbit emits primarily at all integer multiples of the mean orbital frequency. The SPA gives a mapping between the time $t_n^*$ and the frequency of the $n$-th harmonic.

The Fourier transform of the signal is therefore given by \cite{Moore:2018kvz, Chandramouli:2021kts}
\begin{align}
    \tilde{h}_{+,\cross}^{(n)}(f) ={}& - \frac{\mchirp}{2D_L} \frac{(2\pi \mchirp \F(t^*_{n}))^{2/3}}{\sqrt{n \dot{\F}(t^*_{n})} } \nn \\
     &   \times\Big[C_{+,\cross}^{(n)}(t^*_{n}) + i S_{+,\cross}^{(n)}(t^*_{n}) \Big] e^{i\psi_n},  
\end{align}
for the $n$-th harmonic.
The signal is emitted for a finite amount of time, and the equation is valid between some initial frequency $\F_0$ and the final frequency of the Last Stable Orbit (LSO) $\F_\text{lso}$: $n\F_0 < f < n\F_\text{lso}$, before the orbit enters the loss cone.
The phase of the $n$-th harmonic is given by 
\begin{equation}
    \psi_n = 2\pi ft^*_n - n \ell - \frac{\pi}{4}.
\end{equation}

\begin{figure}
    \centering
    \includegraphics[width=\textwidth]{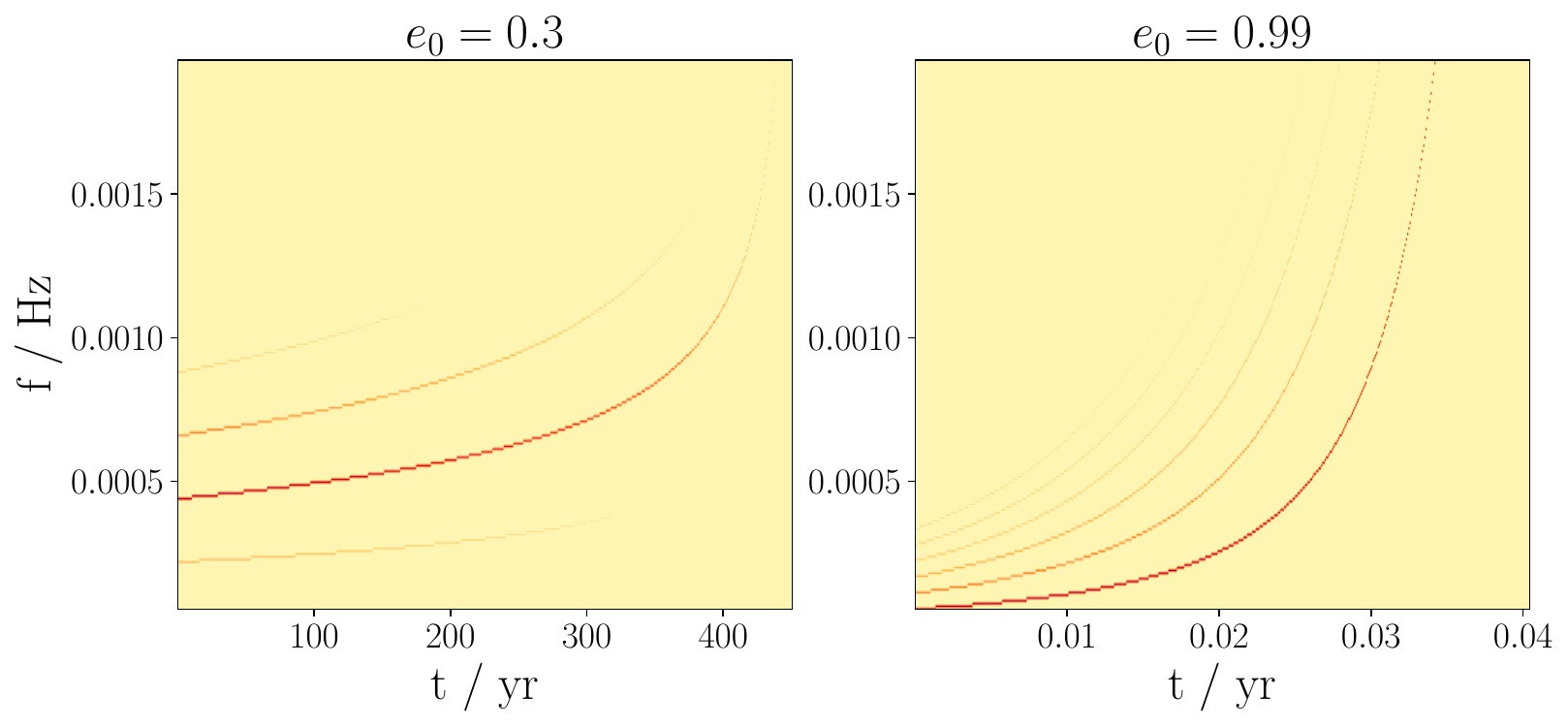}
    \caption{A Q-plot of an inspiraling system with $m_1=10^5 \Msun, m_2=10\Msun$, initial eccentricity $e_0=0.3, 0.99$, and initial semimajor axis $a_0=50r_\isco, 20r_\isco$. This is a plot in time and frequency space that shows the relative strengths of the harmonics. In the low eccentricity case, the second harmonic is dominant, and more so over time as the system is being circularized. For high eccentricities, the first harmonic is the strongest. The different times and frequencies are reflected by the fact that an eccentric inspiral is much faster, as the energy loss scales with the eccentricity.}
    \label{fig:Qplot}
\end{figure}

An example of the GW signal during an inspiral is shown in \figref{fig:Qplot}. The different harmonics of the system are visible with their relative strengths. For low eccentricity orbits, the second harmonic is dominant, for high eccentricity orbits, the first. 

\FloatBarrier
\subsection{Signatures}

\subsubsection{Braking Index}
A useful quantity to describe the system is the \textit{braking index}, which is defined as \cite{Lu:2022oys, Renzo:2021aho}
\begin{equation}
    n_b \equiv \frac{\mathcal{F} \ddot{\mathcal{F}}}{\dot{\mathcal{F}}^2}.
\end{equation}
Here, $\F$ can be the frequency of any harmonic. We can translate this to the semimajor axis with \eqref{eq:kepler_f}
\begin{equation}
    n_b = \frac{5}{3} - \frac{2}{3} \frac{a \ddot{a}}{\dot{a}^2}, \label{eq:braking_index}
\end{equation}
which, according to \eqref{eq:da_dt}, depends on the dissipative forces. For a dominant dissipative force, the braking index can be constant throughout the evolution. For example, a circular inspiral due to GW emission has $\dot{\F} \propto \F^{11/3}$ and therefore $n_b = 11/3$\cite{Cutler:1994ys}. Here, the chirp mass or the orbital radius of the system is irrelevant. In this way, the braking index can characterize populations of inspiraling systems depending on the dominant forces at play.

To get a heuristic for this, we can assume a generic dissipative force $F(r,v)\equiv F_0 r^{-\gamma}v^{\delta}$ acting antiparallel to the velocity.
We can approximate the differential equations 
\begin{align}
    \dot{a} &= \dv{E_\orb}{t} / \pdv{E_\orb}{a}  \\
    &{}= -\frac{2F_0}{\mu} a^{k_1} (1+e^2)^{k_1 - 1/2}m^{(\delta-2)/2}  \nn \\
    & \cross \underbrace{\int_0^{2\pi} \frac{d\phi}{2\pi} (1+e\cos\phi)^{-(2-\gamma)} (1+2e\cos\phi + e^2)^{(\delta+1)/2}}_{\approx 1+k_2e^2}  \nn \\
    \text{with }& k_1 = 3/2 - \gamma - \delta/2  \label{eq:k_1}\\
    &{} k_2 = (3+\gamma^2 - \gamma(3-2\delta) - 2\delta + \delta^2)/4 \label{eq:k_2}
\end{align}
where the approximation is taken at second order in $e$. Taking the time derivative gives
\begin{align}
    \ddot{a} &= \dot{a} \left(k_1\frac{\dot{a}}{a} - 2e\dot{e} \frac{k_1-1/2}{1-e^2} + 2e\dot{e} \frac{k_2}{1+k_2 e^2} \right) \\
    &{}= \frac{\dot{a}^2}{a}\left(k_1 + 2ae\dv{a}{e}(\frac{1/2-k_1}{1-e^2} + \frac{k_2}{1+k_2e^2}) \right)
\end{align}
which results in the equation for the braking index of the semimajor axis
\begin{align}
    \frac{a\ddot{a}}{\dot{a}^2} \approx k_1 + 2ae\dv{e}{a} \left(\frac{1/2-k_1}{1 - e^2}  + \frac{k_2}{1+k_2 e^2} \right) .\label{eq:braking_index_a}
\end{align}

For a system at constant eccentricity, the braking index is therefore just an algebraic combination of the power law indices of the dominant force. This is most interesting for quasi-circular inspirals, where this implies we can differentiate different dissipative forces dominating an inspiral if it is not dominated by GW emission.

As another example, we can consider the case of GW emission in the high eccentricity limit $e \approx 1$. Here, $\dv{e}{a} \approx \frac{1-e}{a}$ \cite{2013degn.book.....M}, and \eqref{eq:braking_index} simplifies to $n_b \approx 4/3$.\footnote{Even though \eqref{eq:braking_index_a} is only valid to third order in $e$, the approximation is in the last term containing $k_2$. Since this term drops out for the given $\dv{a}{e}$, we can still use the equation.}

\subsubsection{Circularization vs. Eccentrification}
As the GW signal depends on the eccentricity (\eqref{eq:h_+}), a clean measurement will also reveal the evolution of the eccentricity.

Looking at \eqref{eq:de_dt}, the eccentricity will grow -- the system will be \textit{eccentrified} -- for \cite{Becker:2021ivq}
\begin{equation}
    \dv{e}{t} > 0 \Longleftrightarrow \underbrace{\frac{1}{E_{orb}} \avg{\dv{E}{t}} + 2\frac{1}{L_{orb}} \avg{\dv{L}{t}} }_{X} < 0
\end{equation}
Assuming a force antiparallel to the velocity, this gives
\begin{equation}
    X = \frac{2(1-e^2)^{3/2}}{\mu} \int_0^{2\pi} \frac{d\phi}{2\pi} (1+e\cos\phi)^{-2} \, F(r,v) \left(\frac{av}{m} - \frac{1}{v} \right).
\end{equation}
If we assume the generic dissipative force with strength $F(r,v) \sim r^{-\gamma} v^{\delta}$, we can insert
\begin{align}
    X &\propto \int_0^{2\pi} (\cos\phi + e) (1+e\cos\phi)^{-2+\gamma} (1+2e\cos\phi + e^2)^{(\delta-1)/2} \\
    &{}\approx \int_0^{2\pi} \left(\cos\phi + e + (-3+\delta + \gamma)e\cos^2\phi \right) \\ 
    &{}=\frac{e}{2}(-1+\delta+\gamma)
\end{align}
which is valid up to second order in $e$. Therefore, a dissipative force will eccentrify the system if $\delta + \gamma < 1$. This is a useful heuristic that we can apply to the environmental effects. For small eccentricities therefore, the circularization or eccentricification rate $\dv{e}{a}$ is given by the properties of the dissipative force.

\begin{figure}
    \centering
    \includegraphics[width=0.9\textwidth]{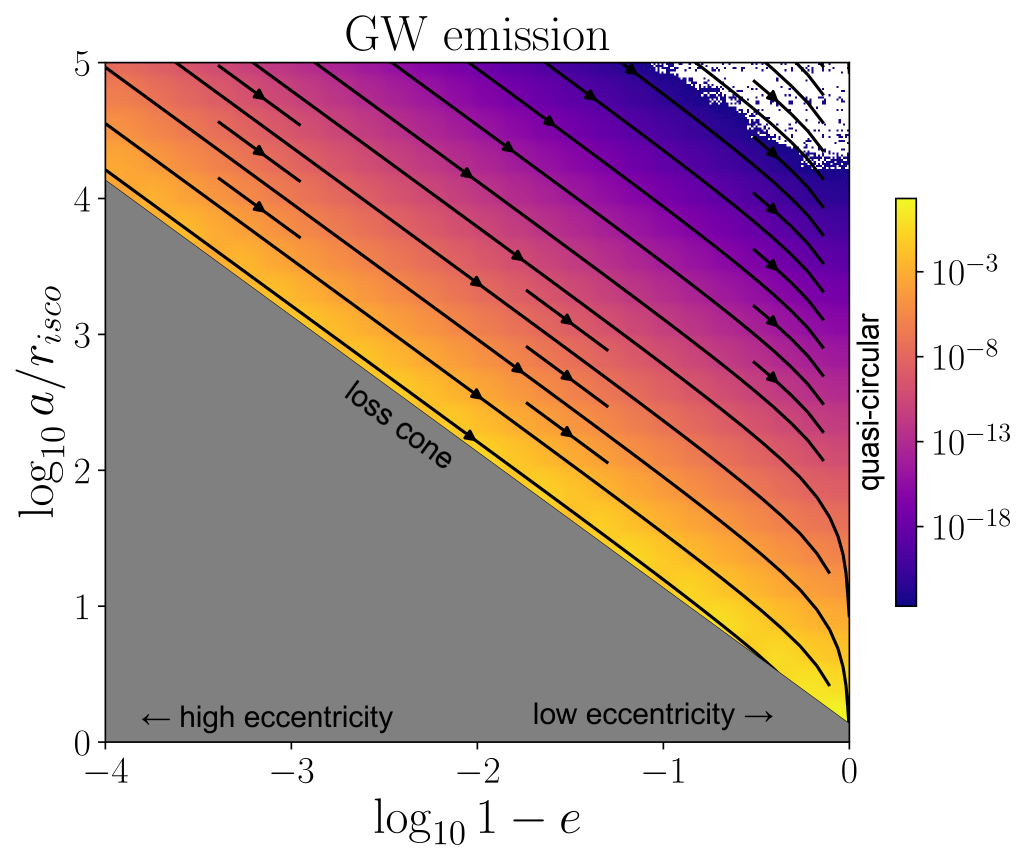}
    \caption{The phase space flow of $a,e$ for a system dominated by GW emission. The lower left half of the phase space is inside the MBH, the so called \textit{loss cone}. The streamlines point to $e\to 0$, which corresponds to a quasi-circular orbit. The circularizing effects of GW emission are clearly visible. Its strength increases with eccentricity and closer to the MBH. GW emission transports parallel to the loss cone, which is an important feature to actually observe I/EMRIs. For high eccentricity $e\approx 1$, we have $(1-e)a \equiv$ const\cite{2013degn.book.....M}. 
    The top right is not colored, as in this region the strength is too low to be compared numerically. }
    \label{fig:gw_psf}
\end{figure}

A related and useful tool is the \textit{phase space flow} of the ODE system. We can plot this for the semimajor axis $a$ and the eccentricity as $1-e$, which shows the direction the ODE system traverses at every point. The loss cone of the MBH in these plots is visible as the gray area. Here, we do not have to worry about any approximations and can numerically evaluate the derivatives. In the following, we will plot the phase space flow in $(a, e)$ for our environmental effects. In \figref{fig:gw_psf} we plot the phase space flow for the energy and angular momentum loss as seen in \eqref{eq:dE_gw}\&(\ref{eq:dL_gw}).

\subsubsection{Dephasing}
Even if GW emission is the dominant dissipative force, the subdominant forces will still affect (and generally speed up) the inspiral. Over the lifetime of the I/EMRI, the system will perform fewer orbits and therefore collect a smaller number of cycles compared to the case without the subdominant forces. This is called \textit{dephasing}.

The number of cycles between some initial $t_\text{i}$ and final $t_\text{f}$ time is given by \cite{Kavanagh:2020cfn, Becker:2021ivq}
\begin{equation}
    N^{(n)}(t_\text{f}, t_\text{i}) = n \int_{t_\text{i}}^{t_\text{f}} \F(t)\mathrm{d}t.
\end{equation}
for each harmonic $n$.
Setting $t_\text{f} = t_\text{c}$ as the time of coalescence, we obtain
\begin{equation}
    \Delta N^{(n)}_\text{DF}(t)= N^{(n)}_\text{GW}(t_\text{c}, t) - N^{(n)}_\text{GW+DF}(t_\text{c}, t).
\end{equation}
where we compare the system with just GW emission with the system of GW emission and the dissipative force DF.

Analogously to the braking index, we can define the \textit{dephasing index} \cite{Becker:2022wlo}
\begin{equation}
    n_d \equiv \dv{\log \Delta N}{\log \F}.
\end{equation}
Since the frequency evolution is additive, the frequency evolution of the total system is given by $\dot{\F}_\text{GW+DF} = \dot{\F}_\text{GW} + \dot{\F}_\text{DF}$. 
If we assume a GW dominated circular inspiral $\dot{\F}_\text{GW} \propto \F^{11/3}$ and again assume $F_\text{DF}\sim r^{-\gamma}v^{\delta}$, then
\begin{equation}
    \varepsilon \equiv \frac{\dot{\F}_\text{DF}}{\dot{\F}_\text{GW}} \propto \frac{a^{k_1-3/2}}{\F_\text{GW}^{11/3}} \propto \F^{-2-2k_1/3}_\text{GW}
\end{equation}
where we assume the evolution of the semimajor axis is given through the GW emission.
The dephasing is then accumulated as 
\begin{equation}
    \frac{1}{2} \dv[2]{\Delta N}{t} = \dot{\F}_\text{GW} - \dot{\F}_\text{GW+DF} = \varepsilon \dot{\F}_\text{GW} = \F^{-2-2k_1/3 + 11/3}
\end{equation}
This can be integrated twice using $\dv{\F_{GW}}{t} \propto \F_\text{GW}^{11/3}$ and leads to
\begin{equation}
    \Delta N \propto \F^{-(11+2k_1)/3} 
\end{equation}
and the dephasing index is
\begin{equation}
    n_d = -\frac{11+2k_1}{3}
\end{equation}
with $k_1$ as defined in \eqref{eq:k_1}. Again, the amount of dephasing at different frequencies depends on the power law behavior of the dissipative force. Also, on circular orbits, the braking and dephasing index are related by
\begin{equation}
    n_b = \frac{16}{3} + n_d. \label{eq:braking_and_dephasing_index}
\end{equation}
Thus, if the braking index is different for different environmental effects, so must be the dephasing index. In this way, different dissipative forces can be characterized and differentiated at different phases of the inspiral.

The same argument can be made for the highly eccentric inspiral $e \approx 1$, where $\dot{\F}_\text{GW} \propto \F^{4/3}$, and the dephasing index would be $n_d = \frac{1-2k_1}{3}$.

\begin{figure}
    \centering
    \includegraphics[width=0.45\textwidth]{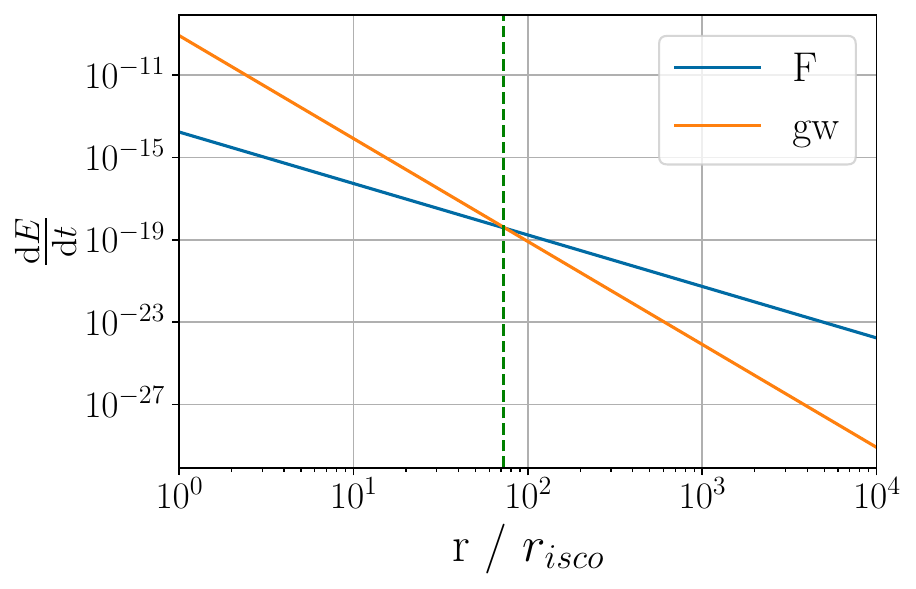}
    \includegraphics[width=\textwidth]{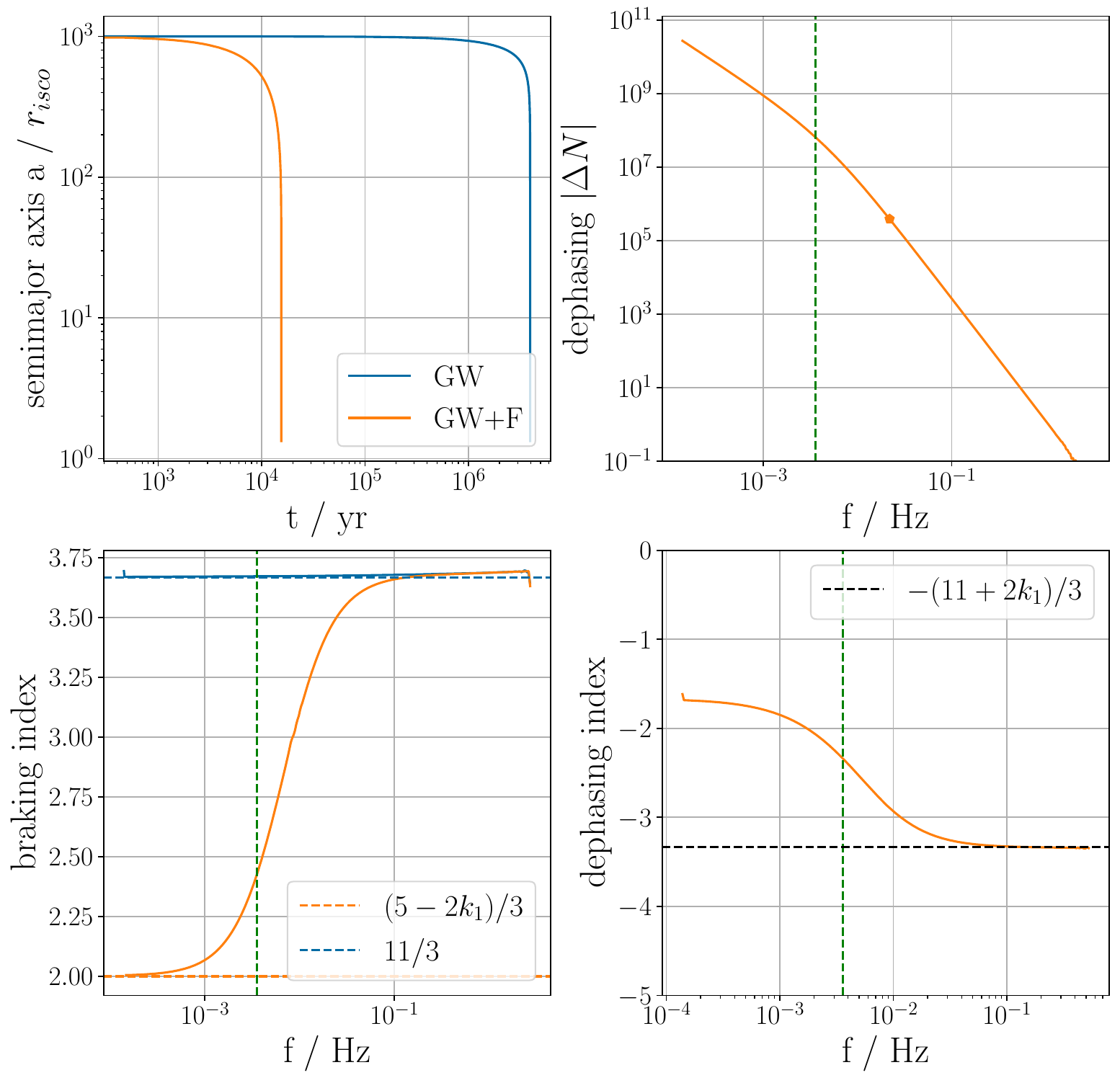}
    \caption{\textbf{Top}: The energy loss of the two dissipative forces, GW emission and a generic force $F$ with $\gamma=3/2, \delta=1$. This implies $k_1 = -1/2$. The point of equality between the two forces is given by the dashed green line.
                \textbf{Middle left}: The semimajor axis evolution over time of the inspiral with initial $a_0 = 10^3 r_\isco$. Once only the GW emission is included, and once both GW emission and the generic force are included. 
                \textbf{Bottom left}: The braking index as calculated numerically versus the frequency. The system with purely GW emission has $n_b=11/3$ as expected. The system with both has initially a braking index given by the dominant generic force $n_b=(5-2k_1)/3$, then an intermediate region, as it moves to the GW emission dominated regime. The dashed green line is the point of equality of the forces, at the frequency corresponding to the respective semimajor axis.
                \textbf{Middle right}: The dephasing that is left observable at a given frequency. We compare the dephasing of the GW+F to the purely GW system. The kink in the evolution can be observed around the point of equality between the two forces.
                \textbf{Bottom right}: The dephasing index as calculated numerically. It can be seen that as the system is dominated by GW emission, it tends to the value given by the approximation $n_d = -(11+2k_1)/3$.  }
    \label{fig:dephasing_index_example}
\end{figure}

To finish this subsection, we give an example of the braking and dephasing index for a circular inspiral, described in \figref{fig:dephasing_index_example}. We choose $\gamma = 3/2, \delta = 1$ for the dissipative force in an IMRI with $m_1=10^3\Msun, m_2=\Msun$, starting from an initial semimajor axis $a_0 = 10^3 r_\isco$. This gives $k_1 = -1/2$. The forces are approximately equal in strength at an orbital radius of $r_{eq} \approx 10^2 r_\isco$. The corresponding frequency of equality is given by \eqref{eq:f_eq}. For the inspiral with both forces, the system is first dominated by the generic force, and later dominated by GW emission, which is reflected in the braking index. There is a long intermediate regime visible, where it tends from one constant region to the next. How large this intermediate region is depends on the shape of F. If F is much flatter than GW emission, this region will be smaller. Also plotted is the dephasing, where we compare the system with GW+F to the system with just GW. The dephasings can be quite large, as the additional force F drastically speeds up the inspiral. The kink in the shape of $|\Delta N|$ is around the frequency of equality and marks the different regimes of dominance. This is also reflected in the dephasing index, where it tends to the value described by \eqref{eq:braking_index_a} in the GW emission dominated regime. 

It can be seen that the heuristic descriptions provided above work reasonably well for the system we explore here. The equations become accurate only in the strictly dominant regimes of either force. The advantage of these indices is the independence of the specific system parameters. Regardless of $m_1, m_2$ and the specific orbital distance, they are constant as long as one force strictly dominates. First, this implies that the effects of different dissipative forces should be distinct for many different systems. Second, this allows one to make statements about populations of inspirals, which we will explore later on in chapter \ref{chap:signatures}.

\subsubsection{Deshifting}
Similarly to the dephasing being the difference in orbital cycles, we can talk about the \textit{deshifting} of the difference in periapse precession. Here, the dissipative force causes a faster inspiral and therefore (usually) less precession. Close to the MBH, the SS precession dominates, as given by \eqref{eq:relativistic_precession}. 

This was explored by \cite{Dai:2021olt, Dai:2023cft} in a different way. In the first reference \cite{Dai:2021olt}, they focus on the comparison between mass precession and SS precession. For large distances, the retrograde mass precession dominates, while for small distances, the prograde SS precession dominates. This gives a turning point for the precession. Unfortunately, this is generally at large semimajor axes, which makes it hard to detect. The second reference \cite{Dai:2023cft}, on the other hand, looks at the relativistic precession caused by a modified Schwarzschild metric due to the presence of a (DM) mass distribution. Even for a Hernquist halo with generally small densities compared to the spikes we will look at, this can give observable dephasings.

We will take another approach here. Close to the MBH, we assume SS precession to be the dominant term. Due to a dissipative force, the inspiral will speed up, giving less time for SS precession to be collected. In this way, the inspiraling system with a dissipative force will have an overall lower amount of precession, and therefore there will be a shift in the periapse angle. We will call this \textit{deshifting}. This is very comparable to the dephasing explored in the previous section. We can define the deshifting as
\begin{equation}
    \Delta \omega = \omega_\text{GW} - \omega_\text{GW+DF} .\label{eq:deshifting}
\end{equation}
Again, this will be normalized such that $\Delta\omega$ is $0$ at the point of inspiral.

We can make a similar calculation, using $\omega \propto a^{-5/2} \propto f^{5/3}$ in the low eccentricity limit. This gives a \textit{deshifting index} of
\begin{equation}
    n_{\omega} \equiv \dv{\log \Delta \omega}{\log \F} = -\frac{9+2k_1}{3} = n_d + \frac{2}{3}. \label{eq:deshifting_index}
\end{equation}
This can be related again to the dephasing and braking index. This shows how closely linked the number of orbital cycles and the number of periapse cycles are, at least in this simplistic description.

\chapter{Environmental Effects\label{chap:environment}}

The local environment of MBH is a large area of research. For SMBHs in the center of galaxies, there is a distribution of stars whose orbit is dominated by the SMBH\cite{Genzel:2003cn}. Additionally, around the two SMBH that have been observed by the EHT\cite{EventHorizonTelescope:2019dse, EventHorizonTelescope:2022wkp}, there are accretion disks. Generally, the accepted model for Active Galactic Nuclei (AGN) is an SMBH with an accretion disk\cite{Padovani:2017zpf}.

For IMBHs, since their origin and distribution is poorly understood, their local environment can fare no better. Similar to their big siblings, they could also have accretion disks and a distribution of stars around them, especially at the center of globular clusters or dwarf galaxies. 

Depending on their growth and merger history, in the \lcdm{} model, MBHs can also host a DM spike\cite{Ullio:2001fb}.

Modeling environmental effects is of significant importance. First, to accurately detect I/EMRIs, we need accurate waveforms, and environmental effects can impact them significantly\cite{Zwick:2022dih}, possibly leading to misidentification or oversight of systems if not properly modeled\cite{Cole:2022fir}. Second, correctly modeling these environmental effects can shine a light on these astrophysical processes\cite{Barack:2018yly, Speri:2022upm}. Therefore, in this chapter, we will discuss the most prominent environmental effects and try to model their impact. We will explore DM spikes, accretion disks, and the stellar distribution, compare and contrast them, and discuss relativistic effects.

This is understood to be a first linear approximation. We will superpose the effects of these models, even though they might have more complex interactions that modify them individually.

\section{Dark Matter Spike \label{sec:environment:spike}}
\subsubsection{Spike Distribution}
Structure formation predicts that the baryonic matter and DM distributions are closely linked. According to the prevalent model, halos form from overdensities in DM in the young cosmos in a bottom-up fashion, into which the baryonic matter follows, creating stars and galaxies\cite{Madau:2014bja}. This means most galaxies are embedded inside a DM halo\cite{Wechsler:2018pic}.

The existence of DM spikes was first postulated in \cite{Gondolo:1999ef}. A central black hole surrounded by a spherically symmetric DM halo can -- as it accretes slowly in mass -- concentrate the distribution into a spike. This requires the growth to be \textit{adiabatical}, slow enough to maintain adiabatic invariants. In the spherical case, these are the angular momentum and the radial action\cite{Ullio:2001fb}.

We can model the spike density as a power law $\rho(r) \propto r^{-\gamma}$. For an initial DM distribution with slope $0<\gamma_i<2$, the adiabatically grown spike slope is $\gamma = (9-2\gamma_i)/(4-\gamma_i)$. This implies $2.25< \gamma < 2.5$. 
For a spike grown from an NFW profile with $\gamma_i=1$, the power law index is $\gamma = 7/3$. The spike extends to a range $r_\sp$ such that the spike mass $m(r_\sp)$ is comparable to the central MBH mass $m_1$.

This relies on a few assumptions \cite{Ullio:2001fb, Coogan:2021uqv}: (i) The MBH is at the center of the spherically symmetric DM Halo, (ii) the MBH grows adiabatically -- on timescales long compared to the dynamical timescale of the halo, and (iii) there are no large scale gravitational disruptions such as mergers or tidal encounters.
If these conditions are not met, the spike can be shallower. If (i) is not given, the MBH has to spiral into the center via dynamical friction, which can take up to a Hubble time from $1$kpc, accreting some of the DM and heating the distribution. The resulting slope would be much shallower at $\gamma = 1/2$. If (ii) is not given, and the MBH grows much faster -- almost instantaneously w.r.t. the dynamical timescale -- the resulting slope would be $\gamma = 4/3$. For gravitational disruptions (iii) it is difficult to make predictions, as they could unbind the spike completely. This will be discussed later in section \ref{sec:outlook:spikes}.

A relativistic analysis of adiabatic growth has been carried out in \cite{Sadeghian:2013laa}. Very close to the MBH $r \lesssim 10 r_\isco$, these effects enhance the density. This can also be applied to rotating MBH in a Kerr spacetime\cite{Ferrer:2017xwm}.

Additionally, inside galactic centers or globular clusters, there would be a distribution of stars around the MBH. This distribution of stars heats up the DM distribution through gravitational encounters in a process called \textit{kinematic heating}. Many studies find, regardless of initial distribution, a power law index of $\gamma = 3/2$ for this case\cite{Gnedin:2003rj, Merritt:2003qk, Bertone:2005xv, Vasiliev:2008uz, Shapiro:2022prq}. This could even happen when there is no initial spike or a depleted one, it can be dynamically \mbox{(re-)generated} through kinematic heating\cite{ Bertone:2005xv, Vasiliev:2008uz}.

For spikes that have formed around PBH, simulations have shown that the power law index would change with radius between $\gamma=3/4 - 9/4$\cite{Boudaud:2021irr, Escriva:2022duf}.

If DM is not collisionless, a sufficiently strong self-interaction can thermalize the DM distribution with $\gamma = 7/4$ \cite{Shapiro:2014oha, Alvarez:2020fyo}. DM self-annihilations on the other hand, can result in a flat density core when the annihilation becomes efficient\cite{Gondolo:1999ef, Aschersleben:2024xsb}. 

If DM is not a particle, there can be more exotic distributions, see for example the gravitational atom\cite{Hannuksela:2019vip, Cole:2022fir}. In this dissertation, we will not explore these effects but assume the standard model of cosmology with DM as a particle.

All of these models leave a range of possible power laws for the DM distribution in the vicinity of the MBH. Until now, these considerations have been theoretical. There have been tentative observations that can be explained by the existence of DM spikes\cite{Chan:2022gqd, Chan:2024yht}, but these are far from conclusive and leave out a comprehensive study of other possible effects. Measurements of the stellar orbits at the center of the Milky Way have been able to put constraints on the local DM spike\cite{Gillessen:2008qv, Shen:2023kkm}, which excludes some of the steeper spike models in the center of our galaxy.

In conclusion, a variety of power law indices is possibly realized $0.5 \leq \gamma \leq 7/3$. 
We will consider two spike profiles, one that is kinematically heated with $\gamma=3/2$, and one that is adiabatically grown $\gamma=7/3$. The former $\gamma=3/2$ is more physically motivated when we combine the DM spike with the stellar distribution, but generally has a smaller density and therefore less observational impact. In the presence of a stellar distribution, we also expect larger rates of I/EMRIs, as the stellar distribution is a prime source for secondaries. The latter $\gamma=7/3$ is more physically motivated in an isolated system, which has had time to grow its spike and has not lost it since. Here, the densities can become large and cause a significant observational impact. We will discuss these choices more later on.
In the following, we will quantify the spike and its impact on the inspiral.

\subsubsection{Description}
We assume a power law distribution for the DM spike and parameterize it as \cite{Coogan:2021uqv}
\begin{equation}
    \rho_\dm(r) = \rho_6 \left(\frac{r_6}{r} \right)^{\gamma} \qquad \text{for } r_{in}< r < r_\sp,
\end{equation}
where $\rho_6$ is the density at $r_6 = 10^{-6}$pc. 

We choose the inner radius to be $r_{in} = 4m_1$ following \cite{Sadeghian:2013laa}, below which the density vanishes. The spike radius can be estimated with the help of $r_h$, which is the radius where the DM distribution has a mass comparable to the MBH $m_1 = M_\dm(r_h) = \int_0^{r_h} \rho_\dm r^2dr$, which for the Milky Way SMBH is on the order of pc. The spike radius is estimated to be $r_\sp = 0.2 r_h$ \cite{Eda:2014kra}. 

We can obtain $\rho_6$ by matching the distribution to an NFW profile at $r_\sp$. This NFW profile can be obtained as described in \cite{Eda:2014kra}.\footnote{This procedure uses assumptions from structure formation that, in light of the new JWST results, might have to be modified\cite{Silk:2024rsf}.} This gives for a flat $\gamma=3/2$ spike $\rho_6=5\cdot 10^{11}\Msun/\text{pc}^3$, and for a steep $\gamma=7/3$ spike $\rho_6=9\cdot 10^{16}\Msun/\text{pc}^3$ for an MBH with $m_1=10^5\Msun$.

The power law spike is a spherically symmetric and isotropic distribution. The corresponding phase space distribution of the spike is given through the Eddington inversion procedure as
\begin{equation}
    f(\E) = \rho_6 \left(\frac{r_6}{m_1}\right)^{\gamma} \frac{1}{(2\pi)^{3/2}} \frac{\Gamma(\gamma+1)}{\Gamma(\gamma- \frac{1}{2})} \E^{\gamma - 3/2}, \label{eq:f_eq}
\end{equation}
with the Gamma function $\Gamma(x)$ and where $\E$ is the (positively defined) specific energy
\begin{equation}
    \E = \Psi(r) - \frac{1}{2} v^2
\end{equation}
with the relative Newtonian potential of the central mass $\Psi(r) = \frac{m_1}{r}$.
The spike density is recovered with
\begin{equation}
    \rho_\dm(r) = 4\pi \int_0^{v_\esc(r)} v^2 f\left(\Psi(r) - \frac{1}{2}v^2\right) \d v,
\end{equation}
with the escape velocity $v_\esc = \sqrt{2\Psi(r)}$.

This will be considered in the mass distribution and therefore contribute to periapse precession (\eqref{eq:mass_precession}). The other impacts of the presence of these DM spikes are given in the following.

\subsubsection{Dynamical Friction}
An object moving through this distribution of DM experiences many small gravitational interactions with the particles. For the involved masses, this is primarily described with dynamical friction\cite{Chandrasekhar:1943ys, 2013degn.book.....M}. As the secondary moves through the halo of DM particles, it attracts them to its position. The particles start clustering, while the secondary moves along its orbit. Thus, there is an overdensity or \textit{wake} behind the secondary, tugging it backwards.

We will use the following equations for dynamical friction \cite{Antonini:2011tu, Dosopoulou:2016hbg, Dosopoulou:2023umg}
\begin{align}
    F_\df(r,v) = {}&-16\pi^2 m_2 \rho_\dm(r) \left[ \ln \Lambda\, \alpha(v) + \beta(v) + \delta(v)\right] \frac{\v{v}}{v^3}, \\ 
    \alpha(v) = {}& \int_0^{v} f(v_\s) v_\s^2 \d v_\s, \\
    \beta(v) = {}&  \int_v^{v_\esc} f(v_\s) v_\s^2 \ln\left(\frac{v_\s + v}{v_\s - v} \right) \d v_\s, \\ 
    \delta(v) = {}& -2 v \int_v^{v_\esc} f(v_\s) v_\s \d v_\s.
\end{align}
For the Coulomb logarithm we use $\ln \Lambda = \ln \sqrt{\frac{m_1}{m_2}}$\cite{Kavanagh:2020cfn}. 

This description can be interpreted as follows. The $\alpha(v)$ describes the DM particles that are moving slower than the secondary object. These particles steal momentum from the secondary and slow it down, which is the dominant contribution. The $\beta(v)$ and $\delta(v)$ describe the particles that are moving faster than the secondary and they can actually boost the momentum, weakening the effect of $\alpha(v)$. These two terms were sometimes disregarded in previous literature, but as \cite{Dosopoulou:2023umg} showed, they can have an important effect, especially for smaller power law indices $\gamma\sim 3/2$. Since we are looking at these spikes specifically, we will include the terms here.

These calculations come from a Newtonian background. For relativistic speeds, i.e.., close to the MBH, we can apply correction terms to the dynamical friction equation.
One such term is
\begin{equation}
    F_\df(r,v) \to F_\df(r) \bar{\gamma}^2 (1+v^2)^2
\end{equation}
where $\bar{\gamma} = 1/\sqrt{1-v^2}$ is the Lorentz factor. These have been derived in \cite{Barausse:2007ph} and considered for I/EMRIs in \cite{Speeney:2022ryg}. They find relatively small additional dephasings on circular orbits with the corrective terms. They also include the relativistic effects of spike growth close to the MBH from \cite{Sadeghian:2013laa}, but these do not apply when we consider heated spike distributions with $\gamma=3/2$, as the spike is not grown through the adiabatic growth mechanism of the MBH. We will ignore the relativistic corrections to the spike density in this dissertation.

\begin{figure}
    \centering
    \includegraphics[width=1.1\textwidth]{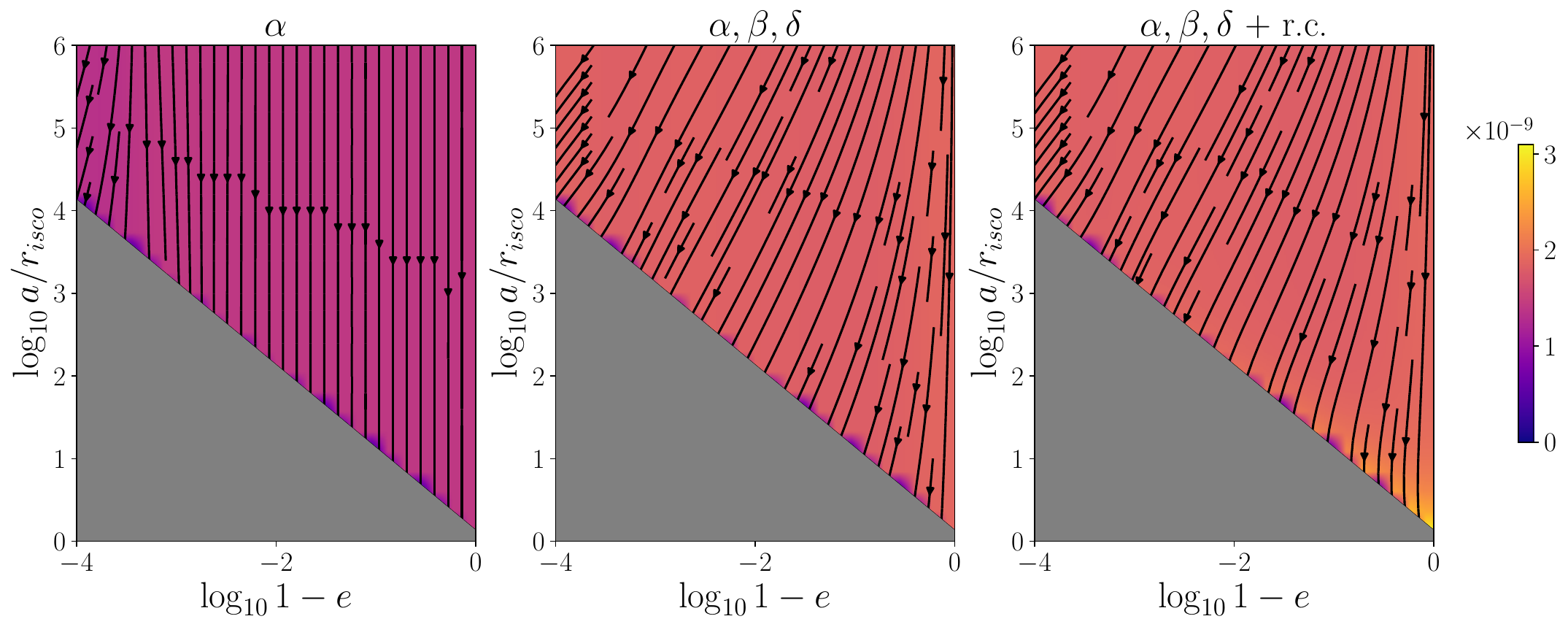}
    \caption{The phase space flow of $a,e$ for dynamical friction inside a DM distribution with power law $\gamma=3/2$. We plot the effects that the terms $\alpha, \beta, \delta$ and the relativistic corrections produce. It can be seen that $\alpha$ by itself does not modify the eccentricity $e$ for $\gamma=3/2$. The introduction of $\beta, \delta$ increases the eccentricity as described in \cite{Dosopoulou:2023umg}. The relativistic corrections only have a small circularizing effect on the phase space flow close to the MBH $m_1=10^5\Msun$. }
    \label{fig:psf_dm_terms}
\end{figure}

The phase space flow for the semimajor axis and specific angular momentum $a,j$ is plotted in \figref{fig:psf_dm_terms}. First, the effects of the different terms are made visible for a power law $\gamma\sim 3/2$. It can be seen that the inclusion of $\beta, \delta$ produces the eccentrification described by \cite{Dosopoulou:2023umg}. The relativistic corrections only slightly moderate this effect, as they increase circularization weakly for high velocities close to the MBH.

\begin{figure}
    \centering
    \includegraphics[width=\textwidth]{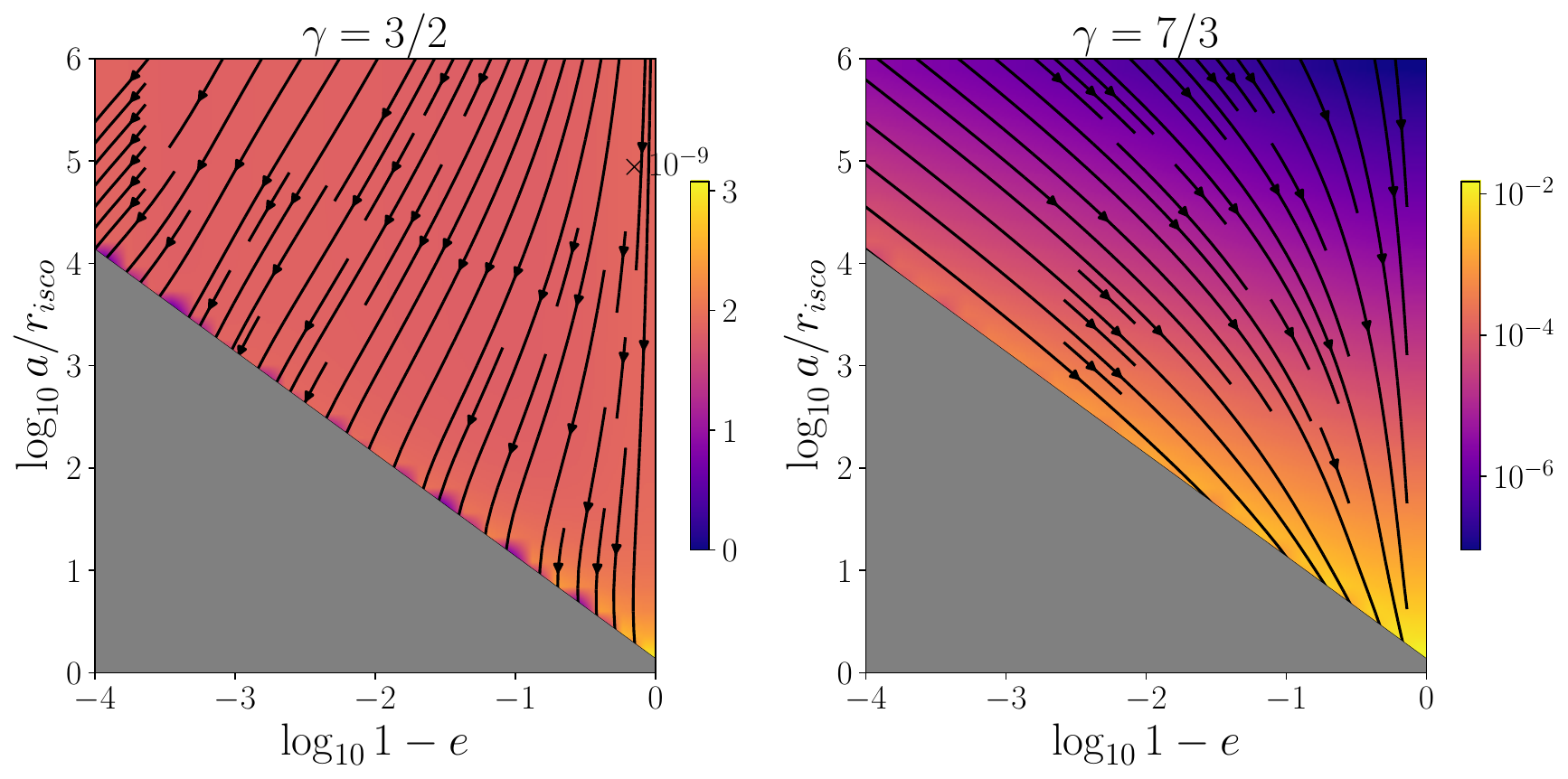}
    \caption{The phase space flow of $a,e$ for dynamical friction inside a DM distribution with different power laws, $\gamma=3/2$ and $\gamma=7/3$. We include all terms described in this section. It can be seen that the smaller power law $\gamma=3/2$ is eccentrifying, while $\gamma=7/3$ is strongly circularizing. Notice also the stark difference in the strengths of the effects, and the flatness and steepness, respectively. }
    \label{fig:psf_dm_pwrlaw}
\end{figure}

We also show the phase space flow (with all terms described here) for two different power laws, $\gamma=3/2$ and $\gamma=7/3$ in \figref{fig:psf_dm_pwrlaw}. It can be seen that $\gamma=3/2$ is eccentrifying, while $\gamma=7/3$ is strongly circularizing, as shown by \cite{Dosopoulou:2023umg}.


\subsubsection{Accretion}
If the secondary is an sBH, it will also cause accretion of the DM particles. If the secondary is a neutron star or normal star, it is unlikely to capture any significant amount of DM particles\cite{Bertone:2007ae}.

For the BH case, the accretion can be modeled as follows. The mass gain acts as a drag term \cite{Yue:2017iwc}
\begin{equation}
    F_\text{acc} = \dot{m}_2 v
\end{equation}
which does not include any recoil effects.
The mass gain is given by 
\begin{equation}
    \dot{m}_2 = \sigma \rho_\dm v
\end{equation}
where $\sigma$ is the cross section of the secondary BH.

For collisionless DM, this can be modeled as \cite{Unruh:1976fm, Yue:2017iwc}
\begin{align}
    \sigma_\text{Collisionless} &= \frac{\pi m_2^2}{v^2}  \frac{512(1-v^2)^3}{4(1-4v^2 + \sqrt{1 + 8v^2})(3 - \sqrt{1+8v^2})^2}  \label{eq:dm_acc_cs}\\
    &\approx_{v\ll 1} \frac{16\pi m_2^2}{v^2}. \nn  
\end{align}

The mass gain can be modeled as an additional equation in the ODE system. At the same time, even though the terms for $a,e$ contain $m_2$, their time derivatives are not affected, see appendix \ref{sec:appendix:accretion}.

This model was explored in \cite{Yue:2017iwc} and they found the accretion effects to be subdominant to dynamical friction. The mass gain of the secondary was on the order of $0.1\%$. Nonetheless, it can still cause large dephasings and for long timescales it should be modeled to produce accurate waveforms.

The cross section \eqref{eq:dm_acc_cs} was derived for an object moving through a homogeneous flow, so it does not take the phase space distribution of the DM into consideration. This has recently been considered in \cite{Karydas:2024fcn}.

\begin{figure}
    \centering
    \includegraphics[width=0.6\textwidth]{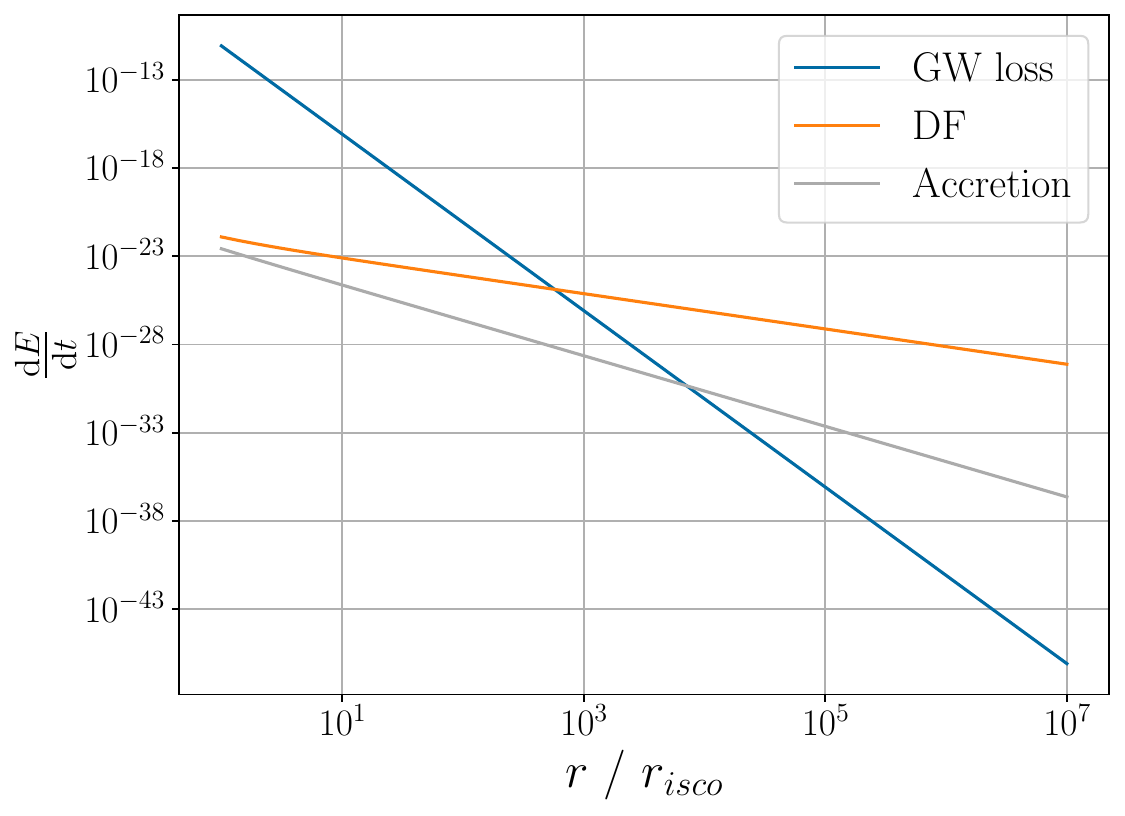}
    \caption{The energy loss $\dv{E}{t}$ for the different dissipative forces on a circular orbit for the given radii. Accretion is always subdominant to dynamical friction. At small separations, the GW emission dominates the energy loss. The system parameters are $m_1=10^5\Msun, m_2=10\Msun, \gamma=3/2, \rho_6=5\cdot 10^{11}\Msun/\text{pc}^3, r_\sp=2$pc. }
    \label{fig:dm_df_vs_acc}
\end{figure}

We plot the energy loss of dynamical friction and accretion and compare it with that of GW on a circular orbit in \figref{fig:dm_df_vs_acc}. The accretion effects are always subdominant to those of dynamical friction. Also, at close distances, the GW emission clearly dominates the energy loss. For eccentric orbits (and semimajor axes instead of radii), the GW emission energy loss would further increase, while the DM effects would stay at similar sizes.

\begin{figure}
    \centering
    \includegraphics[width=\textwidth]{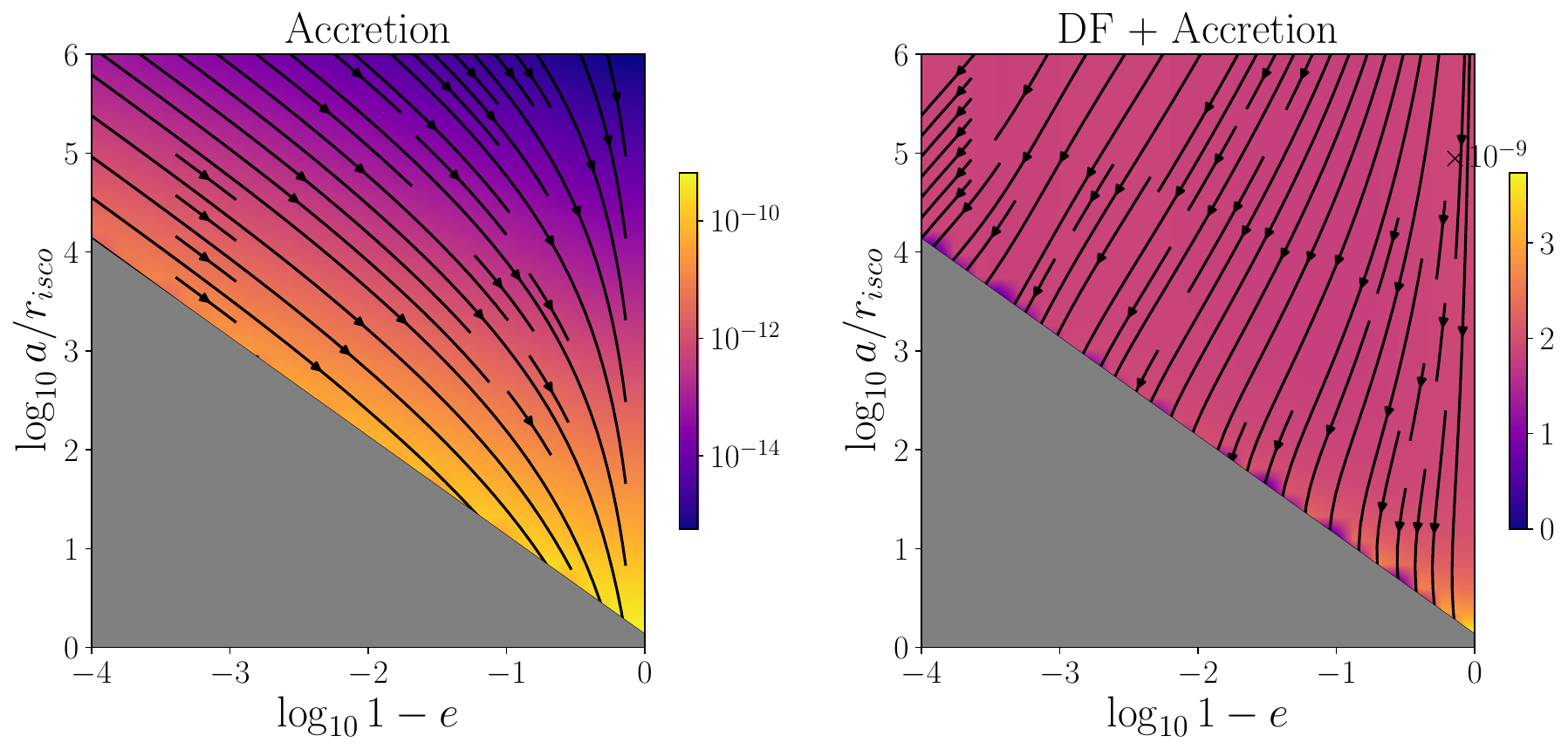}
    \caption{The phase space flow of $a,e$ for accretion and accretion and dynamical friction. Accretion is circularizing, but strongly subdominant to the dynamical friction, so it does not change the total system as compared to \figref{fig:psf_dm_terms}.}
    \label{fig:dm_acc_df_psf}
\end{figure}

We plot the phase space flow of accretion in \figref{fig:dm_acc_df_psf}. Since $F_\text{acc} \propto r^{-\gamma} v^{0}$, accretion is circularizing, as can be seen in the picture. But as it is weaker than the dynamical friction, it barely modifies the total phase space flow of the DM effects.

\subsubsection{Halo Feedback}
As the secondary loses its kinetic energy due to dynamical friction, this energy is injected into the DM halo. Here, it can increase the orbital energy of the orbiting particles and possibly unbind them. The authors of \cite{Kavanagh:2020cfn} have developed a framework to describe these effects, the \textit{halo feedback} model. It promotes the halo from a static spectator to a dynamic actor in the inspiral. The model is based on circular orbits. They found that for larger mass ratios $10^{-4} \leq q$, the secondary can inject enough energy to deplete their local orbital radius of DM, thus reducing the amount of dynamical friction and prolonging the inspiral. As the object inspirals, it refills the depleted regions with particles of lower radii. Close to the MBH, the GW emission dominates and there is not enough time to significantly affect the distribution.

The model was extended in \cite{Nichols:2023ufs, Karydas:2024fcn} to include the accretion effects described above. They remove DM particles from the phase space and cause further depletion. As dynamical friction effects get weaker, the accretion becomes relatively stronger and should be taken into account.

Note that \cite{Mukherjee:2023lzn} has published a similar investigation and found the halo feedback model insufficient on longer timescales. They argue that three-body effects between the primary, secondary, and DM particles become important and can unbind the spike. This is adressed in section \ref{sec:outlook:spikes}.

Initially, the halo feedback model only worked for circular inspirals. During the finalization of this dissertation, the formalism was extended to a moderate eccentricity ($e \lesssim 0.8$)\cite{Karydas:2024fcn, Kavanagh:2024lgq}. Unfortunately, we did not have time to implement this. Instead, we will try to estimate the energy injected into the distribution and compare it to the binding energy of the spike. This gives a measure of whether the halo feedback effects would be relevant.

Once we have a solution to the system $a(t), e(t)$, we can go back and integrate over the energy loss of the dissipative forces. The energy lost due to dynamical friction is simply given by
\begin{equation}
    \Delta E_\df = \int_{t_0}^{t_\text{fin}} \dv{E_\df}{t} \d t.
\end{equation}

The binding energy of the system can be estimated following \cite{Kavanagh:2020cfn}. The potential energy of a shell of DM is given by
\begin{equation}
    \d U_\dm(r) = - \frac{m_1 + m_\dm(r)}{r} \, 4\pi r^2 \rho_\dm(r) \d r.
\end{equation}
For the spike system we are considering, integration from some initial radius $r_{in}$ to $r$ gives 
\begin{equation}
    \Delta U_\dm(r) = - \frac{m_\dm(r) (3-\gamma)}{r} \left( \frac{m_1 - m_\dm{r_{in}}}{2-\gamma} + \frac{m_\dm(r)}{5-2\gamma} \right) - U_{in},
\end{equation}
with 
\begin{equation}
    U_{in} = - \frac{m_\dm(r_{in}) (3-\gamma)}{r_{in}(2-\gamma)} \left( m_1 - \frac{m_\dm(r)(3-\gamma)}{5-2\gamma} \right) .
\end{equation}
The total binding energy is then given by $U_\dm = \Delta U_\dm(r_\sp)$.

We can use the \textit{feedback ratio} 
\begin{equation}
    r_\df = \frac{\Delta E_\df}{U_\dm} \label{eq:feedback_ratio}
\end{equation} 
to estimate how much energy is being inserted into the spike and whether the energy is enough to affect the spike. For $r_\df \ll 1$, the inserted energy is much smaller than the binding energy, and the spike will not be affected by the inspiral. In this scenario, halo feedback can be neglected. For $r_\df \gtrsim 1$ on the other hand, there is significant energy injection into the spike, such that it can be reshuffled or unbound. In this case, modeling halo feedback is important.

\FloatBarrier

\section{Stellar Distribution\label{sec:environment:stars}}
The stellar distribution inside galactic cores has been theoretically studied and observed for some galaxies, most importantly our own Milky Way center. 
The distribution at the center can be approximately described as isotropic, and therefore with a power law in the energy distribution. Through gravitational interactions, the stars can exchange energies and relax into a steady state distribution\cite{Bahcall:1976aa, Frank:1976uy, Bar-Or:2015plb, Hopman:2006xn}. While the stars move into the loss cone of the MBH, they are replaced by outer stars moving inward. This generally gives a power law index of $\gamma = 7/4$. These models assume that the relaxation time is smaller than the age of the galaxy\cite{2013degn.book.....M}.
There are also effects such as mass segregation, where heavier stars tend to migrate inwards and lighter stars outwards\cite{Hopman:2006xn, Raveh:2020jxg, Balberg:2023opq}.

This can be compared with observations. For the Milky Way, the measurements point to a distribution of $\gamma=1.4$ very close to the MBH, and $\gamma=2$ further out \cite{Genzel:2003cn, Gillessen:2008qv}. This can be explained by the fact that the relaxation time is too long for our Milky Way center. Indeed, approximations show that only smaller and fainter spheroids have had sufficient time to relax\cite{2013degn.book.....M}. If we cannot assume a relaxed system, it has to be treated as an initial value problem, and solved with numerical simulations. Here, we will assume a relaxed distribution, which might only be applicable to IMBHs with masses smaller than Sgr A*.

\subsubsection{Distribution}
Thus, we assume a simple power law distribution for the stars with 
\begin{equation}
    \rho_* (r) = \rho_\text{inf} \left( \frac{r_\text{inf}}{r} \right)^{\gamma} \qquad \text{for } r< r_\text{inf},
\end{equation}
with a power law index of $\gamma = 7/4$. We take the radius of influence $r_\text{inf}$ to be the radius where the stellar mass is comparable to the central MBH $m_1 = M_*(r_\text{inf})$. For low mass cusp galaxies, this can be estimated as $r_\text{inf} = 11\text{pc} (m_1/10^8 \Msun)^{0.58}$\cite{Qunbar:2023vys}. From this follows $\rho_\text{inf}$.

We assume all stars in the stellar distribution to be of the same mass $m_*$, to simplify calculations. This precludes mass segregation and simplifies the model. We will typically choose this to be $m_* = 10\Msun$.

This model will also cause periapse precession due to the mass distribution. The second effect we consider is \textit{two body relaxation}.

\subsubsection{Two Body Relaxation}
As our secondary passes through the field of stars, it has many stochastic gravitational encounters. These can be described with the \textit{diffusion coefficients}, which model the average change in velocity. These can be translated to the average change in energy and angular momentum. 

First, we model a single gravitational encounter. We split the change in velocity into parallel $\Delta v_\parallel$ and perpendicular $\Delta v_\perp$ parts, depending on the direction of motion. The average changes are given by \cite{Bar-Or:2015plb, 2013degn.book.....M} 
\begin{align}
    \avg{\Delta v_\parallel} = & \kappa \, \frac{m_* + m_2}{m_*} F_2(v) , \\
    \avg{(\Delta v_\parallel)^2} =& \frac{2}{3}\kappa v (F_4(v) + E_1(v)) , \\
    \avg{(\Delta v_\perp)^2} =& \frac{2}{3} \kappa v (3 F_2(v) - F_4(v) + E_1(v)),
\end{align}
where $\kappa = 16 \pi^2 m_*^2 \log \Lambda $ and the distribution functions are given by
\footnote{Note that the distribution function $f$ from these references has a different normalization, it is normalized to the number density of the stars $n(r) = \int f(v) \d v$, while in the previous section it was normalized to the mass density $\rho(r) = \int f(v) \d v$. They are related by a factor of the (average) field star mass $m_*$.}
\begin{align}
    E_n(v) =& \int_v^\infty \left(\frac{v_*}{v}\right)^n f(v_*) \d v_* , \\ 
    F_n(v) =& \int_0^v \left(\frac{v_*}{v}\right)^n f(v_*) \d v_* .
\end{align}
Here, averaging refers to averaging over the distribution of the field stars.

To translate this into changes in energy and angular momentum, we use
\begin{align}
    \E = \Phi(r) - \frac{1}{2} v^2 , \\
    J = \abs{\v{r}\cross \v{v}} = \sqrt{am_1 (1-e^2)} ,
\end{align}
and obtain
\begin{align}
    -\avg{\Delta \E} =& \frac{1}{2}\avg{(\Delta v_\parallel)^2} + \frac{1}{2}\avg{(\Delta v_\perp)^2} + v \avg{\Delta v_\parallel} ,\\ 
    \avg{(\Delta \E)^2} =& v^2 \avg{(\Delta v_\parallel)^2} ,\\
    \avg{\Delta J } =& \frac{J}{v} \avg{\Delta v_\parallel} + \frac{r^2}{4J} \avg{(\Delta v_{\perp})^2} ,\\ 
    \avg{(\Delta J)^2 } =& \frac{J^2}{v^2} \avg{(\Delta v_\parallel)^2} + \frac{1}{2}( r^2 - \frac{J^2}{4v^2}) \avg{(\Delta v_{\perp})^2} ,\\ 
    \avg{\Delta \E \Delta J } =& - J\avg{(\Delta v_\parallel)^2}.
\end{align}
Note that in the case of $m_2 \gg m_*$ the $\avg{\Delta v_\parallel}$ dominates, and the equations for dynamical friction from the previous section can be recovered. 

To arrive at the secular energy and angular momentum loss, these equations have to be averaged over an orbit\cite{Bar-Or:2015plb}\footnote{Beware of the different conventions for the orbital energy used in the reference $E=\frac{m_1}{2a}$ and in this dissertation $E=-\frac{m_1 m_2}{2a}$.}
\begin{align}
    D_E  ={}&  -\frac{m_2}{T}\int_0^T \avg{\Delta \E} \d t, \\
    D_L  ={}&  \frac{m_2}{T}\int_0^T \avg{\Delta J} \d t, \\
    D_{EE} ={}& \frac{m_2^2}{T}\int_0^T \avg{\Delta \E^2} \d t, \\
    D_{LL} ={}& \frac{m_2^2}{T}\int_0^T \avg{\Delta J^2} \d t, \\ 
    D_{EL} ={}& -\frac{m_2^2}{T}\int_0^T \avg{\Delta \E \Delta J} \d t.
\end{align}
The latter coefficients describe the (co-)variance of the energy and angular momentum changes. These are the coefficients of the Fokker-Planck equations for the phase space $p$\cite{Bar-Or:2015plb}
\begin{align}
    \pdv{p}{t} =& - \pdv{}{E}[D_E p] + \frac{1}{2}\pdv[2]{}{E}[D_{EE} p] \\
    &{} - \pdv{}{L}[D_L p] + \frac{1}{2}\pdv[2]{}{L}[D_{LL}p] + \pdv[2]{}{E}{L}[D_{EL}n] .\nn
\end{align}
This corresponds to the SDE as described by \eqref{eq:fokker_planck}
\begin{equation}
    d\begin{bmatrix}
        E \\
        L
    \end{bmatrix} = 
    \begin{bmatrix}
        D_E \\ 
        D_L
    \end{bmatrix} dt + \\ 
    \bm{\sigma} d\v{W}, \label{eq:EL_SDE}
\end{equation}
where  $\v{W}$ is a 2-dimensional Brownian motion, and $\bm{\sigma}$ is a solution to 
\begin{equation}
    \bm{\sigma}\bm{\sigma}^T = \v{D} = 
    \begin{bmatrix}
        D_{EE} & D_{EL} \\ 
        D_{EL} & D_{LL}
    \end{bmatrix}.
\end{equation}

\begin{figure}
    \centering
    \includegraphics[width=0.6\textwidth]{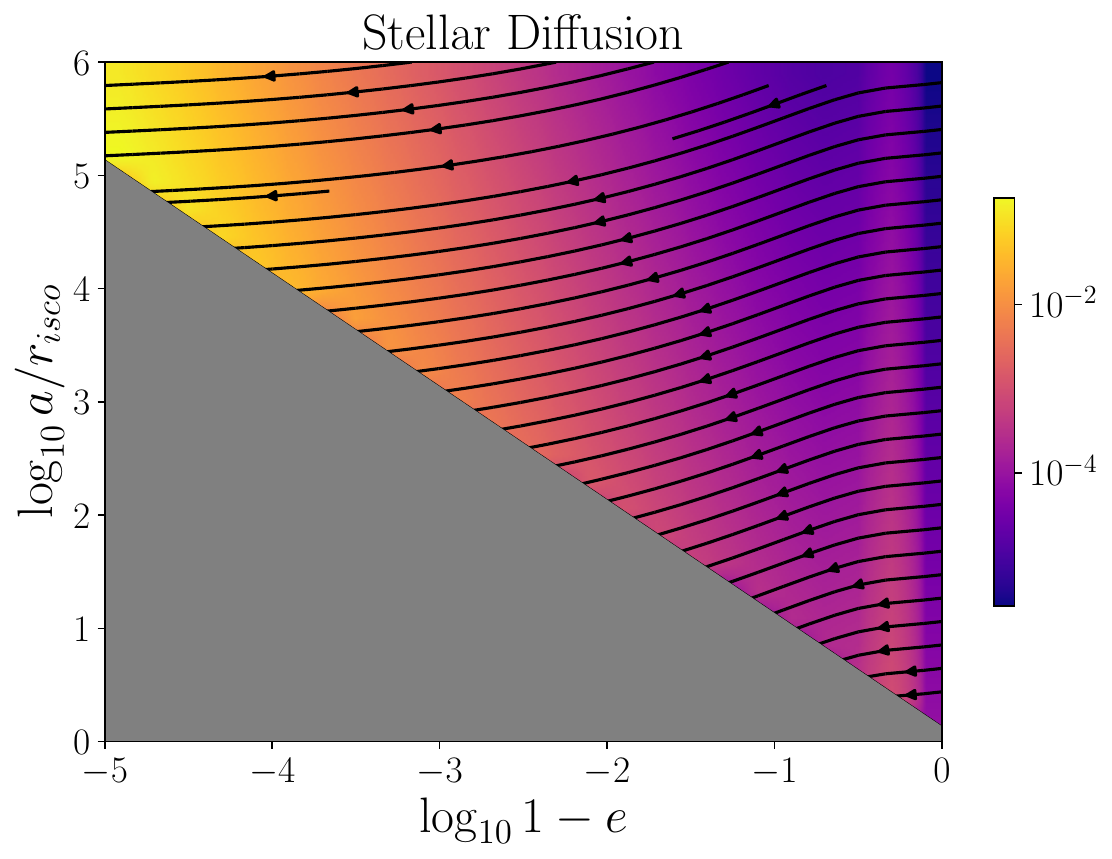}
    \caption{The phase space flow in $a,e$ due to stellar diffusion. The strong eccentrification effects are clearly visible and drive test particles straight into the loss cone. }
    \label{fig:sd_psf}
\end{figure}

The SDE containing $(a,e)$ that is solved in the code can now be obtained with \eqref{eq:sde}.
The phase space flow, corresponding just to the dissipative part of this SDE, is plotted in \figref{fig:sd_psf}. The strong effects on the angular momentum can be seen. The relaxation due to the diffusion processes generally drives objects to high eccentricity and onto capture orbits\cite{Hopman:2005vr}. This is a great source for highly eccentric I/EMRIs, as we will discuss later. 

Of course, this simple description is only a first approximation. There are various other processes involved, such as resonant relaxation\cite{Bar-Or:2015plb}, mass segregation, tidal effects, etc. For a review, see \cite{Amaro-Seoane:2012lgq}.

In this dissertation, we will look at both the SDE \eqref{eq:EL_SDE} and the ODE just containing the dissipative part. The stochastic part is primarily relevant at larger separations and impacts the rate of plunges vs inspirals\cite{Qunbar:2023vys}, more on this in chapter \ref{chap:rates}. Late in an inspiral, firstly, the stochastic terms become inadequate descriptions (since close to the MBH the distribution is difficult to describe as isotropic), and secondly, GW emission dominates the inspiral, minimizing the effects. We will use the SDE description for rates and stick to the ODE system to describe inspirals. 

\FloatBarrier
\section{Accretion Disk\label{sec:environment:disk}}
There is direct observational evidence for accretion disks around SMBHs \cite{EventHorizonTelescope:2019dse, EventHorizonTelescope:2022wkp}. The currently accepted model for AGN includes an accretion disk around SMBHs \cite{Antonucci:1993sg, Urry:1995mg, Padovani:2017zpf}. Thus, MBHs in the centers of galaxies can reasonably host an accretion disk\cite{Greene:2019vlv}.

\subsubsection{Distribution}
There is a considerable amount of literature on the properties of accretion disks\cite{Abramowicz:2011xu, Kocsis:2011dr}. Some of the oldest models were developed in 1972 by Shakura \& Sunyaev \cite{Shakura:1972te}, describing a geometrically thin disk dominated by radiation and gas pressure. Assuming a viscosity description, one can obtain numerical solutions for the disk distribution\cite{Sirko:2002ex}.
Further out, where viscosity dissipation heating is less efficient, these models appear to be gravitationally unstable, and the gas can start forming clumps and structures\cite{1964ApJ...139.1217T}. This could be a source for I/EMRIs. There have been numerous modifications and improvements to these models, applying some mechanism to heat the disk and keep it marginally stable \cite{Sirko:2002ex, Thompson:2005mf, Derdzinski:2022ltb}.
There have been other proposals of disks being dominated by their magnetic fields, such as magentically arrested disks\cite{Narayan:2003by} or `hyper-magentized' disks\cite{Hopkins:2023lgk}. 

We will follow Derdzinski \& Mayer\cite{Derdzinski:2022ltb} and use a thin, Keplerian, steady-state accretion disk. They are based on the model of Sirko \& Goodman\cite{Sirko:2002ex}, with opacity descriptions by Bell \& Lin\cite{Bell:1993qi}. 

The model assumes that the MBH $M=m_1$ accretes at a constant rate 
\begin{equation}
    \dot{M} = 3\pi \nu \Sigma,
\end{equation}
where $\nu$ is the kinematic viscosity, $\Sigma$ the surface density, and $\Omega =\sqrt{M/r^3}$ the Keplerian orbital frequency. The viscosity can be parameterized by the dimensionless viscosity parameter $\alpha$, which describes the relation between the viscosity, the soundspeed $c_s$ and the scale height of the disk $H$ as $\nu = \alpha c_s H$. $\alpha$ can be estimated from observations as $\alpha \sim 0.1$\cite{King:2007cu}. 
The sound speed is given by 
\begin{equation}
    c_s^2 = \frac{p_\text{rad} + p_\text{gas}}{\rho_\b},
\end{equation} 
with the radiation pressure $p_\text{rad}$, the thermal gas pressure $p_\text{gas}$, and the disk density 
\begin{equation}
    \rho_\b = \Sigma/2H,
\end{equation}
where the scale height is also related to the sound speed as
\begin{equation}
    H = \frac{c_s}{\Omega}.
\end{equation}
The effective temperature is determined via viscous dissipation
\begin{equation}
    T_\text{eff}^4 = \frac{9}{8\sigma} \Omega^2 \nu \Sigma,
\end{equation}
whereas the midplane temperature is given by the radiatively efficient transport of photons to the disk surface
\begin{equation}
    T_\text{mid}^4 = \tau_\text{eff} T_\text{eff}^4,
\end{equation}
with the effective opacity $\tau_\text{eff} = \frac{3}{8}\tau + \frac{1}{2} + \frac{1}{4\tau}$, interpolating between the optically thin and thick regimes. The optical depth is given by
\begin{equation}
    \tau = \frac{1}{2} \kappa \Sigma,
\end{equation}
where $\kappa$ is the Rosseland mean opacity of the gas at midplane density. This can be approximated piecewise polynomially as
\begin{equation}
    \kappa = \kappa_0 \rho^a T_\text{mid}^b,
\end{equation}
where we take the values for $\kappa_0, a, b$ from the table provided in \cite{Derdzinski:2022ltb}, derived by \cite{Bell:1993qi}.

To complete the set of equations, the gas pressure is given by 
\begin{equation}
    p_\text{gas} = \frac{k_B}{\mu m_H} T_\text{mid} \rho,
\end{equation}
and the radiation pressure 
\begin{equation}
    p_\text{rad} = \frac{\sigma_B}{2} \tau T_\text{eff}^4,
\end{equation}
with the Boltzmann constant $k_B$, the hydrogen mass $m_H$, the mean molecular weight $\mu=0.62$, and the Stefan-Boltzmann constant $\sigma$.
The Toomre stability parameter $Q$ tests whether the disk is gravitationally stable\cite{1964ApJ...139.1217T}. We assume that 
\begin{equation}
    Q = \frac{\Omega^2}{2\pi \rho}
\end{equation}
is always above a critical value of $Q_0 = 1.4$. This gives a simple scaling on the outskirts of the disk.
We assume the density of the disk to extend into the $z$ direction as \cite{Canto:2012bg}
\begin{equation}
    \rho(r,z) = \rho(r) \exp(-z^2 / H^2).
\end{equation}

\begin{figure}[h]
    \centering
    \includegraphics[width=0.5\textwidth]{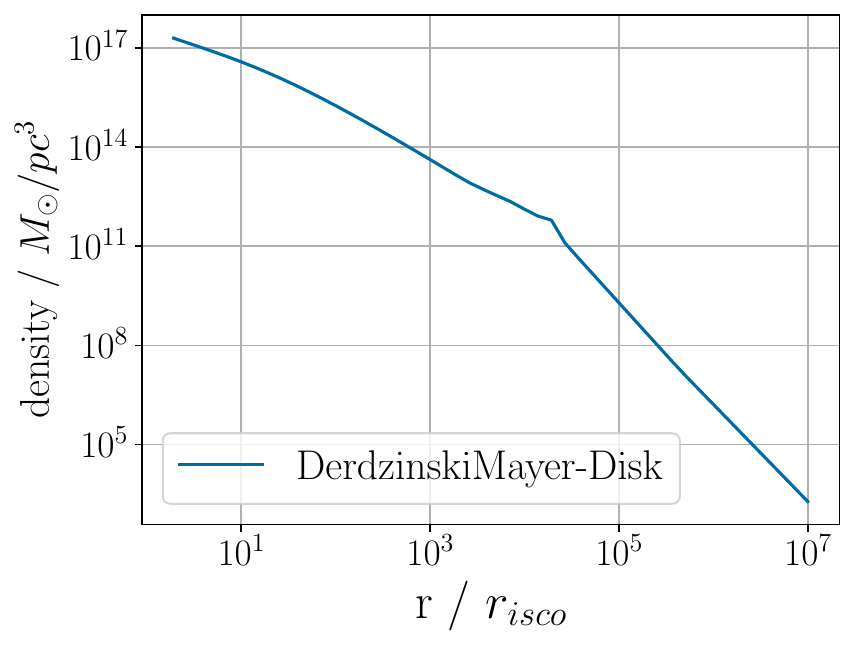}
    \caption{The density distribution of the Derdzinski-Mayer disk around a $m_1=10^5\Msun$ MBH. The kink is the transition to the Toomre stability parameter dominated part of the disk. }
    \label{fig:dm_disk_density}
\end{figure}

The set of equations can be solved numerically, given a choice of central mass $M$, accretion rate $\dot{M}=f_\text{edd}\dot{M}_\text{edd}$ (as a fraction of the Eddington accretion rate), and viscosity parameter $\alpha$. In this dissertation, we use $f_\text{edd} = 0.1$ and $\alpha = 0.1$. The result is shown in \figref{fig:dm_disk_density}.

\subsubsection{Interaction}
So far, it has not been essential to differentiate between sBH and stars as the secondary, or between the prograde vs retrograde motion, as the previous distributions were spherically symmetric. The accretion disk however, is concentrated in a plane and rotates in a given direction. We will use the accretion disk to define the fundamental frame as described in chapter \ref{chap:inspiral}.

For the interaction with the disk, there are several models. It depends on the geometry of the disk, the geometry of the secondary, and the relative speeds. For a black hole as a secondary, there is primarily the Bondi-Hoyle-Littleton accretion and dynamical friction with the gas\cite{Barausse:2007dy,  Sanchez-Salcedo:2019wjx, Sanchez-Salcedo:2020pae, Chatterjee:2023cry, ONeill:2024tnl}. For an extended secondary, such as a star, there is geometric drag from the disk\cite{Fabj:2020qqc}. These effects also depend on the disk geometry, whether we have a thin or a thick disk\cite{Canto:2012bg}. Prograde, quasi-circular motion inside the disk can induce density waves or open gaps in the disk as Type--I and Type--II torques\cite{Kocsis:2011dr, Derdzinski:2018qzv, Derdzinski:2020wlw,Garg:2022nko}. 

There have been a variety of studies looking at I/EMRIs in the presence of an accretion disk\cite{Barausse:2007dy,Kocsis:2011dr, Derdzinski:2018qzv, Derdzinski:2020wlw, Derdzinski:2022ltb, Garg:2022nko,Speri:2022upm, Chatterjee:2023cry, Secunda:2020cdw, Fabj:2020qqc, Nasim:2022rvl, Zwick:2021dlg}, and in comparison to DM effects as in \cite{Cole:2022fir, Becker:2022wlo}. In this dissertation, we will expand our work in \cite{Becker:2022wlo} and look at two models specifically.

One of the most common models originates from planetary formation modeling and is called \textit{Type--I} migration\cite{2002ApJ...565.1257T}. In planetary migration models, the relevant quantity is the torque acting on the secondary.
The equation for this torque is given by \cite{2002ApJ...565.1257T}
\begin{equation}
    \Gamma_\typeI = \Sigma r^4 \Omega^2 q^2 \mathcal{M}_a^2 ,
\end{equation}
which can be translated into a force -- the language of our model -- by
\begin{equation}
    F_\typeI = \Gamma_\typeI \, q / r .
\end{equation}
This can be derived by assuming the creation of density waves in the disk as the secondary orbits, which causes a negative torque on the perturber. This seems to give a decent approximation for the systems we are considering\cite{Derdzinski:2020wlw}, but is only valid for circular orbits.

\begin{figure}
    \centering
    \includegraphics[width=\textwidth]{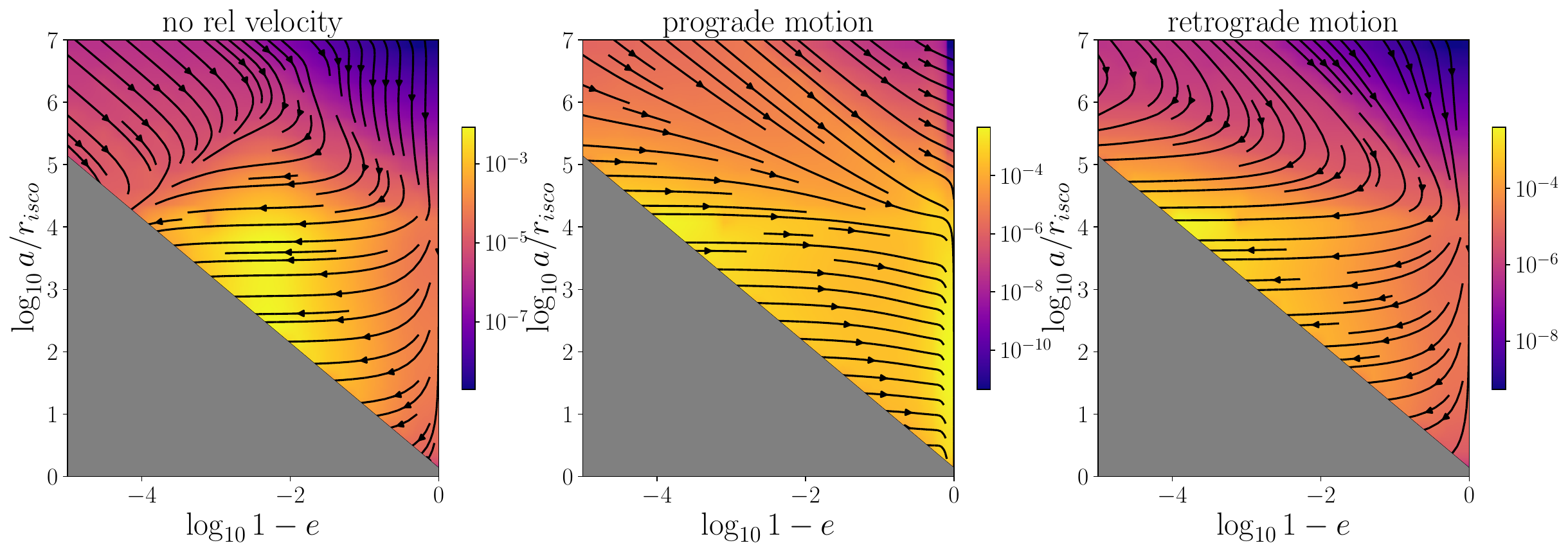}
    \caption{Phase space flow of accretion disk interaction for three different treatments of the relative velocities: Left: No relative velocity is considered. Middle: The secondary is on prograde motion w.r.t. the disk. Right: The secondary is on retrograde motion w.r.t. the disk. The different eccentrification/circularization effects are clearly visible.}
    \label{fig:ad_psf_rv}
\end{figure}

For an sBH on a retrograde orbit, dynamical friction of the secondary with the gas should be the dominant effect \cite{Nasim:2022rvl,Sanchez-Salcedo:2020pae}. We will follow the model of Ostriker\cite{Ostriker:1998fa} and assume for the dynamical friction with the gas 
\begin{equation}
    F_\b =  4\pi m_2^2 \rho_\b(r,z) I  \frac{\bar{\gamma}^2 (1+v_{rel}^2)^2}{v_{rel}^3} \v{v}_{rel},
\end{equation}
with relativistic corrections as in \cite{Barausse:2007dy} and with
\begin{equation}
    I = \frac{1}{2}
    \begin{cases}
        \log\frac{1- v_{rel}/c_s}{1+v/c_s} - v_{rel}/c_s    & \text{subsonic } v_{rel} < c_s\\
        \log (1-(v_{rel}/c_s)^{-2}) + \log \Lambda      & \text{supersonic } v_{rel} > c_s
    \end{cases}
\end{equation}
where $\v{v}_{rel} = \v{v} - \v{v}_{gas}$. 
The relative velocity treatment here is essential, especially when looking at prograde vs. retrograde motion and on inclined orbits. On prograde motion, the same direction as the disk is moving, the dynamical friction will point retrograde at periapse (since the secondary is moving faster than the disk) and prograde at apoapse (where the secondary is slower than the disk). This strongly circularizes the orbit. For retrograde motion of the secondary, the dynamical friction will always point prograde (retrograde wrt to the secondary's motion) and be weaker for larger relative velocity, i.e. at periapse, and therefore eccentrify the orbit. This can be seen in \figref{fig:ad_psf_rv}.

As we will also look at the inclination change, the 3D structure of the disk becomes relevant. Over time, interactions with the disk will align the plane of the orbit of the secondary with the disk. On prograde orbits, the relative velocities are smaller, and therefore the alignment effects are stronger, while on retrograde orbits they can be much weaker.

Initially, we mentioned accretion disks dominated by magnetic fields. These can have high densities and interesting properties. But the interaction models listed here are not necessarily applicable when the disk is highly magnetized, as the derivations do not assume any magnetic fields. To our knowledge there has been no investigation of a secondary within a highly magnetized disk. This has to be left for future work.

\FloatBarrier

\section{Spacetime Effects\label{sec:environment:spacetime}}
Most of the effects described previously assume a Newtonian approximation. In reality, the environment around an MBH is highly relativistic, best described by a Kerr spacetime, and possibly with deformations due to the presence of matter. This introduces relativistic effects, which are important for modeling waveforms and extracting them from observations\cite{Burke:2023lno}. Of course, the most accurate model would be a GRMHD simulation of the I/EMRI system, which has been quite successful for similar mass binaries\cite{DalCanton:2017ala}. Unfortunately, for these mass ratios, the simulations are computationally quite expensive, even without adding any environmental effects\cite{Lousto:2022hoq}. Thus, there has been lots of work to model these relativistic effects, which we will briefly discuss here. Of course, they are not environmental effects in the same sense as the presence of other matter, but they can be considered the spacetime environment and need to be modeled and understood just the same. In this dissertation, we do not consider these effects in the modeling, but they are still listed for completeness. This section can be seen as a To-Do list instead.

\subsubsection{Self-force description}
A first instinct might be to calculate the PN expansion of the energy flux in velocity and mass ratio. As it turns out, the series does not converge in the strong field regime and is not very well suited for I/EMRIs\cite{Yunes:2008tw}. Instead, a more fruitful approach has been the \textit{self--force} model. Here, the Schwarzschild (or Kerr) spacetime is perturbatively disturbed by the secondary within the Teukolsky formalism. The secondary interacts with its own gravitational perturbation and gives rise to this self force and the radiative evolution. For a review, see \cite{Barack:2018yvs}. There is also code being developed to produce these highly accurate waveforms\cite{Katz:2021yft, Vishal:2023fye}. While there has been some significant progress, there are still challenges, with highly eccentric orbits being one of them.

\subsubsection{Orbital Resonances}
Incorporating all these relativistic effects is vital to accurately detecting I/EMRIs and there are certainly interesting consequences. For example, orbits in Kerr spacetime are generally \textit{ergodic} (space-filling), in that they eventually visit all points allowed inside a torus of conserved quantities, in contrast to Keplerian orbits. For some values of conserved quantitites, there are resonance orbits, which are periodic instead of ergodic, meaning they perform an integer number of radial cycles and an integer number of longitudinal cycles at the same time. For these orbits, the self force effects are enhanced and the emitted flux can be much larger\cite{Berry:2016bit,Mihaylov:2017qwn, Isoyama:2021jjd}. The timescale of these resonances is $t_\text{res}\sim m / \sqrt{q}$, which is longer than orbital timescale but much shorter than GW dissipation timescales, and they can therefore dominate the inspiral shortly.

\subsubsection{Too many additional effects}
Incorporating the spin of the primary also impacts the late inspiral considerably. The frame dragging effects change the geodesics and loss cone of the system and can influence the emission of GWs\cite{2013degn.book.....M}. The spin and tidal deformability of the secondary also change the angular momentum evolution and leave important imprints on the waveform \cite{Drummond:2023loz, Burke:2023lno}. 

Another effect that is relevant in I/EMRIs is that of relativistic \textit{quenching}. Loosely speaking, the relativistic precession close to the MBH can dominate the angular momentum loss of the system, basically freezing the angular momentum and slowing the adiabatic inspiral\cite{2013degn.book.....M}.

The presence of another additional massive object can induce perturbations in the spacetime that result in the so called \textit{Kozai Lidov oscillation}. In these three body systems, the tertiary body can induce oscillations in the eccentricity of the orbit of the other two, thus increasing GW emission\cite{Fragione:2018nnl}. Lastly, GW \textit{memory effects} might be detectable\cite{Islam:2021old}.

For a recent review see, \cite{Cardenas-Avendano:2024mqp}.

\subsubsection{Interaction with other environmental effects}
For all of these relativistic effects, the interactions with the environmental effects are of great interest. For example, the presence of additional masses can affect the spacetime\cite{Moore:2017lxy, Polcar:2022bwv} and its resonant orbits\cite{Destounis:2021rko, Strateny:2023edo}. They could `unquench` the inspiral with other sources of energy and angular momentum loss.

The presence of additional mass distributions also changes the emission of GWs. The leading order effect of this is an additional redshift of the signal leaving the mass distribution. For further discussion, see \cite{Duque:2023nrf}.

A first effort to combine the environmental effects with relativistic ones has been made in \cite{Cardoso:2022whc}. They can already describe a coupling between spacetime perturbations and fluid perturbations and see the effects in the GW spectrum. 

\subsubsection{}
Overall, this section might sound like an immeasurable task. But there is significant progress being made, and rapidly\cite{Albertini:2023xcn, Drummond:2023wqc, Cardenas-Avendano:2024mqp}. After all, if we get this right, the payoffs are immense. We can learn about GR, BHs, accretion disks, stellar distributions, DM, galactic cores, galactic mergers, structure formation, and so on. By the time LISA launches in the late 2030s, we can hope for a completely new window into the most extreme regions of \textit{spacetime}.

\chapter{Observational Signatures\label{chap:signatures}}
In this chapter, we will combine the different environmental effects described in chapter \ref{chap:environment}, simulate inspirals, and analyze possible observational signatures. These include the phase space flow, the dephasing, the deshifting of the periapse, and alignment with accretion disks.

This is done with increasing complexity. Firstly, we assume an MBH surrounded solely by a DM spike. This is either a kinematically heated flat $\gamma=3/2$ or an adiabatically grown steep $\gamma=7/3$ spike. Secondly, we add a stellar distribution and consider dry inspirals, which are generally highly eccentric. Thirdly, we add an accretion disk and consider wet inspirals, on a quasi-circular orbit aligned with the disk. Lastly, we look at the alignment process itself.

As an example system, we use an MBH with $m_1=10^5\Msun$, an sBH with $m_2=10\Msun$ at a luminosity distance of $D_L = 50$Mpc. This should be a typical I/EMRI system observable for LISA\cite{Babak:2017tow, Seoane:2024nus}. The corresponding environmental effects can be calculated with the parameters given in \ref{chap:environment}. When the central mass changes, the environmental effects change accordingly. We also plot the dephasing for different central masses, to get an estimate of the dependance on primary mass.

All computations, plots, and further animations are available at \cite{imripy}.

\section{Isolated Spikes \label{sec:signatures:isolated}}
We first look at isolated spikes, as described in section \ref{sec:environment:spike}. These have been investigated classically in the literature of DM spikes\cite{Eda:2014kra}. We look at two spike models. The adiabatically grown steep $\gamma=7/3$ spike is more physically motivated in an isolated system that has had time to grow its spike. The kinematically heated flat $\gamma=3/2$ spike is more physically motivated in the presence of a stellar distribution that heats the DM spike. Regardless of physical motivation, we explore both options in this section in an isolated spike system.

We assume the dissipative forces to be dynamical friction with and accretion of the DM spike, as well as GW emission. First, we will show the phase space flow of the different systems considered. Then, we simulate an example inspiral and discuss the results. This allows us to make inferences for populations of inspirals.

\subsection{Phase Space Flow}
\begin{figure}
    \centering
    \includegraphics[width=\textwidth]{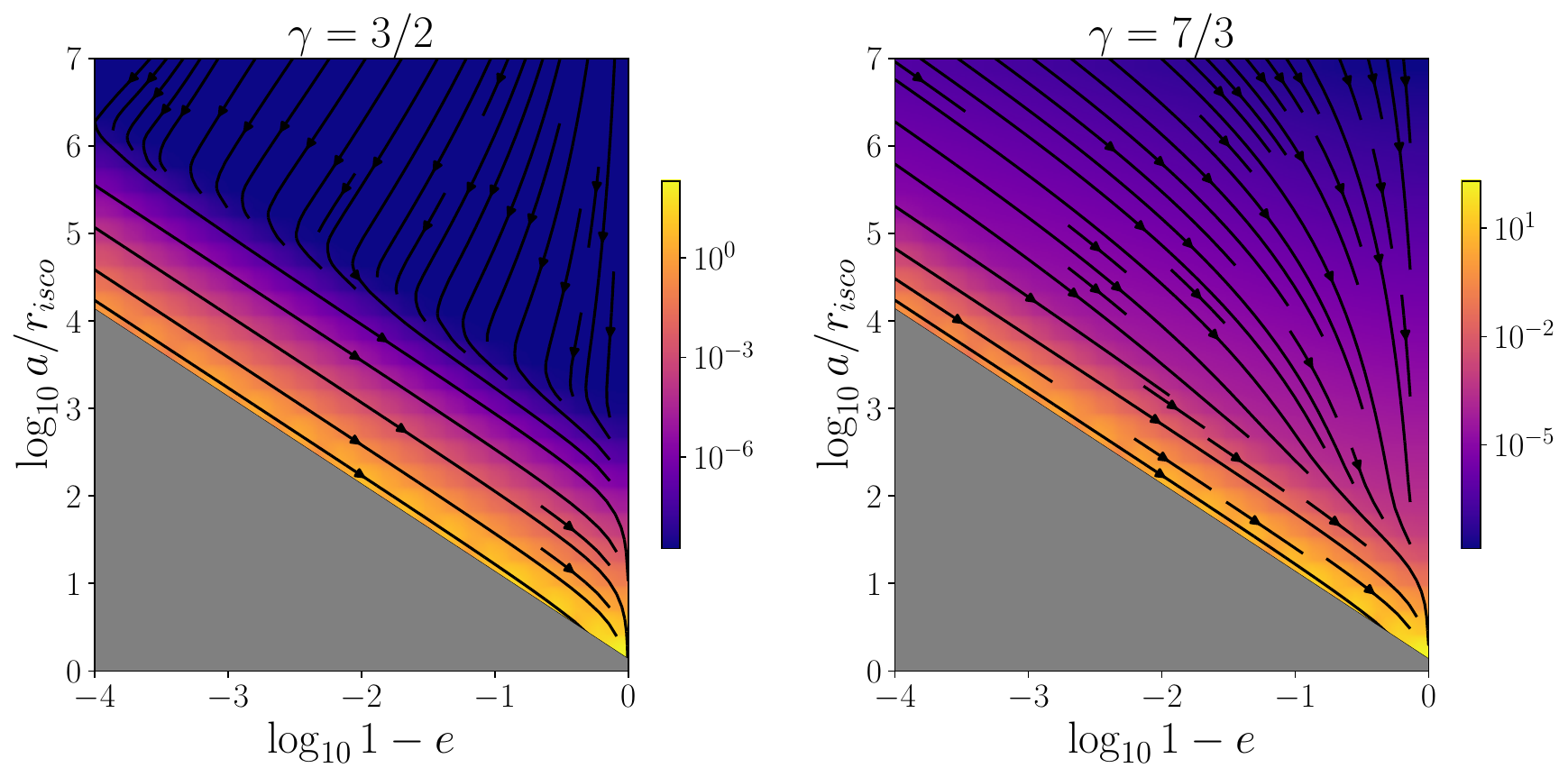}
    \caption{The phase space flow for a system with GW emission and dynamical friction and accretion of a DM spike. The spike laws are a flat $\gamma=3/2$, or a steep $\gamma=7/3$ spike. The eccentrifying and circularizing effects of the flat and steep spikes respectively, are clearly visible. Close to the loss cone, the GW emisssion dominates. There is a clear boundary visible in the phase space, especially in the flat spike case.}
    \label{fig:psf_iso}
\end{figure}

The resulting phase space flow is shown in \figref{fig:psf_iso}. As shown previously, the steep $\gamma=7/3$ spike has circularizing effects, while the flat $\gamma=3/2$ spike is eccentrifying. As expected, for large semimajor axis and low eccentricities, the DM effects dominate, while close to the loss cone the GW emission dominates. As the steep spike gets to much higher densities close to the MBH, its effects tend to dominate more closely toward the loss cone.

The combination of the flat $\gamma=3/2$ spike and GW emission produces a clear boundary in the phase space flow, going from eccentrifying to circularizing. As most objects would be coming in from outside the GW emission dominated zone, they would be transported to and then along the boundary. This gives a typical eccentricity and semimajor axis evolution for these types of inspirals.  Depending on the distance and masses involved, this might happen before the system enters the observable band. In that case we would expect most of these inspirals to happen quasi-circularly. If on the other hand there is a population of closeby systems with a flat spike, we might be able to observe this characteristic behavior. Especially a different normalization of the spike density could move this into the observable band.

The steeper $\gamma=7/3$ spike has the circularizing effects explored in \cite{Becker:2021ivq, Dosopoulou:2023umg}. These inspirals tend to linger in the low eccentricity region, which would make them prime targets for the study of the braking index in the low eccentricity limit, \eqref{eq:braking_index_a}.

Of course, for these isolated systems, \cite{Kavanagh:2020cfn} has shown that the spike is not static and subject to feedback effects. This holds for the flat $\gamma=3/2$ spike as well. Its densities and therefore the energy exchange is less, but the binding energy is comparably lower, and the inspiral time due to lower friction much longer. All in all, this increases the feedback ratio as we will show in the next subsection. Thus, this phase space flow must be modified to take halo feedback effects into account.

\subsection{Sample Inspiral}
\begin{figure}
    \centering
    \includegraphics[width=1.1\textwidth]{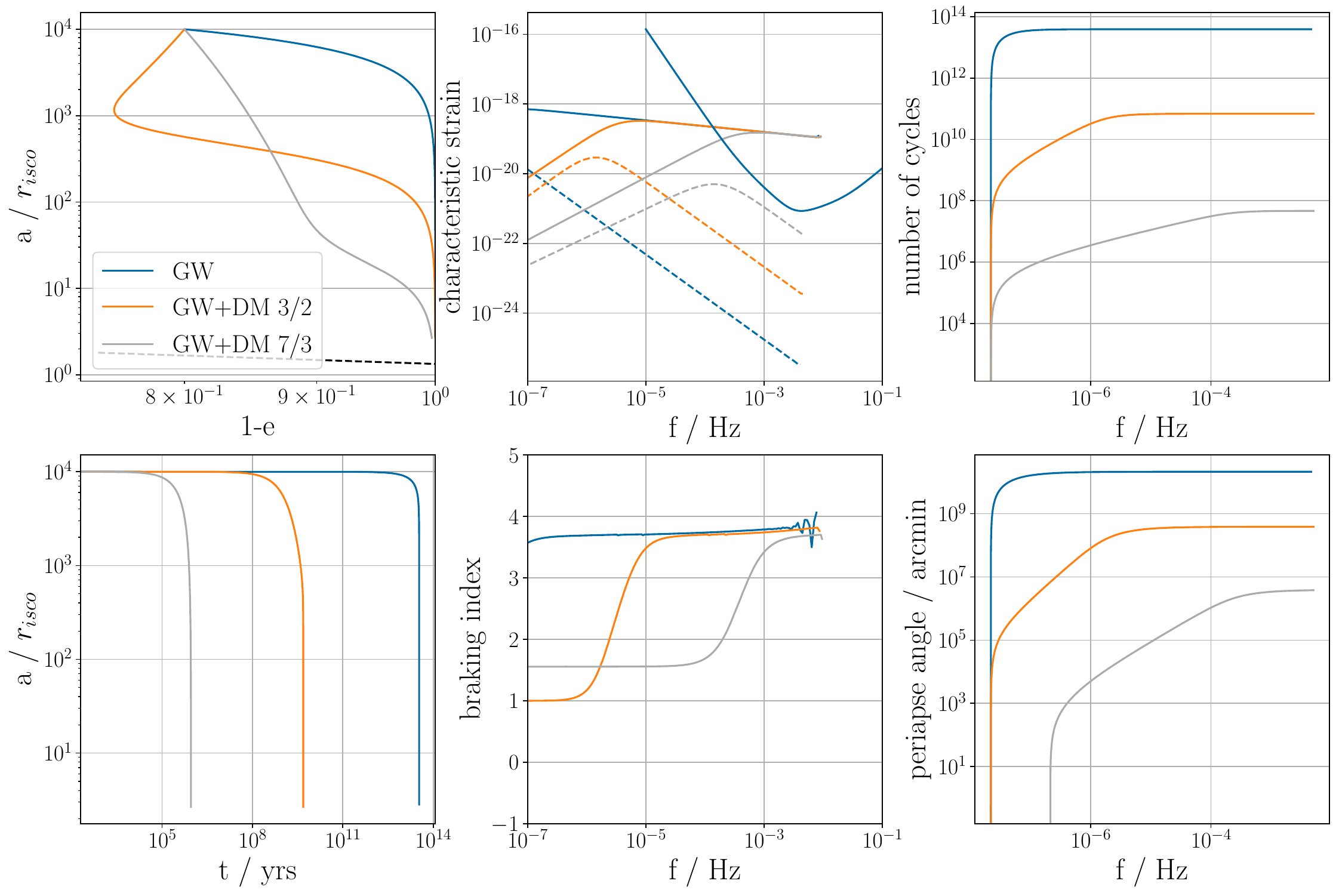}
    \caption{The evolution of the isolated spike system with and without DM spikes for a central               mass of $m_1 = 10^{5}\Msun$ and secondary $m_2=10\Msun$.
            \textbf{Top Left}: The semimajor axis versus the eccentricity evolution. The phase space flow (\figref{fig:psf_iso}) of the different scenarios is reflected here. 
            \textbf{Bottom Left}: The semimajor axis versus the time. The huge time discrepancies can be seen in the different inspirals. 
            \textbf{Top Middle}: The characteristic strain of the systems and the LISA sensitivity. The solid line shows the second harmonic and the dashed line the first. 
            \textbf{Bottom Middle}: The braking index of the system as computed numerically. The different regimes of dominance can be seen.
            \textbf{Top Right}: The number of cycles collected by the systems. This is mostly a reflection of the time it takes to inspiral, but also the different regimes of dominance can be observed.
            \textbf{Bottom Right}: The precession of the periapse. This is generally smaller than the number of cycles, but they have a similar overall structure. Here, the SS precession is dominant.
            }
    \label{fig:ev_iso}
\end{figure}
Now, we can solve the differential equations for the different scenarios. We start with an initial $a_0=10^4 r_\isco$ and an eccentricity $e_0 = 0.2$.
The evolution is shown in \figref{fig:ev_iso}. As expected, the evolution in the $(a,e)$ plane happens as shown by the phase space flow. The different timescales here are of interest. Without any additional dissipative forces, GW emission takes a very long time to inspiral such a system. Accordingly, it also collects a large number of phases, and periapse precession, despite the modest eccentricity. When DM is present, the system inspirals much more quickly, also resulting in a smaller number of cycles and precession. 
The different regimes of the phase space flow can clearly be seen in the characteristic strain, the braking index, and also in the slopes of the cycles and precession. The braking index is almost constant when one force is clearly dominant, characterizing the dominant force.

For the steep $\gamma=7/3$ spike, we have a feedback ratio of $r_{df, 7/3} \approx 0.7$, comparable to the results from \cite{Kavanagh:2020cfn} in the $q=10^{-4}$ case. The flat $\gamma=3/2$ spike has a feedback ratio of $r_{df, 3/2} \approx 16.9$. This means that here, feedback effects can affect the spike and the inspiral, and should be taken into account. Even though the density is lower, the interaction time is much longer and the potential energy of the flat spike is lower, making the feedback ratio higher. Thus, flatter spikes can be subject to stronger feedback effects. They might even be unbound due to the lower binding energy.

\subsection{Dephasing}
\begin{figure}
    \centering
    \includegraphics[width=1.1\textwidth]{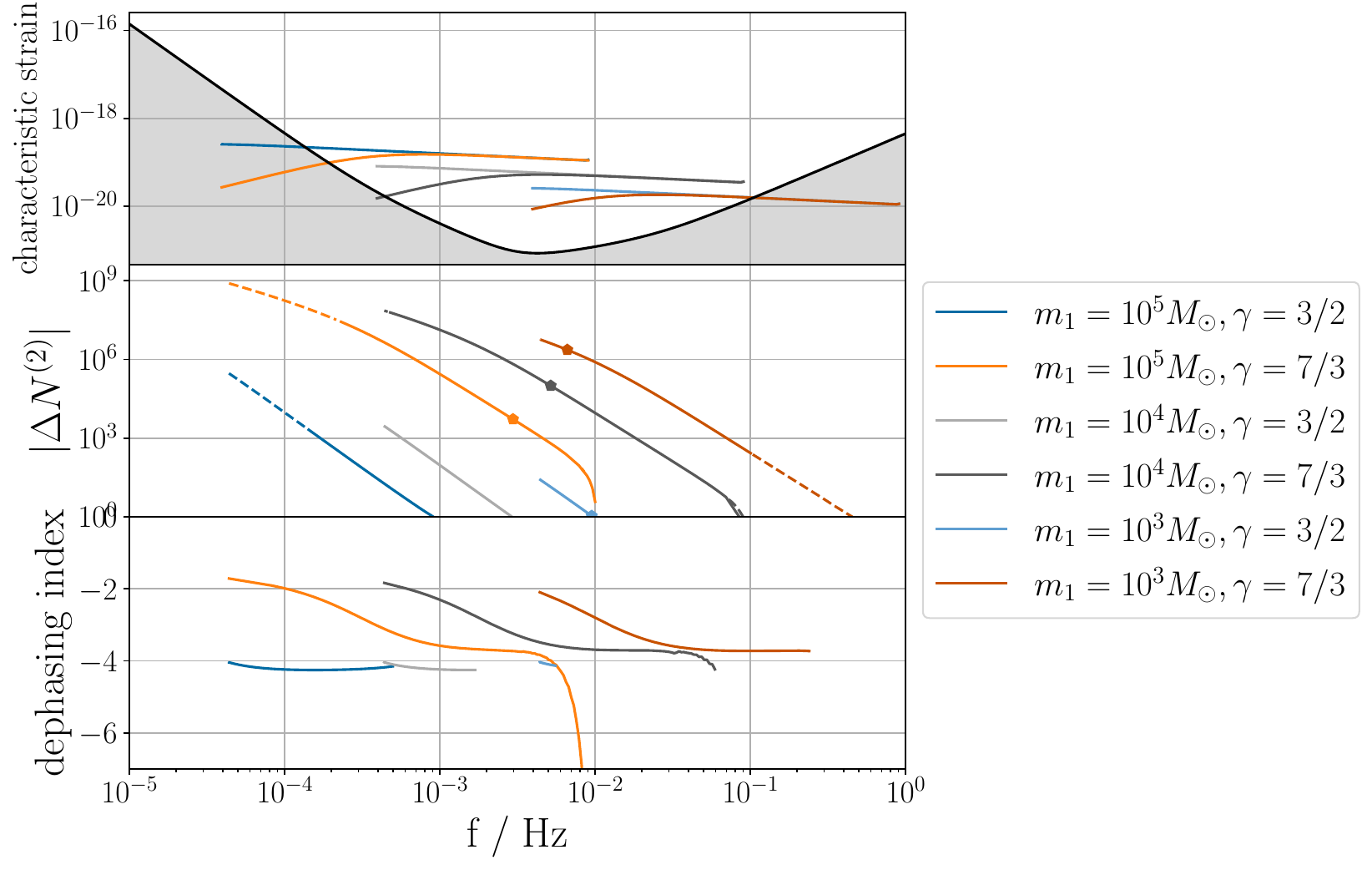}
    \caption{ The results for three different primary masses $m_1=\{10^3, 10^4, 10^5\}\Msun$  with different spike models $\gamma=3/2, 7/3$. \textbf{Top}: The characteristic strain for the inspirals in the second harmonic. 
    \textbf{Middle}: The amount of dephasing in the second harmonic for inspirals in isolated DM spikes. The pentagon marks the point where the inspiral lasts for another $5$yrs, which is potentially observable within the lifetime of LISA. The dashed lines refer to the areas where the second harmonic is outside the observable region. The steep $\gamma=7/3$ spike leaves strong dephasings, also due to the modified eccentricity evolution as seen in \figref{fig:ev_iso}, while the flat $\gamma=3/2$ spike leaves smaller dephasings, and no observable dephasing in the last $5$ years. The flattening of the curves for the steep spike at the end is a numerical artifact. 
    \textbf{Bottom}: The dephasing index, only plotted when there is enough dephasing ($|\Delta N^{(2)}| > 10)$. With the steep spike, the two regimes of dominance can be roughly observed and with the flat spike, the constant value during GW emission domination is visible.  }
    \label{fig:dephasing_iso}
\end{figure}

Now we widen our study to different central masses and study the dephasing of the systems.
Since the GW inspiral takes place on such large timescales, it is difficult to get an accurate number of cycles accumulated. Comparing this to the other inspirals is numerically difficult, as can be seen in \figref{fig:ev_iso}. To assess the dephasing, we reduce our initial semimajor axis to $a_0=10^2 r_\isco$, which is enough to calculate the dephasing for the systems we are interested in. The results for different central masses and the two different spikes are shown in \figref{fig:dephasing_iso}.

The results are comparable to previous inquiries about dephasing\cite{Eda:2014kra, Kavanagh:2020cfn, Becker:2021ivq}. Generally, smaller central masses give smaller timescales for the inspiral. This also means that there is more observable dephasing within the lifetime of LISA. This observable dephasing can be up to $|\Delta N^{(2)}| \sim 10^6$ in the steep spike case. In the flat spike $\gamma=3/2$ case, the dephasings are much smaller. There is technically dephasing, but not in the last $5$ years of an inspiral, which means that the changes would most likely be slow and difficult to detect within the lifetime of LISA. 

For these inspirals, the feedback ratios for the flat spike case are all below $r_{df,3/2} < 1$, meaning feedback effects are small. This is due to starting the evolution at $a_0 = 10^2 r_\isco$, where according to \figref{fig:psf_iso}, the GW emission is already dominant over the dynamical friction. The inspiral does not have enough time to affect the spike so close to the central MBH. 

\subsection{Deshifting}
\begin{figure}
    \centering
    \includegraphics[width=1.1\textwidth]{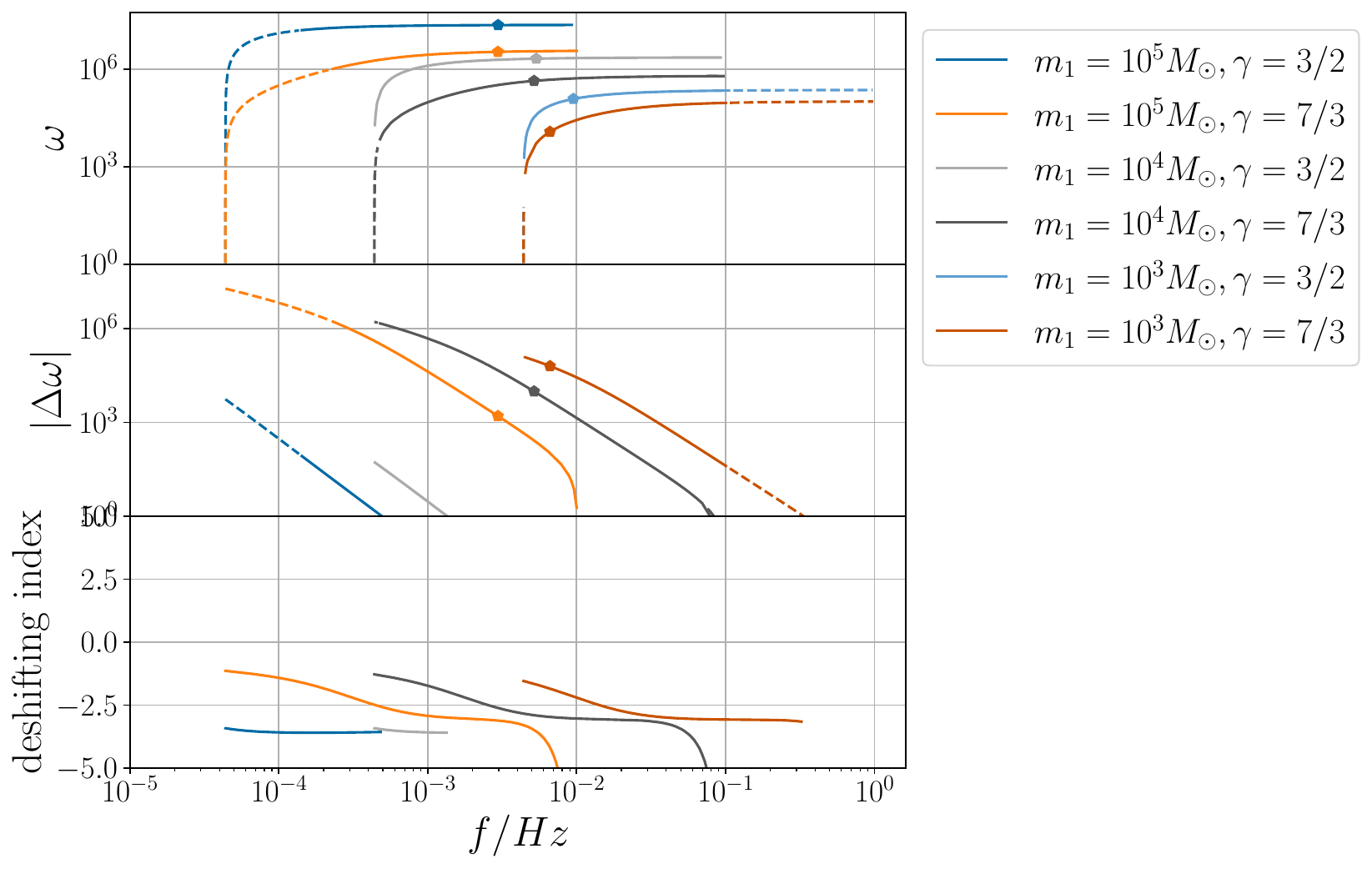}
    \caption{The deshifting for different systems central masses and for the two different spikes.      \textbf{Top}: The total accumulated periapse shift. \textbf{Middle}: The deshifting. The structure is very similar to that of dephasing, only about 2 order of magnitude lower. 
    \textbf{Bottom}: The deshifting index, comparable to the dephasing index. }
    \label{fig:deshifting_iso}
\end{figure}

The total deshifting is shown in \figref{fig:deshifting_iso}. It can be seen that the deshifting is about $2$ orders of magnitude lower than the amount of dephasing observed previously. This means that for the steep $\gamma=7/3$ spike, there is still a considerable amount of deshifting throughout the lifetime of the inspiral and in the observable band. For the flat $\gamma=3/2$ spike, only the heavier systems have deshifting in this timeframe. 

This implies deshifting -- as in the different shift of the periapse due to differences in precession -- has a similar quality to the dephasing. The plots share the same structure. Here, the mass precession is negligible, and the differences observed come down to differences in the SS precession. This is modified due to the accelerated inspiral. This means that dephasing and deshifting measure the same thing in this simplified description. In these low eccentricity cases here, the deshifting is lower than the dephasing. Still, this can be an additional tool in the observation of these systems and identification of the DM spikes.

\FloatBarrier
\section{Dry Inspirals \label{sec:signatures:dry}}
Now we add a stellar distribution and the resulting stellar diffusion on top of our MBH, as described in section \ref{sec:environment:stars}. We ignore DM accretion effects, as these are subdominant, and compare with the DM spikes. 

\subsection{Phase Space Flow}
\begin{figure}
    \centering
    \includegraphics[width=1.1\textwidth]{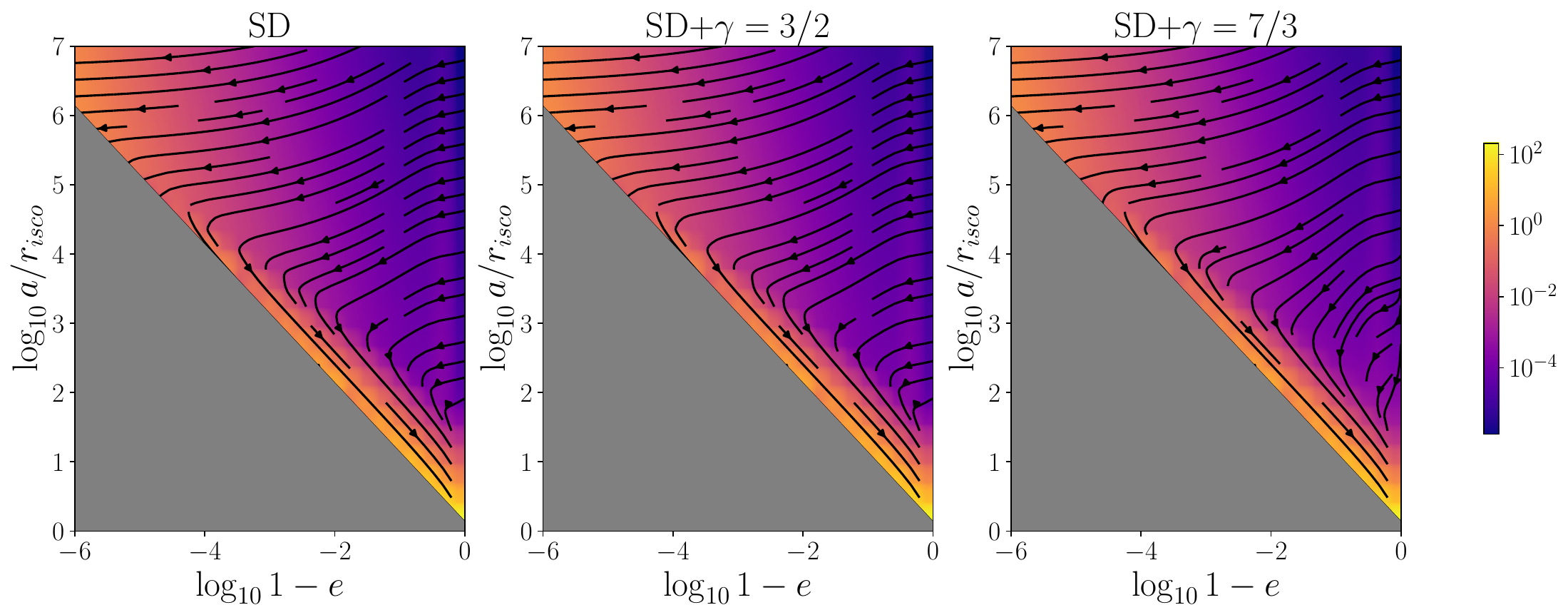}
    \caption{The phase space flow for a system with GW emission, stellar diffusion and dynamical friction of a DM spike. The spike laws are a flat $\gamma=3/2$, or a steep $\gamma=7/3$ spike. The plunge vs dry inspiral separatrix is clearly visible in the stellar diffusion case. The addition of the flat $\gamma=3/2$ does not visibly affect the phase space flow, it is subdominant everywhere. The addition of the steep spike $\gamma=7/3$ results in slightly circularizing effects close to the GW emission dominated zone. }
    \label{fig:psf_dry}
\end{figure}

The resulting phase space flow is plotted in \figref{fig:psf_dry}. The \textit{plunge} vs \textit{inspiral} line is clearly visible. For large semimajor axis, the eccentricity increase is so large, that the secondary will fall directly into the loss cone with eccentricity close to unity -- a plunge. For lower semimajor axis, the GW emission can dominate and circularize the secondary, leading to an inspiral. These generally happen very quickly, as the GW emission is much stronger for larger eccentricity, as seen in \eqref{eq:dE_gw}.

The addition of the flat $\gamma=3/2$ spike does not change the phase space flow visibly, it is subdominant everywhere. The steep $\gamma=7/3$ spike can moderate the strong eccentrification, but its effects seem minor. It does not seem to change the separatrix between plunges and inspirals. The consequences would be dry inspirals but with slightly less extreme eccentricies.

Of course, the more physically motivated system is a kinematically heated flat $\gamma=3/2$ spike inside a stellar distribution. In this case, the phase space flow, and an associated general population of inspirals, would not give hints to its DM spike. Only when observing single inspirals could the DM be inferred through other signatures, such as dephasing or deshifting.

Reference \cite{Zwick:2022dih} has argued that in these systems, due to the high eccentricities and strong circularization, that environmental effects are negligible and the low frequency part `skipped'. While the timescales are much shorter, we show here that the environmental effects can still be detected -- other than the environmental effect that is causing the dry inspiral.

\subsection{Example Inspiral}
\begin{figure}
    \centering
    \includegraphics[width=\textwidth]{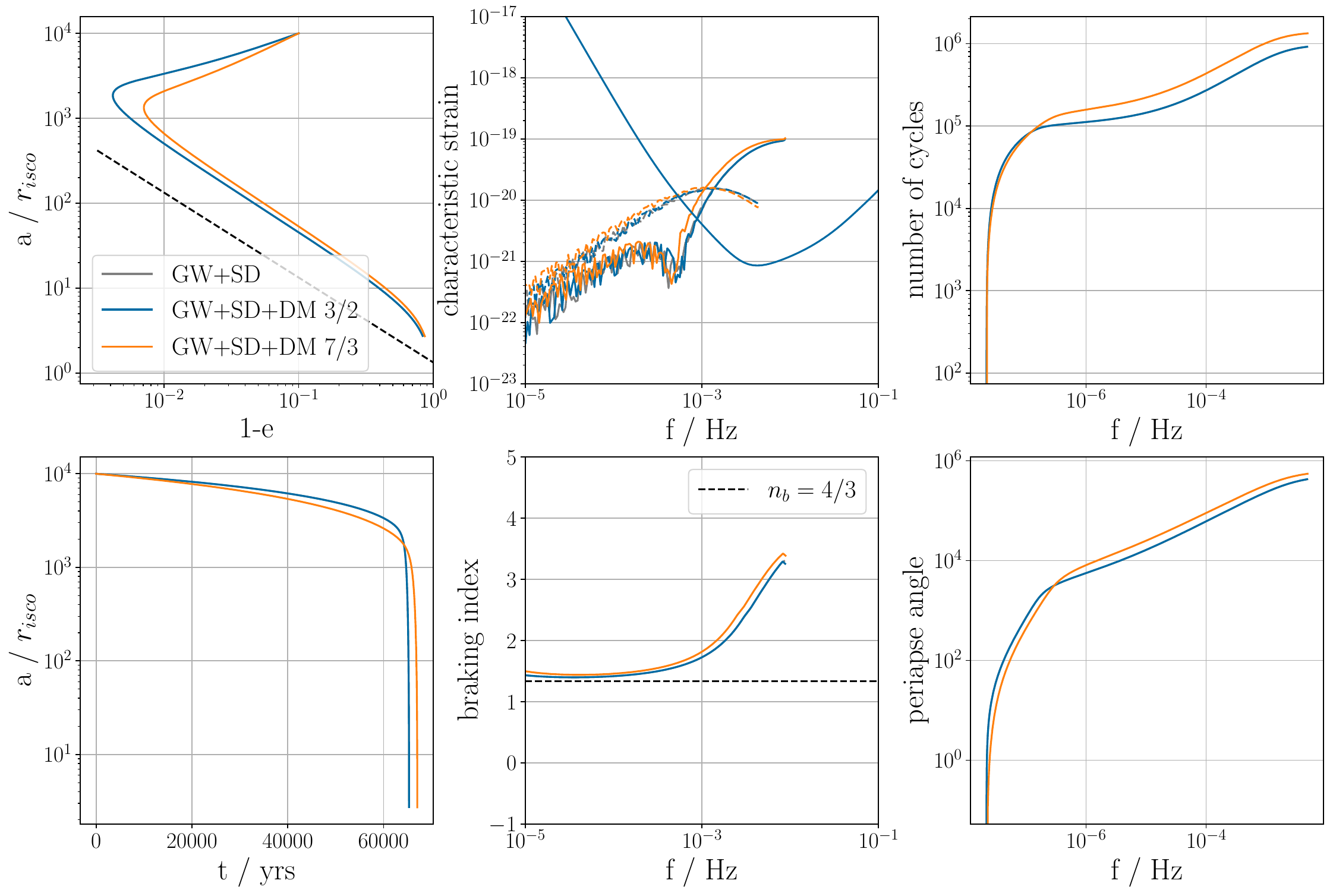}
    \caption{The evolution of a dry inspiral with $m_1=10^5\Msun$, dominated by the stellar diffusion, with and without the flat and steep spike. \textbf{Top Left:} The semimajor axis vs the eccentricity. It can be seen that the system is driven to high eccentricities immediately. For the $\gamma=7/3$, this effect is weakened. The $\gamma=3/2$ and no DM case overlap, as its effects are always subdominant. \textbf{Bottom Left:} The semimajor axis versus time. Despite the presence of an additional dissipative force in the $\gamma=7/3$ case, due to the circularizing effects, the inspiral actually takes longer. \textbf{Top Middle:} The characteristic strain and the LISA sensitivity. In solid lines is the second harmonic, in dashed lines the first harmonic. It can be seen that the first harmonic enters the observable band first and afterwards the second harmonic becomes stronger as the system circularizes. \textbf{Bottom Middle}: The braking index of the system. The high eccentricity case approaches the $n_d=4/3$ limit, before moving toward the circular $11/3$ limit. \textbf{Top Right:} The total number of cycles vs the GW frequency. During the inspiral not many cycles are collected, only once the system has a moderate eccentricity. \textbf{Bottom Right}: The periapse angle due to precession. Initially, the more circular inspiral collects less precession (as SS precession is $\propto (1-e^2)^{-1}$), but the inspiral takes more time, eventually overtaking the eccentric one. \\
    The feedback ratio gives $r_{df,3/2} =8\cdot 10^{-5} $ and $r_{df,7/3} = 4\cdot 10^{-2}$ for the $\gamma=3/2$ and $\gamma=7/3$ cases respectively. }
    \label{fig:ev_dry}
\end{figure}

For our example inspiral, we choose $a_0 = 10^4 r_\isco$ and $e_0 = 0.9$. This assures that the initial semimajor axis is below the plunge separatrix. The initial eccentricity is somewhat arbitrary, as the system will immediately drive it to much higher eccentricities on short timescales. We plot the evolution for the stellar diffusion case, without DM, with a $\gamma=3/2$, and with a $\gamma=7/3$ spike. The result is shown in \figref{fig:ev_dry}.
First, the different timescales compared to the isolated case are sticking out. The inspirals here are on the order of $10^4$yrs, compared to the $>10^6$yrs in the isolated case. This is due to the high eccentricity that is produced by the stellar diffusion. Interestingly, the presence of an additional dissipative force as in the $\gamma=7/3$ case, can slow down the inspiral with its circularizing effects. The braking index can be seen to approach the high eccentricity limit $n_b=4/3$, before the system gets circularized at the end of the inspiral. The first harmonic actually enters the LISA band before the second harmonic, and only then becomes subdominant to it. In these cases the eccentricity evolution can be mapped out clearly. The system as whole collects less cycles during its lifetime compared to the isolated case. While it is in the high eccentricity inspiral phase, almost no cycles get collected, which only moderates toward the high frequency end.

Compared to the isolated case, the feedback ratios are smaller. For the steep spike $r_{df,7/3}\approx 0.04$, which is a reduction of about $\sim 100$ times, while for the flat spike $r_{df,3/2}\approx 8\cdot10^{-5}$, which is a reduction of an order of $10^6$. This is because the inspiral here is dominated by the stellar diffusion and GW emission. This means that in this case we do not have to worry about feedback. The question is then, how much dephasing can we actually observe if its impact is so small?

\subsection{Dephasing}
\begin{figure}
    \centering
    \includegraphics[width=\textwidth]{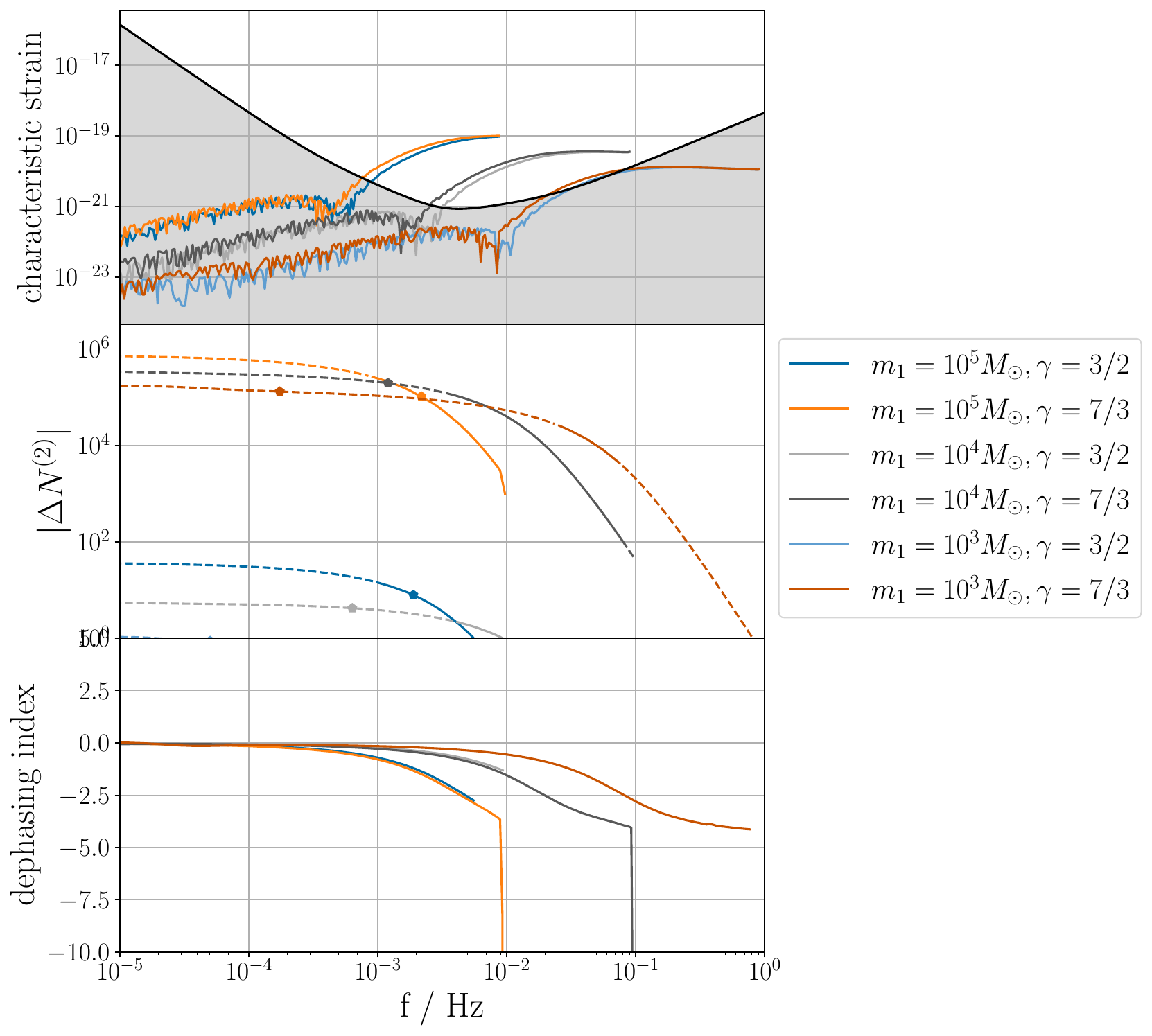}
    \caption{ The dephasing for different central masses  $m_1=\{10^3, 10^4, 10^5\}\Msun$ and different spike models $\gamma=3/2, 7/3$. \textbf{Top}: The characteristic strain of the inspirals in the second harmonic. 
    \textbf{Middle}: The amount of dephasing in the second harmonic for dry inspirals. The pentagon marks the point where the inspiral lasts for another $5$yrs, which is potentially observable within the lifetime of LISA. The dashed lines refer to the areas where the second harmonic is outside the observable region. The steep $\gamma=7/3$ spike leaves strong dephasings, also due to the modified eccentricity evolution as seen in \figref{fig:ev_dry}, while the flat $\gamma=3/2$ spike leaves smaller dephasings $\lesssim 10^2$, but within the 5yr to inspiral. 
    \textbf{Bottom:} The dephasing index. Due to the strong eccentricity evolution, it does not attain a constant value (other than $0$, which means there is no dephasing). }
    \label{fig:dephasing_dry}
\end{figure}

We plot the dephasing for different central masses $m_1=\{10^3, 10^4, 10^5\}\Msun$ in \figref{fig:dephasing_dry}. As seen in the example, most of the cycles are collected at the higher frequencies, and therefore most of the dephasing as well. At lower frequencies, the GW emission dominates and the timescales are very short. The dephasing can be seen to plateau at lower frequencies. The steep $\gamma=7/3$ spike leaves large amounts of dephasing, while the flat $\gamma=3/2$ spike causes a much smaller amount, comparable to the isolated case. 

The plot seems to suggest something positive for the observation of DM: While the overall dephasing is smaller, the dephasing that does happen is on shorter timescales, and therefore more likely to be observable. This can be seen by the pentagons, which mark the 5yr to inspiral point. These are generally higher compared to the isolated spike case. They were not visible for the flat spike in \figref{fig:dephasing_iso}, but for the dry inspirals they move into the observable region. So while the overall dephasing is smaller, the \textit{observable} dephasing due to DM is larger in dry inspirals.

For the flat spike, halo feedback seems to be of little concern. The feedback ratio is always below $r_{df,3/2}<0.04$. For the steep spike, the feedback ratio grows up to $r_{df,7/3}\approx 27$ when the primary mass is $m_1=10^3\Msun$ and $q=10^2$. This is still less by a factor of $\sim10$ compared to the isolated case, but still relevant. Still, in the more physically relevant scenario, halo feedback should be of little concern.

Of course, in dry inspirals, the eccentricity evolution makes it necessary to track several harmonics. Initially the first harmonic dominates, while in the end the second one does. To observe such an inspiral in its totality, all these changes need to be tracked. Also, regarding the dephasing index, it is not close to a constant value, it is a mix of the change in semimajor axis and eccentricity. This could imply that different models are harder to differentiate. Add on top of this the strong relativistic effects that were mentioned in section \ref{sec:environment:spacetime}, which become relevant at these high velocities where the dephasing is observable.

So while there might be more observable dephasing, it could be difficult to extract from the mess of spacetime around MBH and attribute to DM.

\subsection{Deshifting}
\begin{figure}
    \centering
    \includegraphics[width=\textwidth]{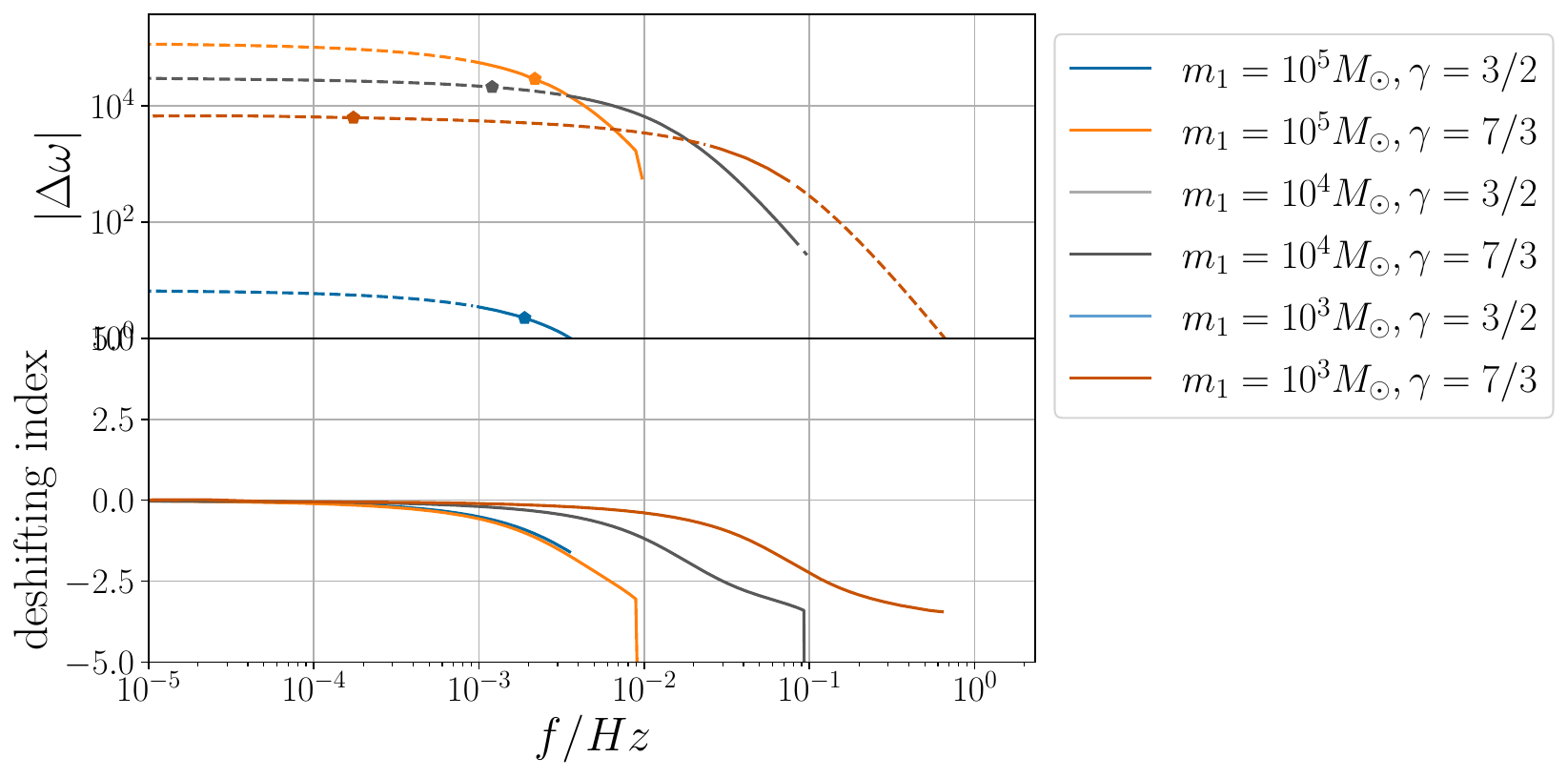}
    \caption{The deshifting for different central masses  $m_1=\{10^3, 10^4, 10^5\}\Msun$ and different spike models $\gamma=3/2, 7/3$. \textbf{Top}: The deshifting here is similar to the dephasing in \figref{fig:dephasing_dry}, albeit at slightly lower values. \textbf{Bottom}: The deshifting index. Here, there is also no constant value attained. }
    \label{fig:deshifting_dry}
\end{figure}

The dry inspirals with their large eccentricities also experience very strong SS precession.
As the $\gamma=7/3$ spike circularizes the orbit (and the SS precession is $\sim1/(1-e^2)$), the relativstic precession is weaker. At the same time, the inspiral is prolonged, which eventually causes more precession. This can be seen in \figref{fig:ev_dry}. We plot the difference in precession in \figref{fig:deshifting_dry}. The plot is overall very similar to the deshifting. Unlike in the isolated spike case, here, the deshifting and dephasing are differ at about one order of magnitude. Similar caveats as in the previous subsection apply, so a measurement of both dephasing and deshifting together can shine more light on the inspiral.

\section{Wet inspirals\label{sec:signatures:wet}}
In this section we add the effects of a Derdzinski-Mayer accretion disk, as discussed in section \ref{sec:environment:disk}. First, we assume the secondary to be in plane of the accretion disk orbiting either prograde or retrograde, later we discuss the alignment process.

\subsection{Phase Space Flow}
\begin{figure}
    \centering
    \includegraphics[width=\textwidth]{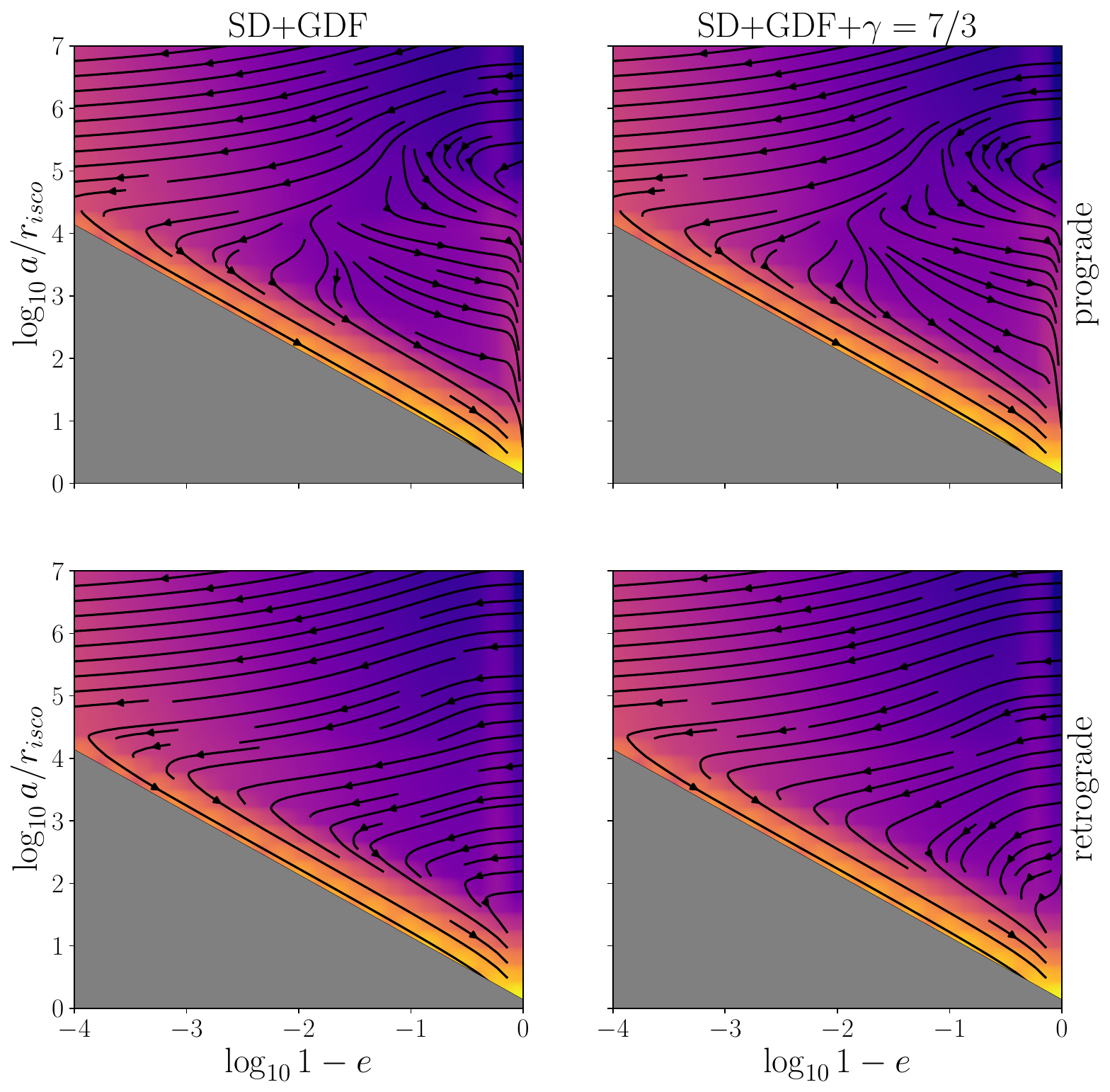}
    \caption{The phase space flow for a system with GW emission, stellar diffusion, gas dynamical friction with an Derdzinski-Mayer accretion disk and dynamical friction with a DM spike. We only consider the steep $\gamma=7/3$ spike, as the flat $\gamma=3/2$ spike effects are already subdominant to the stellar diffusion. On top is prograde rotation, on the bottom retrograde rotation with regard to the disk. On prograde orbits, the addition of the accretion disk effects change the phase space flow in a region with intermediate distances and low eccentricity. Here, the strong eccentrification is turned to circularization. The addition of the steep spike adds to this.\\
    On retrograde orbits, the higher relative speeds and therefore reduced dynamical friction appears to not affect the phase space flow at all, only the addition of the steep spike is marginally impactful for low eccentricities. }
    \label{fig:psf_wet}
\end{figure}

We plot the phase space flow for four different scenarios in \figref{fig:psf_wet}. We look at pro- and retrograde orbits inside accretion disks and stellar halos, with and without a steep $\gamma=7/3$ spike. As we have seen before, the flat $\gamma=3/2$ spike is negligible here. We assume gas dynamical friction with the accretion disk for now, as described in section \ref{sec:environment:disk}.

On retrograde rotation, the gas dynamical friction does not alter the phase space flow visibly. It seems that retrograde dynamical friction is subdominant to stellar diffusion. As before, the steep $\gamma=7/3$ spike counteracts this eccentrification with its circularization, but overall not changing much about the nature of the phase space flow.

For prograde rotation, there is an additional region in the phase space flow, dominated by the circularizing effects of the gas dynamical fricton. The steep $\gamma=7/3$ spike seems to add to this. 

The strong circularization effects of the gas dynamical friction drive the secondary to a quasi-circular orbit. Here, unfortunately, the description of the gas dynamical friction breaks down, as the relative velocity of the secondary and the disk vanishes. It has been shown that on circular orbits, Type--I torques provide a better description of the interaction\cite{Derdzinski:2020wlw}. Type-I torques are comparatively weaker and subdominant to steep DM spikes for small semimajor axis\cite{Becker:2022wlo}. As this is valid for circular orbits only, we cannot plot the phase space flow involving the eccentricity. Instead, we explore the quasi-circular case.

\subsection{Braking Index}

\begin{figure}
    \centering
    \includegraphics[width=\textwidth]{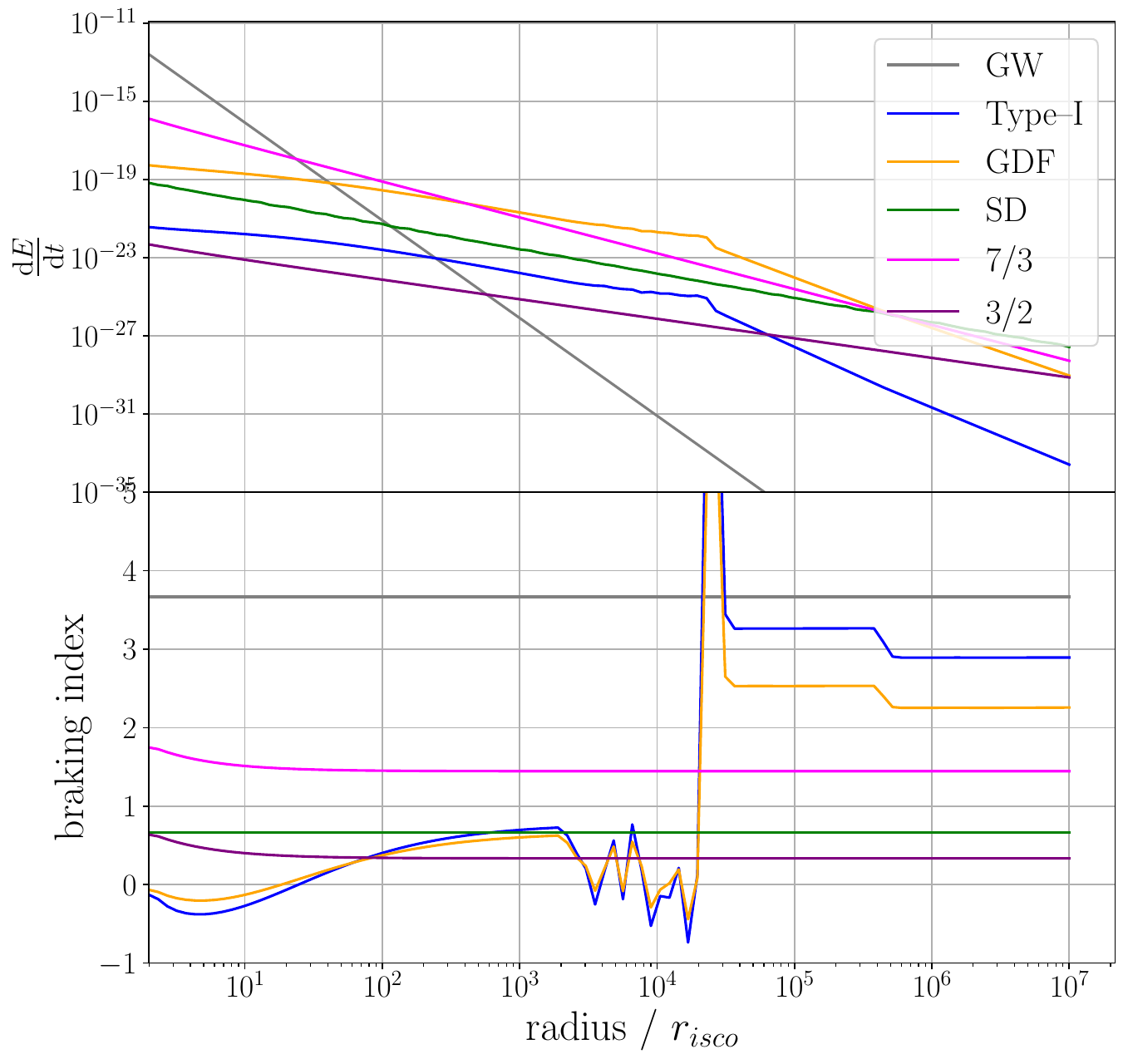}
    \caption{\textbf{Top}: The energy loss of all the different environmental effects explored in this chapter. This is on (prograde wrt to the disk) circular orbits with different radii. The spike indices refer to dynamical friction with the spike.
    \textbf{Bottom}: The associated braking index $n_b$ of the environmental effects. For all effects except the accretion disk interactions the braking index is constant and sufficiently distinct. This means the effects are differentiable. By extension, using \eqref{eq:braking_and_dephasing_index}, the dephasing index $n_d$ is also different for all these forces. 
    The spike in the accretion disk braking indices are due to the kinks in the disk density profile, where the numeric differentiation fails. }
    \label{fig:dEdt_all}
\end{figure}

Here, we plot the energy loss for all the components on circular orbits in \figref{fig:dEdt_all}. This is a nice summary of this chapter, which gives an overview of the relative strengths of the environmental effects. As expected, GW emission dominates all other forces close to the MBH ($\lesssim 10 r_\isco$). Next, the steep spike has a strong effect on the energy loss. Closely following is the gas dynamical friction inside the Derdzinski-Mayer disk, but its description is not physically well motivated. This is followed by the stellar diffusion, then the Type--I torque and lastly the flat spike.

Where one force dominates, the evolution will be described with its braking index, which is given by the slope of its curve. We can see some of the braking indices that were observed in the previous sections. If the environmental effects have different slopes, the frequency evolution of the inspiral should reveal them. In turn, measuring the braking index can show which environmental effect is dominating the inspiral. This of course assumes that the braking index is sufficiently different between the effects. As can be seen in \figref{fig:dEdt_all}, the effects have similar slopes and therefore similar braking indices, but still distinct. Therefore, an accurate measure of the braking index and a good theoretical understanding of the forces involved can pinpoint the exact environmental effect. That is, on circular orbits. For eccentric orbits, the additional information of the eccentricity evolution, i.e. the phase space flow, can help in understanding the dissipative force at play. For low eccentricity this works according to \eqref{eq:braking_index_a}.

As the braking indices are distinct, so must be the dephasing and deshifting indices, according to section \ref{sec:inspiral:GWsignal}. This means that we can also identify the environmental effects when GW emission dominates the system. 

The close relationship between Type--I torques and gas dynamical friction is visible in the braking index. Even though the description is mathematically very different, the slopes of the curves are the same on the inner part of the disk. For these effects, the braking index is not (piecewise) constant, as the density profile is very different from any power law approximation. Only in the latter part of the disk where a power law scaling is valid, it approaches a constant value. The spike in between is given due to numerical difficulties resolving the kink in the density distribution.
In the following, we will show an example inspiral with the Type--I torque.

\subsection{Example Inspiral}
\begin{figure}
    \centering
    \includegraphics[width=\textwidth]{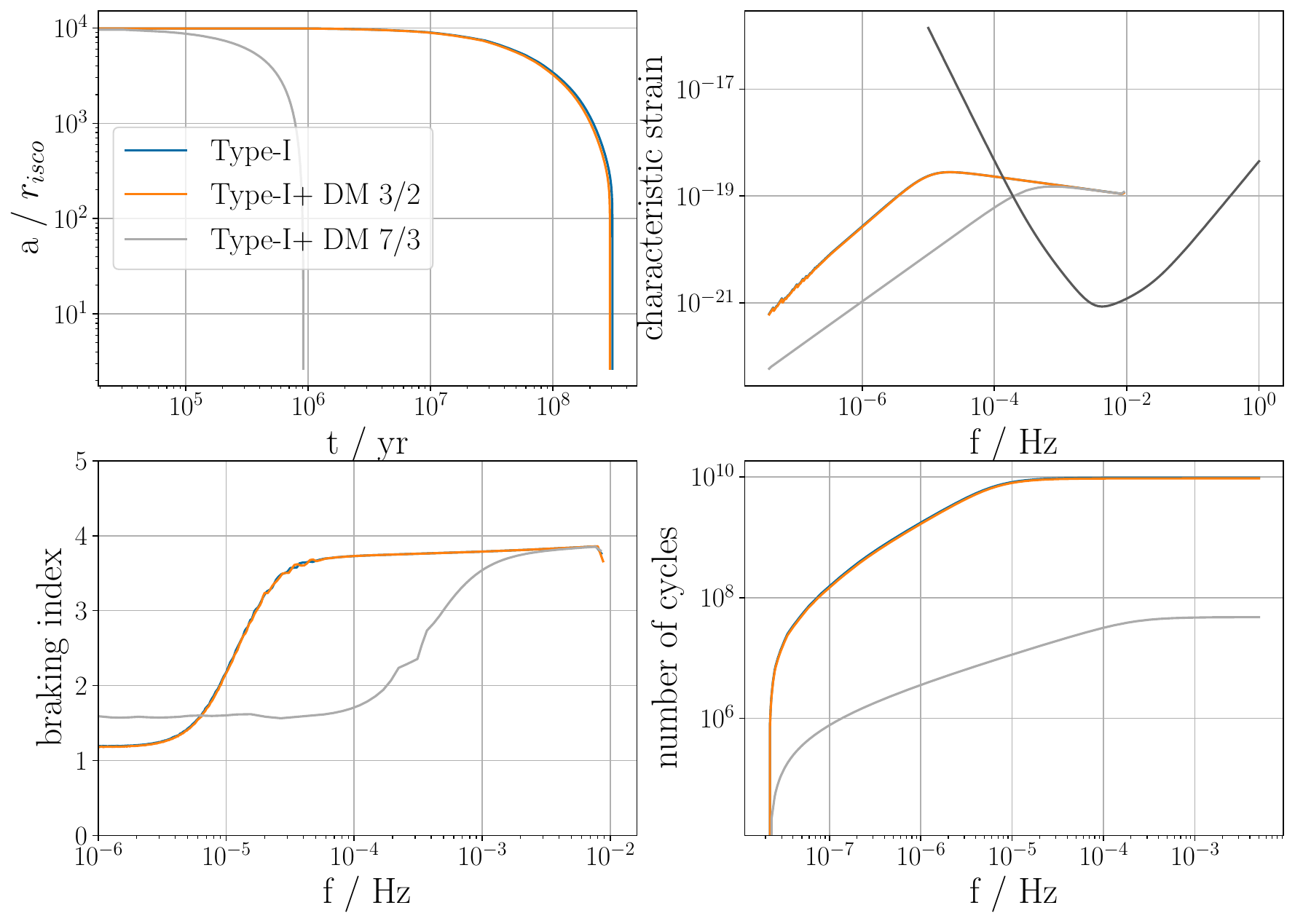}
    \caption{A quasi-circular wet inspiral, with a Type--I interaction with the Derdzinski-Mayer accretion disk around an MBH with $m_1=10^5\Msun$, without and with a $\gamma=3/2$, or a $\gamma=7/3$ DM spike. \textbf{Top Left:} The semi-major axis vs time. The presence of the steep $\gamma=7/3$ spike speeds up the inspiral, while the flat spike does not significantly affect the inspiral time. \textbf{Bottom Left}: The braking index. The different regimes of dominance are visible. \textbf{Top Right:} The characteristic strain of the second harmonic and the LISA sensitivity. \textbf{Bottom Right}: The number of cycles collected.
    The periapse precession is not plotted, as a circular orbit cannot precess.
    }
    \label{fig:ev_wet}
\end{figure}

Wet inspirals are dominated by the circularizing effects of the accretion disks, at least on prograde orbits. Therefore, the inspirals happen quasi-circularly when they enter the observable band. As mentioned previously, accretion disk effects are very much uncertain and are a topic of active research. Here, we will rely on Type-I as described in section \ref{sec:environment:disk} and assume a quasi-circular inspiral. We have looked at this previously in \cite{Becker:2022wlo}. 

According to \figref{fig:dEdt_all}, the stellar diffusion would be stronger than the Type--I interaction, while the secondary is inside the disk. This is difficult to believe physically. If this were the case, the disk could quickly be unbound due to the stochastic interactions with the stellar distribution. Reality is likely more complicated. It might be that the simple linear combination of the two breaks down and the interaction between them needs to be studied further. This has been done in the context of EMRI rates in \cite{Pan:2021ksp}, but to our knowledge the effect on the stellar distribution has not been studied. For now, we will ignore the dissipative effects of the stellar diffusion.

An example inspiral is shown in \figref{fig:ev_wet}. We choose a circular orbit with $e_0=0$ and an initial semimajor axis of $a_0=10^4 r_\isco$. The inspiral times are slightly shorter compared to the isolated case, due to the Type--I interaction. The presence of the steep $\gamma=7/3$ spike speeds up the inspiral as in the isolated case. The steep spike effects dominate the Type--I interaction, as seen in \figref{fig:dEdt_all}. The flat $\gamma=3/2$ spike slightly speed up the inspiral, it is of similar magnitude as the Type--I interaction.

The feedback ratios are $r_{df, 7/3}\approx 0.7$, which is similar to the isolated case, and $r_{df,3/2}\approx 1.6$, which is about $\sim 10$ times smaller than in the isolated case, but still significant. This implies that in this simplified accretion disk + spike model, the DM should be impactful, and the interactions between the distributions should be studied further.

\subsection{Dephasing}
\begin{figure}
    \centering
    \includegraphics[width=\textwidth]{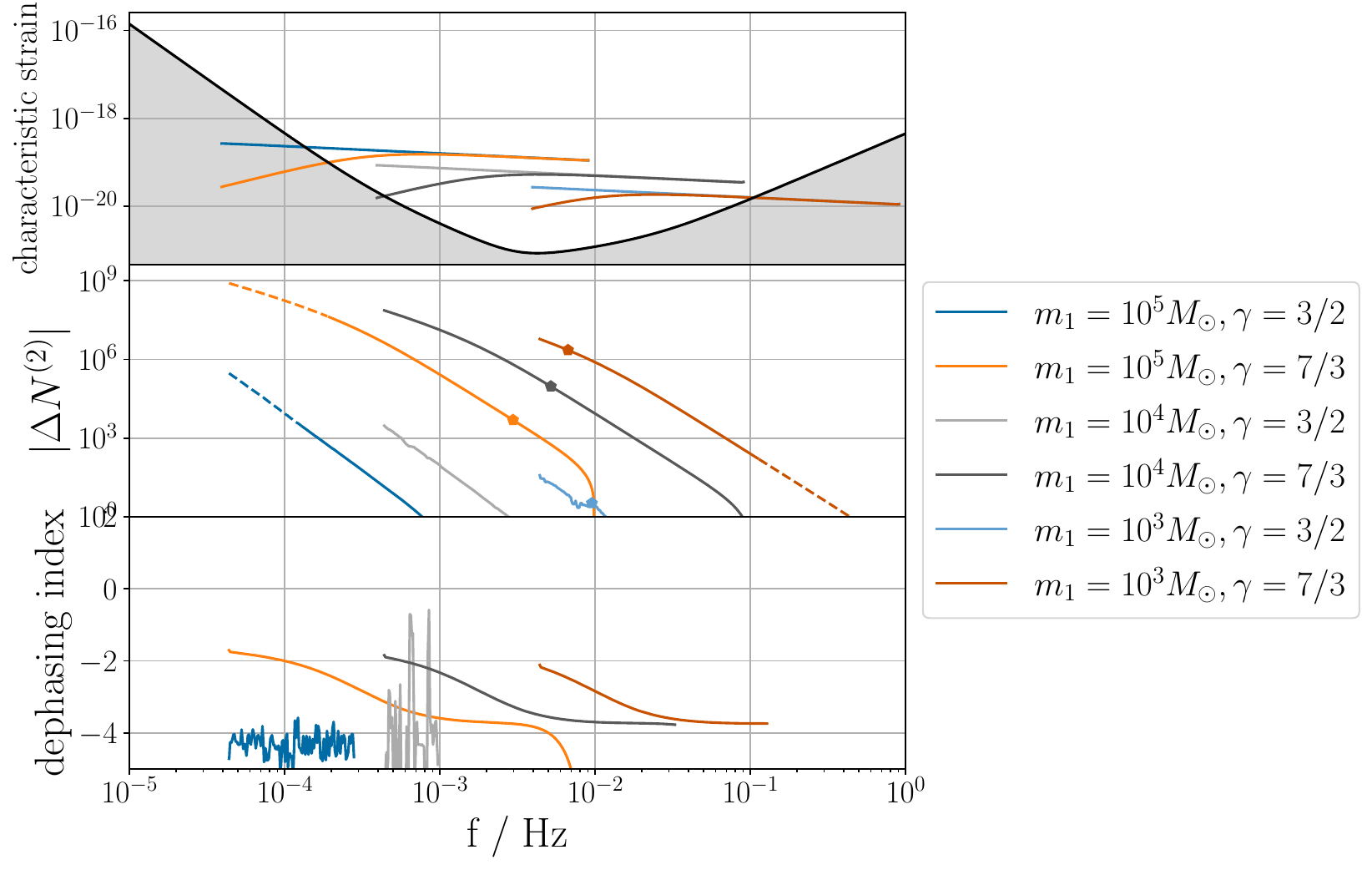}
    \caption{The amount of dephasing for wet inspirals for different central masses $m_1=\{10^3, 10^4, 10^5\}\Msun$ and different spike models $\gamma=3/2, 7/3$. The pentagon marks the point where the inspiral lasts for another $5$yrs, which is potentially observable within the lifetime of LISA. The dashed lines refer to the are where the second harmonic is outside the observable region. The steep $\gamma=7/3$ spike leaves strong dephasings, while the flat $\gamma=3/2$ spike leaves smaller dephasings, and nothing observable in the last 5 yrs of the inspiral. It seems that the numerical resolution of the flat spike is insufficient here, as the dephasing index is not resolved very well. }
    \label{fig:dephasing_wet}
\end{figure}

In \figref{fig:dephasing_wet} an overview over the expected dephasings is plotted for different central masses. A similar picture as in the isolated case emerges, with a comparable amount of dephasing. The steep spike has a similar feedback ratio as in the isolated case, but the flat spike has a feedback ratio of order $\sim 10^{-2}$, with similar reasoning as in the isolated case. This implies that the feedback effects in the late-stage inspiral are subdominant or negligible.

In the case of wet inspirals, the inspiral should of course be dominated by the disk dynamics. This is not the case in the adiabatically grown steep $\gamma=7/3$ case, but this spike is physically not well motivated here. The flat spike has more physical justification. But of course, the disk interaction is still subject of ongoing research. As it is now, Type-I interaction looks very similar to the flat spike case from the perspective of the energy loss and braking index. First, the disk dynamics have to be understood better, to ascribe anything left to DM. 

\subsubsection{}

Here, we have previously studied the deshifting, but this only makes sense for eccentric inspirals, as there is no difference between periapse and apoapse on circular orbits, and therefore no precession. Instead we can look at the alignment with the disk.

\FloatBarrier
\subsection{Alignment}
\begin{figure}
    \centering
    \includegraphics[width=\textwidth]{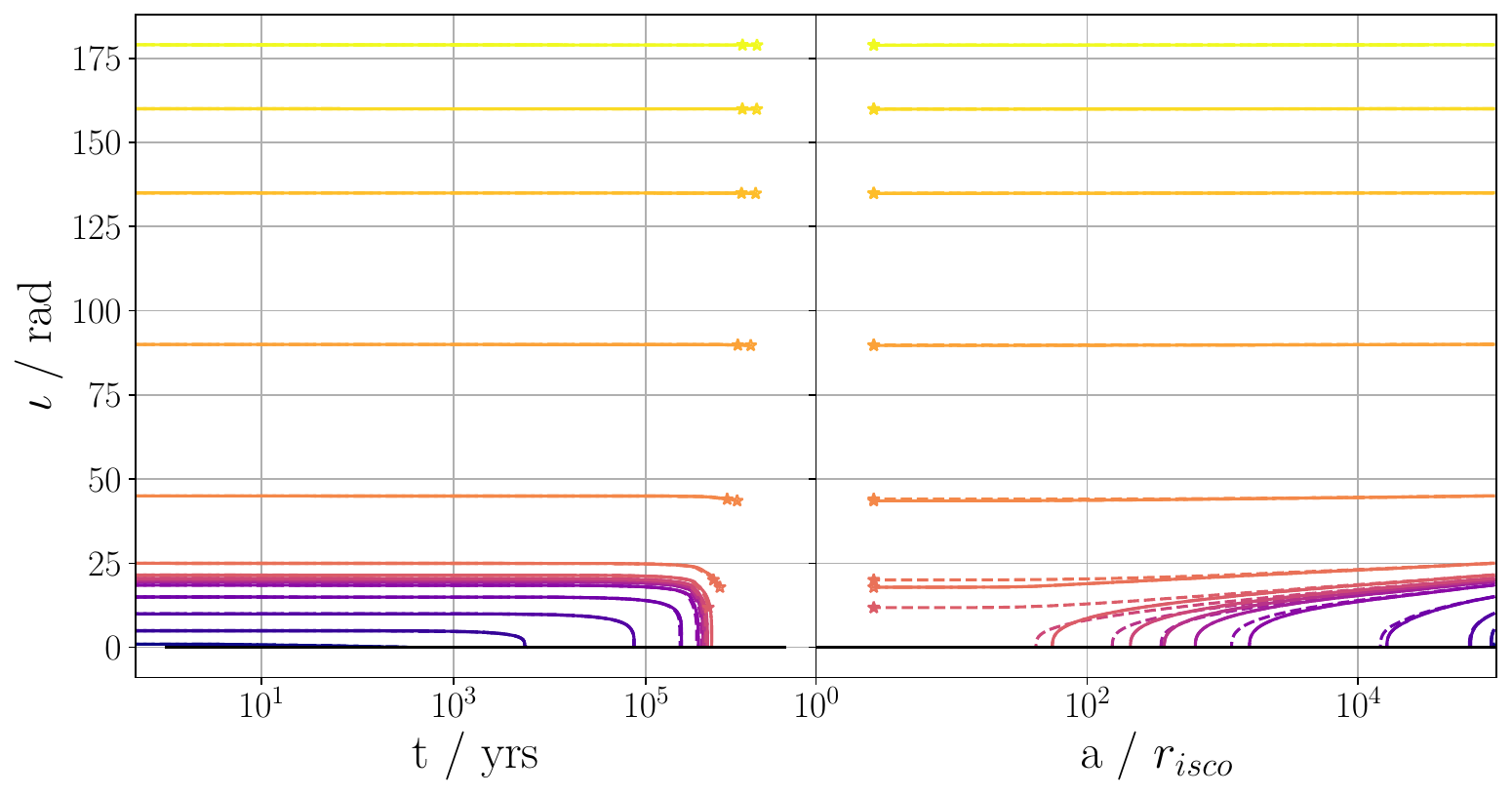}
    \caption{Several inspirals with different initial inclinations for a system with $\{m_1, m_2, a_0, e_0\} = \{10^5\Msun, 10\Msun, 10^4r_\isco, 0\}$, employing gas dynamical friction with the accretion disk, and stellar diffusion. The dashed lines mark the addition of the $\gamma=7/3$ spike interactions, and the dash-dotted lines that of the $\gamma=3/2$ spike. Left and right display the same inspirals, once shown with regard to time, and once with regard to semimajor axis. If an inspiral ends before it is aligned, it is marked with a star, filled for the $\gamma=7/3$ spike and clear for no spike. It can be seen that the existence of the steep spike alters the alignment procedure. Due to the additional interaction, the inspiral speeds up and some of the systems have no time to align. For the flat spike there are no observable effects. }
    \label{fig:inc_sd_gdf}
\end{figure}

If the initial orbit is not inside the disk, an orbit that is inclined with regard to the disk plane will align its orbital plane over time due to interactions with the disk. This is the transition from a dry to a wet inspiral. The evolution of the inclination angle is given by \eqref{eq:dinclination_angle}. 

This process was explored more in depth in \cite{Fabj:2020qqc, Nasim:2022rvl}. Their findings show that alignment depends on the disk model and disk interaction model. For a high density QSO disk all objects align with the disk throughout, while for a lower density TQM disk, there is a critical angle of inclination above which the plane will not align. At the same time, an sBH with dynamical friction can actually align on a retrograde, while a star with geometric drag will always align on a prograde orbit.

Here, we will do the same procedure with a Derdzinski-Mayer disk, which has a slightly lower density than QSO disks, and for a smaller MBH of $m_1=10^5\Msun$. We assume gas dynamical friction on circular orbits, which is valid as the relative velocity does not vanish for inclined orbits. We additionally assume the presence of a stellar distribution. We will also add two different spike models $\gamma=3/2, 7/3$ and compare the alignment effects. A look at \figref{fig:dEdt_all} shows that inside the disk, the gas dynamical friction is the strongest. This implies that it actually has a chance to align the object with the disk, compared to stellar diffusion or the steep $\gamma=7/3$ spike.

An example is shown in \figref{fig:inc_sd_gdf}. The inclination angle of the system is plotted vs the temporal and radial evolution for different initial inclination angles. An inclination of $0^{\circ}$ is inside the disk on a prograde orbit and $180^{\circ}$ corresponds to a retrograde orbit inside the disk. Below a critical initial angle, the inclination angle tends to zero, and thus the orbit aligns with the disk. Above the critical alignment angle, the system does not have enough time for alignment. In the evolution of the radius it can be seen that the inclination angle barely changes below $10 r_{isco}$. Here, the GW emission is dominant and the alignment is inefficient. The presence of a steep dark matter spike $\gamma=7/3$ can affect the alignment process visibly. The additional dissipative force speeds up the inspiral, which can be seen by the star markers in the plot. This leaves less time for alignment. The flatter $\gamma=3/2$ spike does not affect the alignment significantly.


\begin{figure}
    \centering
    \includegraphics[width=\textwidth]{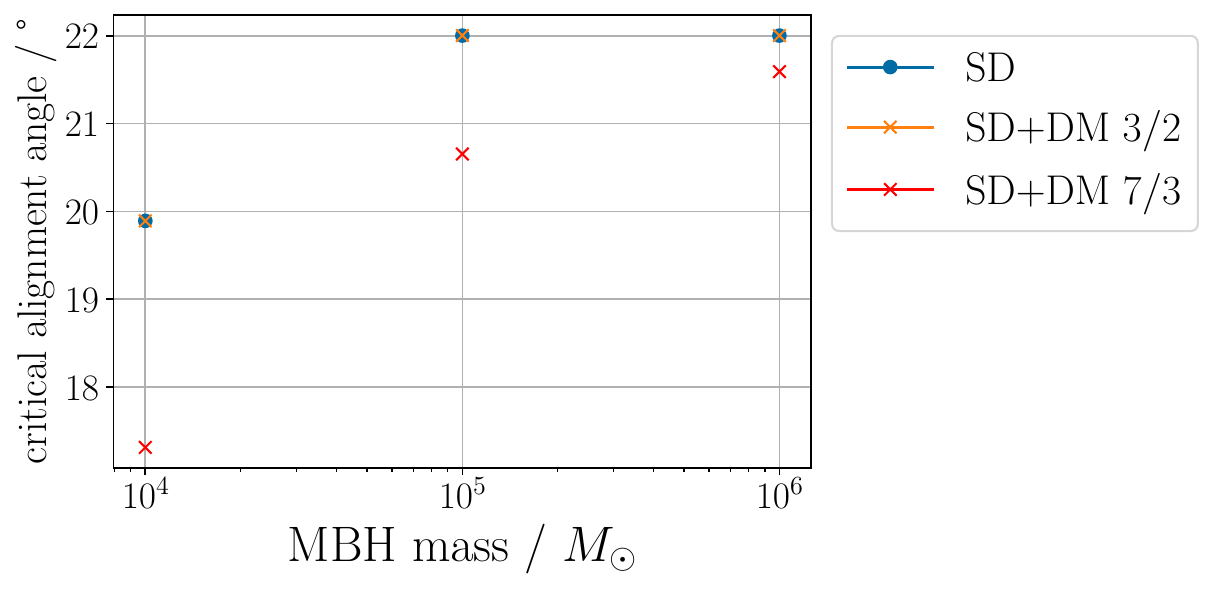}
    \caption{The critical alignment angle for different central MBH masses and spike models. The secondary is assumed to be an sBH with $m_2=10\Msun$. The SD and SD+DM $\gamma=3/2$ case are indistinguishable. The steep $\gamma=7/3$ spike lowers the critical alignment angle, as it speeds up the inspirals and leaves less time for alignment. This effect weakens for larger central masses. }
    \label{fig:caa}
\end{figure}

Our results different to \cite{Fabj:2020qqc, Nasim:2022rvl}. They find that sBHs are always captured by the disk before SMBH capture. Compared to them, we have given the system less opportunity to align, with a lower density disk and the additional stellar diffusion. Also, we look at a comparatively smaller MBH with faster inspiral times. What exactly it boils down to and what are realistic assumptions will have to be left for future work. We can measure the DM alignment effects in our case, and try to extrapolate.

In \figref{fig:caa} the critical alignment angles for different systems are plotted. In the absence of DM, the critical alignment angle increases with central mass. This was calculated for initial semimajor axis of both $a_0=10^4 r_\isco, 10^5 r_\isco$, which did not change the result.\footnote{If the critical alignment angle is dependant on initial semimajor axis, it would have to be averaged over the source distribution.} Generally, larger central masses increase the timescales of the inspiral, giving more time for alignment to happen.

It can be seen that the effects of DM are more pronounced for smaller central MBH. This is in line with the dephasing effects, where smaller MBH generally leave more room for DM effects. The effects decrease for increasing central mass, implying that for masses $m_1 > 10^6\Msun$ no effect would be visible. If we were to leave out the effects of the stellar diffusion, DM effects would be more impactful in comparison.

This subsection has to be understood as a first exploration of the alignment efficiency in the presence of a DM spike. There is a lot of uncertainty about the disk model, CO-disk interaction, possibly spike-disk interactions. For example, inclined orbits can induce specific kinds of waves inside the accretion disk such that the interaction is not given by dynamical friction\cite{2004ApJ...602..388T}. 

Also, the model should account for eccentric inspirals. Even in the Milky Way, the S-stars in the galactic center have high eccentricies\cite{2013degn.book.....M}. These eccentricities can drastically speed up the inspirals as shown in previous sections. But including eccentric effects self-consistently means we have to account for periapse and nodal precession as well, which complicates the problem, expands the parameter space and increases computational time. We leave this for future work. Anticipatory, the inclusion of eccentricity will speed up the insipral due to stellar diffusion, leave less time for alignment and decrease the critical alignment angle. Then, the circularizing effects of a $\gamma=7/3$ spike can actually prolong the inspiral, and give more time for alignment.

The observable effects would most likely be of statistical nature if the alignment happens outside the observable band. A population of wet inspirals would be aligned with the disk, while above the critical alignment angle the inspirals would be dry. The observation of the wet inspirals illuminates disk properties and interactions, while the dry inspirals give the critical alignment angle. Any mismatch could be attributable to DM. Unfortunately, for a realistic $\gamma=3/2$ spike, the effects seem to be negligible.

\FloatBarrier

\subsubsection{}
Concluding this chapter, we can summarize what we have learned.

In the isolated spike case, the steep $\gamma=7/3$ spike would be detectable through its dephasing and deshifting, but is subject to halo feedback effects. The flat $\gamma=3/2$ spike also causes dephasing and deshifting, but the timescale are too large for it to be effectively detectable by LISA. Also, halo feedback is stronger, and leaves the question whether this spike could be unbound.

In the dry inspiral case, most of the dephasing and deshifting is collected at the end of the inspiral. This leaves less overall dephasing, but more \textit{observable} dephasing within the lifteime of LISA. This is because dry inspirals happen on much shorter timescales due to their large eccentricity. This, and the associated higher velocities, also make it more susceptible to relativistic effects, which need to be modeled.

In the wet inspiral, it is difficult to make predictions. We saw that the accretion disk interaction and the stellar diffusion have similar strenghts, and their interaction with each other is most likely more complicated. Nevertheless, for Type--I interactions, we saw a similar picture as in the isolated spike case, where the steep spike is visible and the flat spike difficult to detect, both subject to halo feedback processes.
We also considered the alignment process and found there to be a critical alignment angle in the presence of stellar diffusion and the gas dynamical friction of the disk. The inclusion of DM effects can reduce that critical alignment angle due to faster inspiral times. We argued that this is not a great signature, since the alignment mostly happens outside the observable band, but possibly a statistical tool.

Lastly, we have looked at the braking index of all these environmental effects and found them to be distinct on circular orbits. Therefore, the dephasing index is also distinct, which makes the forces differentiable at late times in the inspiral, if they leave sufficient dephasing. On eccentric orbits this is more complicated and the modeling needs to be improved.

Of course, underneath these findings are a lot of assumptions. The distributions and interactions could change qualitatively. The simplest change could be a different normalization of the distributions involved. This can already affect the qualitative behavior when the environmental effects are superposed. Regardless, the underlying tools and signatures we have explored here are still applicable.

\chapter{Rates \label{chap:rates}}
In this chapter we cover the rate estimates of I/EMRIs, following \cite{Babak:2017tow}. Subsequently, we try to estimate the effect of DM spikes on these rates.

\section{Estimates \label{sec:rates:estimates}}
\subsection{MBH population}
The model developed in \cite{Barausse:2012fy, Sesana:2014bea, Antonini:2015cqa, Antonini:2015sza} assumes the initial formation of light MBH seeds $\sim 150\Msun$ from the collapse of Population III stars\cite{Madau:2001sc}. It follows the evolution of them and their host halos, accounts for accretion of baryonic matter, halo mergers, and the associated MBH mergers afterwards. They find an approximate mass function of 
\begin{equation}
    \dv{n}{\log M} = 0.005 \left(\frac{M}{3\cdot 10^3 \Msun} \right)^{-0.3} \text{  Mpc}^{-3}, \label{eq:MBH_mass_function}
\end{equation}
almost independent of redshift. This is consistent with observational constraints.

The model also predicts a large spin for these MBH. This is mostly due to the fact that MBH align with their accretion disk and grow considerably with it, therefore adopting its spin. This is unfortunate, as we have not modeled spin for the central MBH in this dissertation.

\subsubsection{Stellar Cusps}
A cusp-like stellar distribution is needed to effectively form I/EMRIs. We have assumed a $\gamma\sim 7/4$ Bahcall-Wolf cusp as the steady state solution around the MBH. However, as galaxies merge, their MBHs do as well, depleting their respective cusps and leaving low density cores\cite{Milosavljevic:2001vi, Antonini:2015cqa}. Here, the formation of I/EMRIs is rare and we assume it to be negligible. Instead, the cusp has to regrow in a time $t_\text{cusp}$. This can be estimated from the semi-analytic models referenced before, to be
\begin{equation}
    t_\text{cusp} \approx 6 \left(\frac{M}{6\cdot 10^6\Msun} \right)^{1.19} q^{0.35} \text{ Gyr}
\end{equation}
where $M$ is the total mass and $q$ the mass ratio of the merger.

The timescales of equal mass mergers can be a significant fraction of the Hubble time, especially for SMBH $M \geq 10^6\Msun$. Smaller MBH can regrow their cusps much quicker.

To estimate the number of cusps around MBHs, we have to calculate the probability of a merger for an MBH during their cusp regrowth time $p_0(M,z)$ at a given redshift $z$. The semi-analytic model gives $\frac{\text{d}^3 n_m}{\text{d}M\text{d}q\text{d}z}$, the differential number density per mass, mass ratio and redshift.
The mean number of mergers in the cusp regrowth time is then
\begin{equation}
    N_m(M,z) = \int \mathrm{d}q \int_z^{z_\text{cusp}(M,q)} \mathrm{d}z' \dv[3]{n_m}{M}{q}{z} \left(\dv{n}{M} \right)^{-1}, 
\end{equation}
where $z_\text{cusp}(M,q)$ is the redshift corresponding to the cusp regrowth time $t_\text{cusp}(M,q)$, given by the cosmology.
The probability for no merger to have occurred is then, assuming Poissonian statistics,
\begin{equation}
    p_0(M,z) = \exp(-N_m(M,z)).
\end{equation}
For low mass MBH, this probability is close to unity, as their cusp regrowth is small. For heavier MBH, this can get lower to $\mathcal{O}(\%)$ level.

\subsubsection{Plunge vs Inspiral \label{sec:rates:plunge}}
\begin{figure}
    \centering
    \includegraphics[width=0.6\textwidth]{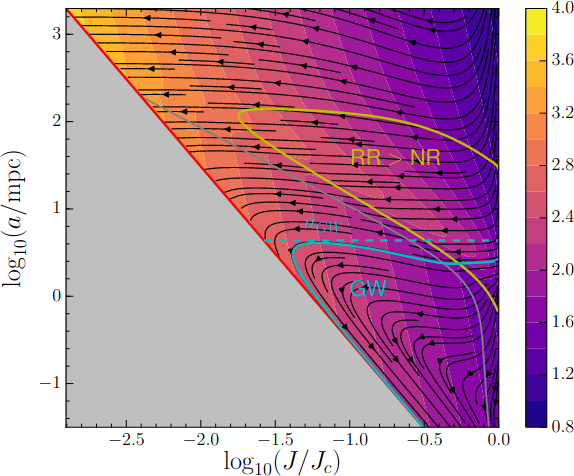}
    \caption{The phase space flow around an MBH with the cyan line as the GW separatrix. Above the separatrix, the COs fall in with high eccentricity due to stellar diffusion in a \textit{plunge}. Below the speratrix, the COs are being circularized by the GW emission and perform an \textit{inspiral}. Plot taken from \cite{Bar-Or:2015plb}. Note that on the $x$-axis is $J\sim \sqrt{1-e^2}$, unlike the other $1-e$ plots. }
    \label{fig:plunge_inspiral}
\end{figure}
Now that we have an estimate of the number of MBH with stellar cusps, we need to calculate the rate of the COs being captured. Looking at the phase space flow, this is the amount that flows into the loss cone in the given relaxation time $t_\text{relax}$ \cite{Qunbar:2023vys}
\begin{equation}
    \dot{N}_\text{capture} = 4\pi\int_0^{a_{\inf}} \frac{a^2 n_*(a) }{\ln(J_{c}/J_{lc})t_\text{relax}} \d a  , \label{eq:N_capture}
\end{equation}
with the stellar number density $n_*(r)$, the critical angular momentum of the loss cone $J_{lc}=4m_1$, and the maximal circular specific angular momentum $J_c = \sqrt{a m_1}$. The relaxation time is given by \cite{Babak:2017tow}
\begin{equation}
    t_\text{relax} = \frac{5}{\ln \Lambda} \left(\frac{\sigma}{10 \text{km s}^{-1}}\right) \left(\frac{r_h}{1pc}\right)^2 \text{ Gyr}, 
\end{equation}
with the velocity dispersion $\sigma$. This can be estimated with observable $M-\sigma$ relations as \cite{Gultekin:2009qn}
\begin{equation}
    M = 1.53\cdot 10^6 \left(\frac{\sigma}{70 \text{km s}^{-1}} \right)^{4.24} \Msun. 
\end{equation}

This capture can happen in two ways. As can be seen in \figref{fig:plunge_inspiral}, the relaxation processes in the stellar cusp drive the individual COs to low angular momentum, causing them to fall in from far away with large eccentricities. These are called \textit{plunges} and will cause a short, difficult to detect GW flares. If on the other hand, the COs end up in the part of the phase space dominated by GW emission, they will inspiral and cause the characteristic I/EMRI GW signal. We can estimate the ratio of plunges to inspirals by finding the \textit{separatrix} in the phase space flow\cite{Bar-Or:2015plb}. This is shown in \figref{fig:plunge_inspiral} as $a_\text{GW}$. Here, the flow transitions from falling into the loss cone to inspiraling to low semimajor axes. This gives the simple description for the number count of plunges and inspirals as a modification of \eqref{eq:N_capture}
\begin{align}
    \dot{N}_\text{plunge} &= 4\pi\int_{a_\text{GW}}^{a_{\inf}} \frac{a^2 n_*(a) }{\ln(J_{c}/J_{lc})t_\text{relax}} \mathrm{d}a,\label{eq:Nplunge}\\
    \dot{N}_\text{inspiral} &= 4\pi\int_0^{a_\text{GW}} \frac{a^2 n_*(a) }{\ln(J_{c}/J_{lc})t_\text{relax}}\mathrm{d}a, \label{eq:Ninspiral}
\end{align}
where $a_\text{GW}$ can be obtained by comparing the timescales of relaxation and GW emission as given in \cite{Hopman:2005vr}. The ratio of plunges to inspirals is then simply given by $N_{p/i} = \dot{N}_\text{plunge}/\dot{N}_\text{inspiral}$.

An additional thing to consider is the stochastic nature of the stellar diffusion. Due to the Brownian motion in the phase space flow, COs can get scattered below the separatrix, which can transform a plunge into an inspiral. This was first explored in \cite{Qunbar:2023vys} and called a \textit{cliffhanger}. These can change the ratio $N_{p/i}$, which we will explore in section \ref{sec:rates:plungevsinspiral}.

\subsubsection{Depletion of stellar cusps}
The rate of COs falling into the MBH can then be expressed as a mass accretion rate for the MBH, which can be estimated as \cite{Amaro-Seoane:2010dzj}
\begin{align}
    \dot{M} ={}& m_* R_0(1+N_{p/i}) ,\\
    \text{with } R_0 ={}& 300 \left(\frac{M}{10^6 \Msun} \right)^{-0.19} \text{ Gyr}^{-1} ,
\end{align}
with $m_*$ the characteristic mass of the CO. 

Unfortunately, this accretion rate is too large, it would exclude the existence of MBH with masses $< 10^{5}\Msun$, and also assumes an infinite supply of COs to draw from \cite{Babak:2017tow}. The time it takes to deplete the stellar cusp form this accretion can be estimated as \cite{Babak:2017tow}
\begin{equation}
    t_d = \frac{20}{1+N_{p/i}} \left(\frac{m_*}{10\Msun} \right)^{-1} \left(\frac{M}{10^6\Msun} \right)^{1.19} \text{ Gyr}.
\end{equation}
This has to be compared with the relaxation time that supplies new COs to the sphere, and leads to a duty cycle for I/EMRIs
\begin{equation}
    \Gamma = \min\left(\frac{t_d}{t_\text{relax}}, 1\right) .
\end{equation}
The mass growth is then
\begin{equation}
    \Delta M = m_* \Gamma R_0 \underbrace{\int \d z \dv{t}{z}p_0(M,z) \mathrm{d}z }_{t_\text{EMRI}} . 
\end{equation}
The resulting accretion rate is still too large in some parameter regions. Therefore, an artificial cap can be placed on it to dampen the accretion. We can assume that the MBH can only grow by an $e$-fold during its lifetime from CO accretion, giving a damping factor
\begin{equation}
    \kappa = \min\left(\frac{1}{e} \frac{M}{\Delta M}, 1 \right) .
\end{equation}
Finally, the effective I/EMRI rate is given by
\begin{equation}
    R = \kappa \Gamma R_0 .
\end{equation}

The influence of the $\Gamma$ factor strongly depends on $N_{p/i}$. For $N_{p/i}\approx 0$, the factor is $\Gamma\approx1$, implying that $R_0$ is already close to the supply rate of COs. 
The $\kappa$ factor primarily suppresses the rate for MBH with masses below $10^5\Msun$, making the estimation conservative in this region\cite{Babak:2017tow}.

The overall rate can then be integrated in the mass range observable by LISA, redshift out to $z=4.5$, and spin distribution. This gives a rate of $R \sim 1600$ for $N_{p/i}\sim 10, m_*=10\Msun$. For a table of different values, see \cite{Babak:2017tow}.

\subsubsection{Dry vs Wet}
So far we have assumed the inspiral due to the relaxation processes of the stellar cusp. These dry inspirals generally have a high eccentricity as they enter the observable band, as seen in section \ref{sec:signatures:dry}.

As explored in section \ref{sec:signatures:wet}, the presence of an accretion disk can align the orbits of surrounding objects and cause a wet inspiral. In this scenario, the accretion disk can create an inspiral out of a plunge. This has been explored in \cite{Pan:2021ksp, Pan:2021oob}. They find that effectively $N_{p/i}\approx 0$ for those systems with an accretion disk, which they assume to be $1\%$ of the active MBHs. 

This is somewhat in contrast to our results, where we have found a critical alignment angle. Only the fraction of plunges below the critical alignment angle $\iota_c$ will be converted into inspirals. If we assume an isotropic distribution, this is the fraction $f=\frac{\iota_c}{\pi}$. The new $N'_{p/i}$ is then given by $N'_{p/i} = N_{p/i} \frac{1-f}{1+f N_{p/i}}$, where $N_{p/i}$ is the rate without the accretion disk. Of course, there are a lot of assumptions involved here, and most likely, the stellar distribution is not isotropic in the presence of an accretion disk. This will have to be left to future studies.

Lastly, there are other possible formation mechanisms for these inspirals, which we have not considered so far. Worth mentioning are the in-situ formation of COs inside accretion disks, as explored in \cite{Derdzinski:2022ltb}. This can provide a source for COs inspiraling outside of the limits of the depletion of the stellar cusp.

\subsubsection{Detection}
Finally, having estimated the rate of I/EMRI events in the local universe, the last step is to determine how many of them can be observed by LISA. If the underlying waveform is known, the SNR $\rho$ can be calculated with \eqref{eq:SNR}. Assuming a detection threshold of $\rho > 20$, \cite{Babak:2017tow} find a sizable number of detections per year. Depending on the model assumptions, these can range from $\mathcal{O}(1) - \mathcal{O}(2000)$yr$^{-1}$. LISA can reliably detect inspirals out to a redshift of $z\sim 1$, and observe a sizable fraction up to redshift $z\sim 3$. The mass ranges for the MBH are $3\cdot 10^4 \Msun - 3\cdot 10^6\Msun$, where the most common detectable mass is between $10^5 - 10^6\Msun$.

Generally, a lower number of plunges per inspiral $N_{p/i}$ leads to a larger number of inspirals and therefore observations. Larger masses for COs also increase the signal strength and allow larger distances to be observed. 

\section{Dark Matter Spikes \label{sec:rates:darkdresses}}
Discussing the presence of DM spikes in the context of rates is difficult, as it comes into play at all stages throughout the description. It affects the MBH evolution, mergers, the stellar cusp, the plunges and the inspirals. There has been no comprehensive study including their effects at all stages. We can sketch some consequences here.

\subsubsection{MBH Mergers}
The study of MBH mergers is often done in $N$-body simulations in the context of structure formation. Here, large DM halos with central MBH are `let loose' and their merging dynamics studied\cite{Merritt:2006hn, 2023arXiv231008079P}. These simulations are generally not sufficient to resolve below the parsec scale on which a DM spike could reasonably be expected.

DM spikes have been invoked in the context of the 'final parsec problem'. Here, two merging MBH can come very close due to dissipation processes in the nucleus, but not close enough for GW emission to be strong enough, and therefore stalling at about $\sim1$pc\cite{Milosavljevic:2002ht}. DM spikes could help by providing additional dynamical friction and thus bridge this gap\cite{Alonso-Alvarez:2024gdz}.

Nonetheless, in these SMBHB mergers, the spike will most likely not survive. Just as the stellar distributions are depleted, so are the DM distributions\cite{2023arXiv231008079P}. This can leave behind scoured cores of low density. From there, any DM spike would have to be regrown dynamically, possibly in conjunction with the stellar distribution as mentioned in section \ref{sec:environment:spike}. This could change the stellar cusp regrowth timescales as an additional energy sink. 

\subsubsection{Plunge vs Inspiral \label{sec:rates:plungevsinspiral}}
One effect on the rates that we can quantize is the plunge vs inspiral dynamic. This could change in the presence of a DM spike.

For a kinematically heated DM spike $\gamma\sim 3/2$, the dissipative effects are always subdominant to the stellar cusp. The presence of the DM spike in this case would not affect any of the mechanisms discussed before. For a DM spike with larger power law index $\gamma\sim 7/3$, the DM energy loss might be stronger in some area of the parameter space, but the angular momentum loss due to the relaxation processes is still much larger, thereby only mildly altering the phase space flow and not the GW separatrix. What could happen is that due to stochastic motions, objects that get thrown into the modified region of the phase space are more likely to stay there. We can explore this following \cite{Qunbar:2023vys}.

We can solve the stochastic differential equations derived in section \ref{sec:inspiral:inspiral} in a system starting at different initial distances $a_0$ and with eccentricity $e_0=0.1$. Then, we can check the ratio of inspirals vs plunges and get an estimate of $N_{p/i}$. To this end, we perform a hundred realizations of the stochastic differential equations for each starting value of $a_0$ and see whether the secondary plunges or inspirals. 

The resulting effect on $N_{p/i}$ can be evaluated by modifying Eqs.{}(\ref{eq:Nplunge})\&(\ref{eq:Ninspiral}). We define a function $S(a)$ as the inspiral fraction starting from a given $a$. The rates are then given by \cite{Qunbar:2023vys}
\begin{align}
    \dot{N}_\text{plunge} &= 4\pi\int_{0}^{a_{\inf}} \frac{a^2 n_*(a) (1-S(a))}{\ln(J_{c}/J_{lc})t_\text{relax}} \mathrm{d}a, \\
    \dot{N}_\text{inspiral} &= 4\pi\int_0^{a_{\inf}} \frac{a^2 n_*(a) S(a)}{\ln(J_{c}/J_{lc})t_\text{relax}}\mathrm{d}a.
\end{align}

\begin{figure}
    \centering
    \includegraphics[width=\textwidth]{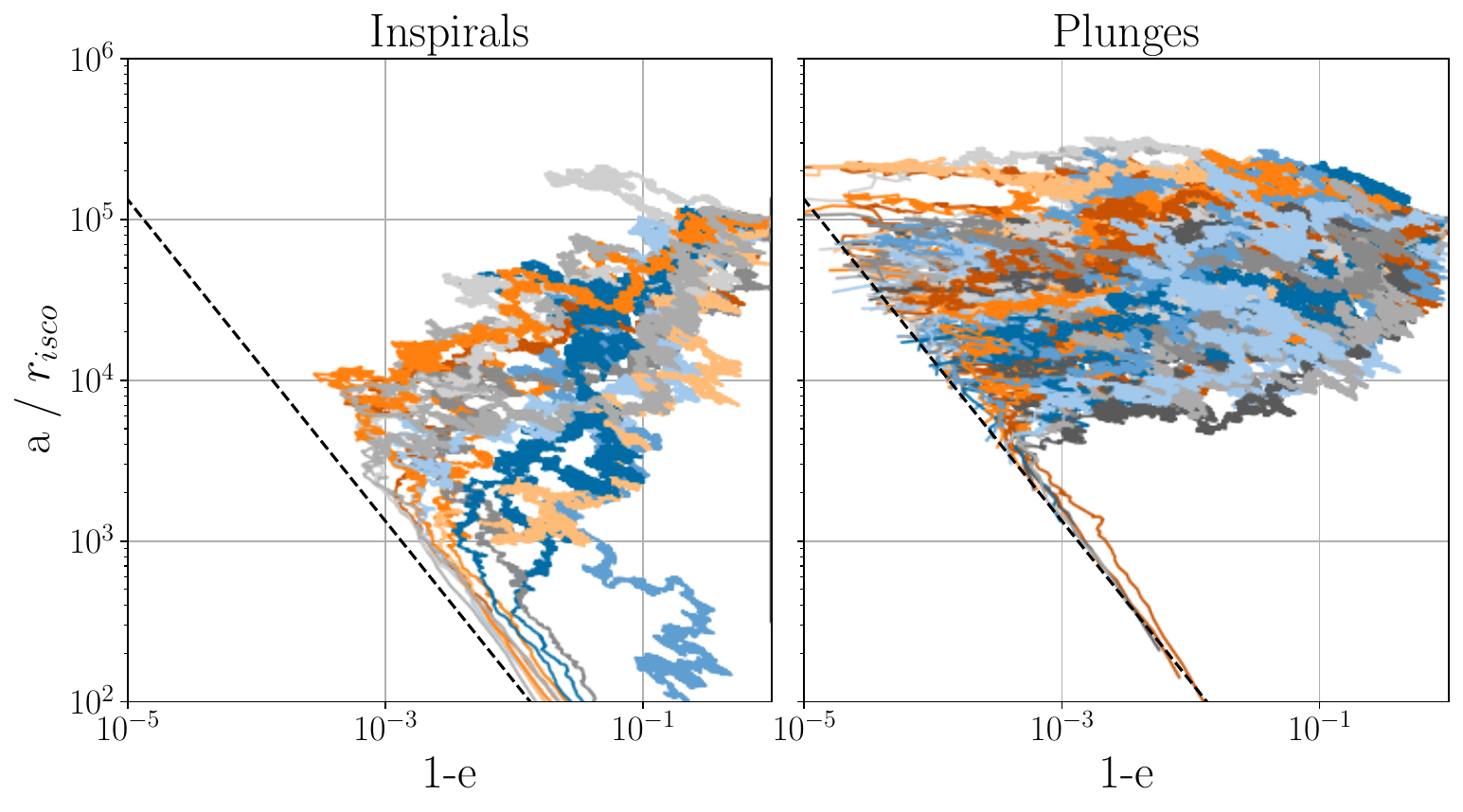}
    \caption{Different realizations of the stochastic differential equations resulting in a random walk in the phase space. Here, $m_1=10^5\Msun$, with $a_0=10^5 r_\isco$, and only the stellar diffusion mechanism. On the left are the inspirals, on the right the plunges.}
    \label{fig:inspiral_vs_plunge}
\end{figure}

A set of example realizations are plotted in \figref{fig:inspiral_vs_plunge}. Some realizations never make it during the evolution time and some result in numerical errors. An inspiral is counted when the semimajor axis dips below $10^2 r_\isco$. In the end, we count the number of inspirals vs the total number of inspirals and plunges to obtain an estimate of $S(a)$.

\begin{figure}
    \centering
    \includegraphics[width=0.7\textwidth]{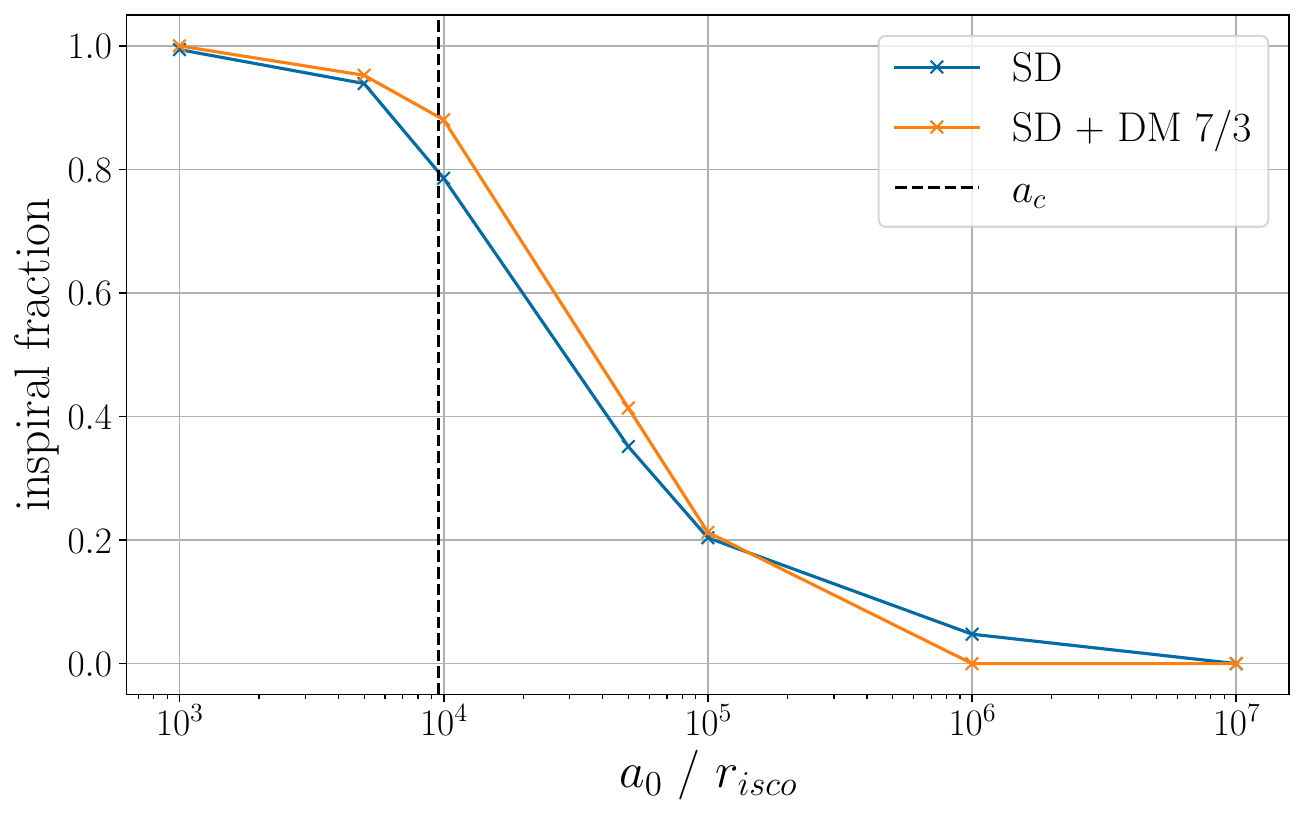}
    \caption{The inspiral fraction for different initial semimajor axes. Here, the fraction of inspirals out of $100$ realizations of the SDEs is shown for a system $m_1=10^5\Msun$ with stellar diffusion and one with a steep DM spike on top. The dashed line gives the approximate critical semimajor axis $a_{GW}$, or the GW separatrix between plunges and inspirals. }
    \label{fig:inspiral_fraction}
\end{figure}

We repeat the whole procedure with the steep spike model spike on top. As seen in \figref{fig:ev_dry}, the effect on the phase space flow is small but measurable for the steep spike, but not for the flat one. The resulting inspiral fraction is plotted in \figref{fig:inspiral_fraction}. It can be seen that the DM spike does not affect the inspiral fraction, within statistical errors. Therefore, at this level, the steep spike should not have any measurable effect on the rate of inspirals.

This inquiry is a first step to asses the rate impact of DM. There are other effects at play in the rate calculations, such as resonant relaxation and mass segregation\cite{Bar-Or:2015plb}. But if DM is already negligible here, there is not much of an effect expectable when the model is complicated.

The I/EMRI rate in the presence of a DM distribution has been considered before in \cite{Yue:2018vtk}. Since they leave out effects from stellar diffusion, they find a large increase in the merger rate, compared to the vacuum case. This can be seen from the phase space flow in the isolated spike case. In our modeling, DM is a subdominant effect to the stellar diffusion, so its effects seem to be negligible. 

\FloatBarrier
\section{Discussion}
The rate calculations in this chapter are based on many assumptions, since the population of MBHs is not well understood or measured. The previous sections have primarily focused on the population of MBHs that is at the center of nuclear clusters surrounded by stars. But there is a plethora of other models that we will briefly mention here.

There are processes that can enhance the rate of I/EMRIs, for example supernova kicks that scatter COs onto small orbits\cite{Bortolas:2019sif}, capture or in-situ formation in accretion disks\cite{Pan:2021ksp, Derdzinski:2020wlw}, and binary disruption events, where binary stars are disrupted close to the MBH\cite{ColemanMiller:2005rm}.
The calculation also assumed that during and after the merger process of MBH there are no I/EMRIs, but the presence of an MBH binary can actually enhance the I/EMRI rate with the help of Kozai-Lidov oscillations\cite{Naoz:2022rru,Naoz:2023hpz}. 
The process of these MBH mergers also results in many unbound MBH after they experience strong recoil kicks\cite{2023arXiv231008079P}. Some of these might 'steal' their environment and have surrounding bound structures that leave the galaxy, resulting in extragalactc I/EMRIs, or find new homes. There could also be a population of PBHs, possibly with DM spikes\cite{Cole:2022ucw}, dispersed throughout the universe. If they are relatively light they could fall into an SMBH to form an I/EMRI\cite{Guo:2017njn, Kuhnel:2018mlr, Wang:2019kzb, Barsanti:2021ydd}. If they are heavy they would need to find a CO source for I/EMRIs, or be the progenitors of the SMBHs we see today. And if they come in different sizes they could inspiral into each other in PBH clusters\cite{Huang:2024qvz, LISACosmologyWorkingGroup:2022jok}.

If the I/EMRI rates are too large, they might even form a confusion background that overshadows LISA sensitivity\cite{Bonetti:2020jku}.

Ultimately, trying to predict the rates of I/EMRIs when there is little known about the underlying MBH population is very difficult. We can try to predict the rates with the MBH population that is expected at the centers of most galaxies, but outside of this, there is much left to discover. Add to this recent results from JWST, which are messing with our models of structure formation\cite{Silk:2024rsf} and calling into question the model presented in the first section. Trying to measure DM effects on top of this has to be taken with a grain of salt then.

On the other hand, this means that there is a lot to learn when the first data comes in. It can illuminate the history of MBHs, structure formation, and more exotic ideas such as PBHs.

\chapter{Discussion \& Outlook \label{chap:discussion}}

\section{Discussion \label{sec:outlook:discussion}}
I/EMRIs are one of the prime targets of LISA\cite{LISAConsortiumWaveformWorkingGroup:2023arg, Babak:2017tow}. Their extended nature allows for stringent tests of GR and the host environment. But they are not the only sources visible with LISA. Other expected sources are galactic binaries, MBH binaries, sBH binaries, and possibly a stochastic GW background from cosmological sources such as cosmic strings or FOPTs\cite{LISAConsortiumWaveformWorkingGroup:2023arg,LISACosmologyWorkingGroup:2022jok, Kamionkowski:1993fg}.

All these sources might be so numerous that their individual protagonists are not resolvable, and a hum of superposed waves is observed. The approach of the data analysis pipeline is primarily a global fit, where all the populations and the most prominent individual sources are measured at the same time. For these fits, one needs accurate waveforms but also a way to cover the huge parameter space that is given for these populations. This is a challenging endeavor.

One way to help detect I/EMRIs might be their electromagnetic counterparts. There are several proposed mechanisms that can lead to EM signals in I/EMRIs, such as tidal disruption\cite{Wang:2019bbk}, fast radio bursts\cite{Li:2023dst}, or Roche lobe overflow\cite{Zhao:2021swe}. These could help narrow the parameter space and allow for individual detections.

In the absence of EM counterparts, one has to rely on GW data alone. We believe that some of the tools presented in this dissertation can help in this effort. First, on the level of populations, the braking index and the phase space flow are useful. In the case of dry inspirals for example, which are highly eccentric, the braking index of the frequency evolution is given by the simple $n_b = 4/3$. If there is a frequency component in the LISA data that evolves in this manner, it would point strongly to dry I/EMRIs, as the other sources are expected to be at lower eccentricity. This can be abstracted to the phase space flow in $(a,e)$, as shown with the different DM power laws in \figref{fig:psf_dm_pwrlaw}. This $(a,e)$ flow can be translated into a flow in frequency and the strength of the harmonics. In this way, different populations with different environmental effects can be extracted from data.

The possible measurement of the braking index has been addressed in the context of galactic binaries in \cite{Takahashi:2002ky, Renzo:2021aho}. As a next step, this can be applied to I/EMRIs. Of course, this is difficult for I/EMRIs in their early stages, where the frequency changes immeasurably during the observation time\cite{Seoane:2024nus}. But later on in the inspirals, especially the dry ones, they are on much shorter timescales, so the measurability should improve.

\subsection{Signatures}
When an individual source can be resolved, there are more possible signatures of the environment. We have looked at several possible signatures of DM, but these are, in principle, applicable to any environmental effect.

\subsubsection{Dephasing}
Dephasing was the first proposed signature of DM spikes\cite{Eda:2013gg, Eda:2014kra}. The systems we have explored can accumulate large dephasings throughout their lifetime. For some of them, the dephasing might be observable in the lifetime of LISA. The broader question is whether this dephasing can be differentiated from other processes that cause dephasing and accurately ascribed to DM. To this end, we have derived the dephasing index and shown that it is closely related to the braking index, and can therefore distinguish between different dissipative effects as well. We can successfully differentiate between different types of DM spikes and accretion disks. It is most useful in the case of quasi-circular inspirals, for highly eccentric inspirals it is not necessarily sufficient to identify the dissipative effects. Here, a generalization is needed. Nevertheless, it can most likely be resolved with a Bayesian analysis akin to \cite{Cole:2022fir}.

\subsubsection{Deshifting}
Similar to dephasing, the difference in the periapse precession can also be a signature of DM\cite{Dai:2021olt, Dai:2023cft, Mukherjee:2023lzn}. This is due to the additional mass precession and also due to the different eccentricity evolution in the presence of DM. In our late-stage inspirals, mass precession is negligible. We have shown that deshifting can also be significant throughout the whole lifetime of inspirals, most importantly in the late stage. In section \ref{sec:inspiral:GWsignal}, we have derived the deshifting index, similarly to the dephasing index, and shown that this late time deshifting and dephasing are closely linked. Deshifting is an additional tool in the case of eccentric inspirals, especially in dry inspirals it can be as large as the dephasing. So far, we have just looked at Schwarzschild precession. As the discussion in section \ref{sec:environment:spacetime} makes clear, this is a first approximation. To accurately assess the impact, we need to model the Kerr spacetime and its geodesics. A first step in this direction has been \cite{Dai:2023cft}, with geodesics in a Schwarzschild background. They also find an observable deshifting for much lower DM densities. While more modeling is needed, this can be a promising additional signature.

\subsubsection{Alignment}
We have, for the first time, looked at the alignment process of a secondary with the accretion disk as a DM signature. Modeling the alignment process is still in its infancy\cite{Fabj:2020qqc,Nasim:2022rvl}. While the authors of these papers have generally found alignment for most of the systems described, in our case, we have found a critical alignment angle. This comes down to several differences in the modeling. We have assumed a Derdzinski \& Mayer disk, with smaller densities. And we have included the energy loss due to stellar diffusion, which speeds up the inspiral. Adding DM on top of this, we have shown that the critical alignment angle decreases further. 

There still remains a lot to be done, firstly the generalization to eccentric orbits. The presence of a stellar distribution can then strongly eccentrify the system and speed up the inspiral. This would make alignment even more difficult. Whether DM can then influence this process significantly remains to be seen. The flat $\gamma=3/2$ spike appears to have no influence on the alignment process.

Lastly, the alignment happens primarily outside of the observable band. We would most likely only observe the resulting distribution and not the process itself. This would mean we could only make statistical inferences.

Overall, the alignment process is of significant physical relevance and needs to be modeled. It will allow us to understand disk properties and the secondary's interaction with the disk. But it does not seem to be a good tool to infer DM.

\subsubsection{Dry vs Wet}
We have looked at both dry and wet inspirals and explored the influence of DM spikes.

For wet inspirals, which are quasi-circular inside an accretion disk, the inspirals are dominated by the accretion disk physics. In the model applied here, DM can leave large dephasings, for a steep $\gamma=7/3$ spike, but also for a more physically motivated flat $\gamma=3/2$ spike. Thanks to the dephasing index, we can differentiate between disk and DM effects. Here, the halo feedback model has to be included for small MBHs. In the long term, in the presence of the disk, the disk-spike interaction also has to be taken into account, as the DM spike cannot be spherically symmetric anymore. Therefore, the main challenge in inferring DM lies in understanding the disk physics, the disk-spike interaction, and halo feedback processes.

For highly eccentric dry inspirals, the inspirals are dominated by GW emission. For a steep $\gamma=7/3$ spike, the DM effects can be visible through a different eccentricity evolution, but for a flat $\gamma=3/2$ spike, the dephasing is relatively small and concentrated at the end. Here, there is also a deshifting of the periapse due to altered precession, which is an additional tool. Fortunately, the halo feedback process seems to be negligible. The main challenge for inferring DM lies in modeling the relativistic processes around the MBH correctly, as the secondary can reach high velocities and is subject to large precession.

\subsubsection{Dark Matter}
In this dissertation, we assumed DM to be cold and collisionless and looked at its signatures. If DM is not described by the CDM paradigm on these scales, the distribution and interactions with the secondary will be different. Generally, though, these interactions tend to be stronger, or, as in the case of the gravitational atom, have a clear structure\cite{Cole:2022fir}. Arguably, this CDM description can be seen as a conservative lower bound, meaning that if DM is not CDM, it will be more observable.

\subsection{Existence of Spikes \label{sec:outlook:spikes}}
We have already discussed some of the spike properties in section \ref{sec:environment:spike}. Here, we want to focus on the existence of spikes. The process of adiabatic growth described by Gondolo \& Silk \cite{Gondolo:1999ef}, assumes an isolated MBH at the center of an NFW profile. While Ullio\cite{Ullio:2001fb} has mapped out the consequences of rapid growth or an off-center MBH, the largest problem is that MBH are not necessarily isolated. Most models of MBH formation assume them to be in the center of baryonic activity. Here, the accretion disks, gas clouds, and stellar populations would interact with the DM. This can be seen in galaxy formation simulations where the stellar feedback processes create a DM core instead of a cusp\cite{Mashchenko:2007jp, Read:2015sta}, and the DM profile is not NFW. Similarly, theoretical models with stellar kinematic heating effects cause a flatter cusp\cite{Gnedin:2003rj, Merritt:2003qk}. 

Secondly, and more importantly, galaxies happily merge, and with them, their central MBH. These can completely unbind the spikes\cite{2023arXiv231008079P}. They have to be formed from scratch with some dynamical friction or diffusion processes, also resulting in a flatter spike. 

Thirdly, at least in the Milky Way, there is no evidence of the existence of a spike in the center, only bounds that exclude steeper profiles\cite{Shen:2023kkm}. The adiabatically grown $\gamma=7/3$ spike is excluded, but the kinematically heated $\gamma=3/2$ spike is still possible.

So how realistic is a steep spike? These require a system that is largely isolated, that has enough material to adiabatically grow to large masses, but not enough to disrupt the spike with baryonic processes. This might most reasonably realized in a PBH configuration\cite{Cole:2022ucw}, but we have no evidence of their existence, and there are bounds for the systems massive enough to be in the LISA band. While these pristine isolated systems might exist, their numbers must be quite low. Add to this the need for a secondary object to inspiral, and it seems unlikely that we will observe this with LISA.

Now on to the flatter spike. While these have more physical justification to exist in galactic centers, we have no evidence for them, only constraints\cite{Shen:2023kkm}. Another process worth mentioning here is that of halo feedback. While it has been shown that for a steep $\gamma=7/3$ spike, it is only `reshuffled' and not unbound, there has been no such simulation for a flatter spike. These have lower binding energies and might be more easily unbound. If one single secondary can unbind the spikes, we should not expect them to last very long.

In the same vein, and more critically, the recent publication \cite{Mukherjee:2023lzn} has a similar setup to that of the halo feedback simulations but evolved for longer. They see the emergence of three body interactions of the secondary with the DM particles, at the end of which the DM particles are ejected from the system. So the halo feedback model might be incomplete and has to take into account these more complex 3-body interactions. But if these results hold, the spike would not just be reshuffled but completely unbound as well.

In these cases, there might be a period of replenishment of DM particles from outside. This could give duty cycles for spikes, similarly to the stellar ones assumed in the rate estimations. DM spikes and stellar cusps could grow in unison after their depletion. This could keep the DM at low densities throughout. More research is needed to assess these interactions.

All in all, this does not paint a great picture for the detection of spikes with most of the inspirals observable. Nonetheless, the formalism developed here can be applied to different possible environmental effects and thus be useful even without DM spikes.

\subsection{Stochastic Effects}
In this dissertation, we have, for the first time, applied SDEs to the problem of inspiral modeling. Previous inquiries have implicitly used a related scheme to solve the Fokker-Planck equations. Mathematically, there is not a difference in the description. But for SDEs, there is a sizable amount of software packages available, mostly coming from the area of machine learning\cite{kidger2021on, kidger2021neuralsde}. Utilizing these libraries could vastly speed up numerical inquiries. Unfortunately, our implementation is not numerically optimized for this. This has to be seen as a proof of concept.

While the stochastic description has mostly been used in the context of rates\cite{Qunbar:2023vys, Bar-Or:2015plb}, arguably it is also useful in a population modeling context, akin to \cite{Seoane:2024nus}. In the case of stellar distributions, the secondaries can perform a kind of random walk in the phase space. This could make them change their frequency evolution stochastically and it would contribute a background signal to LISA with hard-to-resolve individual sources. But the population as a whole has a characteristic drift in the phase space, which might make it identifiable. To this end, one could model the population with SDEs and calculate the resulting GW signal. This requires a careful treatment of the SPA due to the nondifferentiability of SDEs.

While the description of the stellar diffusion in section \ref{sec:environment:stars} has negligible effects close to the loss cone, its stochastic description breaks down. Close to the MBH, the presence of a single massive object can influence the inspiral in a complex three-body interaction. While it is difficult to describe the individual inspiral, this might be done on a population level. Also, in section \ref{sec:environment:disk}, we have described the accretion disk as a static entity. But observations of Sag A* point to variabilities in disk luminosity in a matter of hours\cite{Baganoff:2001kw}. Generally, the smaller the MBH, the faster its variability in the accretion process can be. Add to this in-situ formation of objects in the accretion disk\cite{Derdzinski:2022ltb} with which the secondary can interact gravitationally. It is clear that describing these effects requires a stochastic description. 

Lastly, to track individual waveforms for years, what was described on a stochastic level needs to be modified to include individual gravitational encounters. Here, the waveform can suddenly change drastically. Detection algorithms will have to be able to adapt to these changes to not lose track of the inspiral.

\subsection{This work}
While many research works go into detail on one specific process, in a kind of depth first analysis, we see this work in a breadth first approach. While the individual models certainly remain to be improved, the combination of them can already give hints as to which effects dominate when and where. In this way, one can focus the exploration on the dominant effects first, instead of polishing ever more detailed models whose effects are not observable in the end. This is in line with the proposal of \cite{Zwick:2022dih}, focusing on environmental effects instead of more accurate vacuum templates. 

To this end, we have tried to do the implementation of the code in a modular fashion. In this way, new models can easily be integrated and their effects assessed. Also, the old models can be refined and made more detailed. Unfortunately, this is in a primarily Newtonian fashion. While we can include PN effects, we cannot model the intrinsically relativistic effects such as orbital resonances. Nevertheless, this can be seen as a first tool for exploration and estimation of different effects. As a next step, the inclusion of spin effects should be prioritized.

The library \lib{imripy} is written in \textsc{Python}, uses \lib{numpy}\cite{harris2020array}, \lib{scipy}\cite{2020SciPy-NMeth}, \lib{matplotib}\cite{Hunter:2007}, and a modified version of \lib{torchsde}\cite{li2020scalable, kidger2021neuralsde} for the SDEs. The code is publicly available at \cite{imripy}, along with examples and the notebooks to create all the results shown in this work.

\section{Outlook}
In this dissertation, we have explored I/EMRIs and the environmental effects they are subject to. We have provided descriptions for diffusion inside stellar distributions, for interactions with accretion disks, and dynamical friction within DM spikes. We have explored their phase space flow, their combined effects, and tried to detect the influence of DM. We have discussed possible signatures of DM, dephasing, and deshifting, a difference in periapse precession. We have also looked at the alignment processes in the presence of accretion disks. We have studied dry and wet inspirals and analyzed their different characteristics. We have also provided a description of the stellar diffusion with SDEs and explored the stochastic effects on rates of I/EMRIs.

We have found that, generally, the stellar diffusion processes are dominant. We have found that steep DM spikes can change the phase space flow of the inspiral, while flat spikes can only be observed through their secondary effects. While these flat spikes might be hard to detect, for some regions of parameter space, a combination of dephasing and deshifting should make the spike inferable. Dry inspirals might be an easier target for the detection of DM spikes, as they do not require halo feedback modeling, and happen more quickly, thus being within the lifetime of LISA. Wet inspirals are more affected by the disk physics, which are still uncertain and can be subject to large variabilities. The stellar distributions in the presence of accretion disks need to be explored.

We have argued that the inclusion of stochastic effects is vital to understand both individual inspirals and whole populations. It might be possible to map out populations of inspirals with the help of characteristic indices, such as the braking index.

The field is moving toward more precise relativistic descriptions of spacetime and the inspiral process, which of course is necessary to resolve the final stages of the inspiral. But further out, some kind of stochastic treatment might still be vital, especially to resolve the waveform for years.

As we have discussed, the existence of spikes is questionable. Certainly, the observation of those pristine, isolated spikes is unlikely. But if these spikes exist, they should be detectable, and with the tools listed here, identifiable. We hope that the tools are useful outside of a DM setting as well.

After all, we have about a decade left before the era of space-based GW observation begins. And with the right modeling, we might be able to learn about accretion disks and DM, stellar distributions and the galactic center, MBHs and their origins, galaxy mergers and their history, Kerr spacetime, and GR. Recent discoveries of the GW era have already revolutionized our understanding of the universe. One can only wonder what space based interferometers will have in store for us.

\newpage 

\appendix

\chapter{Appendix}
\section{Differential Equations with Accretion \label{sec:appendix:accretion}}
The total derivative of \eqref{eq:E} is $\dv{E_{orbit}}{t} = \pdv{E_{orb}}{a}\dv{a}{t} + \pdv{E_{orb}}{m_2}\avg{\dv{m_2}{t}}$. In deriving \eqref{eq:da_dt}, we have disregarded the second part. Here, we want to justify this in the case of accretion. This boils down to momentum conservation for the secondary object $\dot{v}m_2 = -v\dot{m}_2$. If we seperate the orbital energy into its constituents, the kinetic and potential energy, the derivative gives (on instantaneous and not secular timescales)
\begin{align}
    \dv{E_{orb}}{t} =& \dv{}{t} \left(\frac{1}{2}\mu v^2 - \frac{m_1 m_2}{r}\right) \nn \\ 
     = {}&  \frac{1}{2}\dot{\mu} v^2 + \mu v\dot{v} - \frac{m_1}{r} \dot{m}_2 \nn \\
     = {}& \frac{1}{2} \frac{\mu^2}{m_2^2} v^2 \dot{m}_2 - \frac{\mu}{m_2}v^2 \dot{m}_2 - \frac{m_1}{r}\dot{m}_2 \nn \\
     = {}& \dot{m}_2 \frac{m\mu}{m_2} \left( \frac{1\mu}{2m_2} \left(\frac{2}{r} - \frac{1}{a}\right) - \left(\frac{2}{r} - \frac{1}{a}\right) - \frac{1}{r} \right) \nn \\
     = {}& \dot{m}_2 \frac{m\mu}{m_2}\left( - \left( \frac{3m_2 - \mu}{2m_2} \right) \left(\frac{2}{r}-\frac{1}{a}  \right) - \frac{1}{2a}  \right) \nn \\
     = {}& - \underbrace{\dot{m}_2 \frac{\mu}{m_2} \frac{3m_2-\mu}{2m_2} v^2}_{= F_{acc}v  } - \frac{m_1}{2a}\dot{m}_2
\end{align}
which means for the derivation of \eqref{eq:da_dt} (on secular timescales)
\begin{equation}
\label{eq:da_dt_expanded}
    \pdv{E_{orb}}{a}\dv{a}{t} + \cancel{\pdv{E_{orb}}{m_2}\avg{\dv{m_2}{t}}} = \dv{E_{orb}}{t} = -\avg{F_{acc}v} - \cancel{\frac{m_1}{2a}\avg{\dot{m}_2}}.
\end{equation}
Similarly, for the angular momentum, we have
\begin{align}
    \v{L} ={}& \mu \v{r}\cross \v{\dot{r}} \\
    \dv{\v{L}}{t}  = & \dot{\mu}\, \v{r}\cross \v{\dot{r}} + \mu \,\v{r}\cross \v{\ddot{r}} \nn \\
     = {} & \v{r}\cross \v{\dot{r}} \left(\dot{\mu} - \frac{\mu}{m_2} \dot{m}_2 \right) \nn \\
     = {} &  \v{r}\cross \underbrace{\v{\dot{r}}\text{ } \dot{m}_2 \frac{\mu}{m_2} \big( -\frac{3m_2 - \mu}{2m_2}}_{\v{F}_{acc}}  + \frac{m_2 + \mu}{2m_2}\big)  
\end{align}
Therefore, the terms that do not include the accretion force cancel out in the derivation of \eqref{eq:de_dt} 
\begin{align}
&\frac{1}{E}\dv{E}{t}|_{F_{acc}=0} &+& 2 \frac{1}{L}\dv{L}{t}|_{F_{acc}=0} &-& \frac{1}{m^2\mu^3}\dv{(m^2 \mu^3)}{t} && \nn \\ 
=& \avg{\dot{m}_2}\frac{1}{m_2} &+& \avg{\dot{m}_2} \frac{m_2 + \mu}{m_2^2} &-& \avg{\dot{m}_2} \frac{2m_2 + 3m_1}{m m_2}  &=& 0 
\end{align}
and we do not have to consider them. Thus, the change of energy due to the accretion can be modeled by the inclusion of $F_{acc} = \dot{m}_2 \frac{\mu}{m_2} \frac{3m_2-\mu}{2m_2} v \approx \dot{m}_2v$ alone.

\section{Translating the three frames \label{sec:appendix:orbtoobs}}
In section \ref{sec:inspiral:GWsignal} we have used the two angles $\iota'$ and $\beta'$ to describe the alignment between the orbital plane and the observer plane. However, in the preceding sections, we have described our system within the fundamental frame, with the parametrization explained in section \ref{sec:inspiral:kepler_orbit} and depicted in \figref{fig:kepler_orbit}. Here, we want to derive the transformation between these three different frames. 
Generally, a 3D rotation can be described by the Euler angles $(\varphi, \theta, \psi)$, as a sequence of three rotations, first $\varphi$ around the $z$-axis, then $\theta$ around the $x$-axis, and finally $\psi$ around the $z$-axis.

This is captured by the three rotation matrices
\begin{align}
    R_1(\varphi) &= \begin{bmatrix}
                        \cos(\varphi) & \sin(\varphi) & 0 \\
                        -\sin(\varphi) & \cos(\varphi) & 0\\
                        0 & 0 & 1
                    \end{bmatrix}
\\
    R_2(\theta) &= \begin{bmatrix}
                        1 & 0 & 0 \\
                        0 & \cos(\theta) & \sin(\theta)  \\
                        0 & -\sin(\theta) & \cos(\theta) 
                    \end{bmatrix} 
\\
    R_3(\psi) &= \begin{bmatrix}
                        \cos(\psi) & \sin(\psi) & 0 \\
                        -\sin(\psi) & \cos(\psi) & 0\\
                        0 & 0 & 1
                    \end{bmatrix}
\end{align}
The overall rotation is then given by the matrix multiplication $R(\varphi, \theta, \psi) = R_3(\psi)R_2(\theta)R_1(\varphi)$.

To go from the orbital frame to the fundamental frame, we can apply the inverse rotation with the angles described in section \ref{sec:inspiral:kepler_orbit} as $\bar{R} = R(-\omega, -\iota, -\Omega)$\cite{poisson_will_2014}. 
If we choose the longitude of the ascending note $\Omega$ such that the X direction of the fundamental frame points to the observer, we just need one more rotation to go from the fundamental plane to the orbital plane. We can parameterize this with the inclination $\tilde{\iota}$, which gives the last rotation as $\tilde{R}=R(0, \tilde{\iota}, 0)$. The whole transformation is then given by $R' = \tilde{R}\bar{R}$, from which we can extract the two angles needed, $\beta'$ and $\iota'$, with
\begin{align}
    \iota'  &= \arccos(R'_{33})\\
    \beta'  &= -\arctantwo(R'_{31}, R'_{32}) \\
    \Omega' & = \arctantwo(R'_{13}, R'_{23})
\end{align}
where $\arctantwo$ is $\arctan$ with the additional consideration of the quadrant it is in. The third angle $\Omega'$ describes the angle between the polarization axis in the observer plane to the ascending node of the orbit. 
This gives the equations
\begin{gather}
\begin{aligned}
    &&\cos{\iota'} &= \mathrlap{\cos\iota \cos\tilde{\iota} - \sin\iota \sin\tilde{\iota} \cos\Omega} \\
    &&\beta' &= \arctantwo \bigl( && (\sin\Omega \cos\omega + \sin\omega \cos\Omega \cos\iota ) \sin\tilde{\iota} + \sin\iota \sin\omega \cos\tilde{\iota}, \\
             && & {} && (\sin\Omega \sin\omega - \cos\omega \cos\Omega \cos\iota )  \sin\tilde{\iota} - \sin\iota \cos\omega \cos\tilde{\iota} \bigl)   \\
    &&\Omega' &= -\arctantwo\bigl( && \sin(\Omega)\sin(\iota) , \\
                &&   &    && \sin\tilde{\iota}\cos\iota + \sin\iota \cos\Omega \cos\tilde{\iota} \bigl),   
\end{aligned}
\end{gather}
which are valid assuming $\iota' \in (0, \pi)\setminus \{\pi/2\}$. For $\iota' \in \{0,\pi/2,\pi\}$, $\beta'$ can be derived from $R'_{11}$ and $R'_{12}$ instead.

These equations become especially interesting in the case that the longitude of the ascending node changes in time $\dot{\Omega}\neq 0$, for example if the accretion disk interaction can change this angle. The signatures of this remain to be studied.

Of course, if the orbit is inside the fundamental plane, we have $\iota=0$ and the equations simplify to 
\begin{align}
          \cos{\iota'} = \cos{\tilde{\iota}} \\ 
          \beta' = \Omega + \omega 
\end{align}
and we can reabsorb the definition of $\Omega$ in $\omega$.

\backmatter

\bibliographystyle{JHEP}
\bibliography{biblio}{}

\end{document}